\documentclass[12pt,preprint]{aastex}

\usepackage{epsfig}

\def\Term#1 #2 #3/{\mbox{$\,^{#1}\!#2_{#3}$ }}
\def\Termo#1 #2 #3/{\mbox{$\,^{#1}\!#2^o_{#3}$ }}
\def\sterm #1 #2 #3/{\mbox{$\,_{#3}\!^{#1}\!#2$}}

\newcommand{\Ha}{H$\alpha$}

\newcommand{\Hb}{H$\beta$}

\newcommand{\sB}{S(\rm H$\beta$)}

\newcommand{\cmq}{cm$^{-3}$}
\newcommand{\cms}{cm$^{-2}$}
\newcommand{\pers}{s$^{-1}$}

\newcommand{\perarc}{arcsec$^{-2}$}
\newcommand{\Te}{$T_e$}
\newcommand{\Ne}{$N_e$}
\newcommand{\cHb}{c$\rm _{H\beta}$}
\newcommand{\thC}{$\rm \theta ^{1}$~Ori~C}

\begin{document}

\title{{\it Spitzer} reveals what's behind Orion's Bar}

\author{Robert H. Rubin\altaffilmark{1,2},
Janet P. Simpson\altaffilmark{1,3},
C. R. O'Dell\altaffilmark{4},
Ian A. McNabb\altaffilmark{5},
Sean W. J. Colgan\altaffilmark{1},
Scott Y. Zhuge\altaffilmark{1},
Gary J. Ferland\altaffilmark{6}
and
Sergio A. Hidalgo\altaffilmark{1}}

\email{robert.h.rubin@nasa.gov}

\altaffiltext{1} {NASA/Ames Research Center, Moffett Field, CA
94035-1000, USA}
\altaffiltext{2} {Orion Enterprises, M.S. 245-6, Moffett Field, CA
94035-1000, USA}
\altaffiltext{3} {SETI Institute, 515 N. Whisman Road, Mountain View, CA
94043, USA}
\altaffiltext{4} {Physics \& Astronomy Department, Vanderbilt University, 
Box 1807-B, Nashville, TN 37235, USA}
\altaffiltext{5} {Kavli Institute for Astronomy \& Astrophysics, 
Peking University, Beijing, China}
\altaffiltext{6} {University of Kentucky, Department of Physics \&
Astronomy, Lexington, KY 40506, USA}

\def\Termo#1 #2 #3/{\mbox{$\,^{#1}\!#2^o_{#3}$ }}

\date{\today}

\begin{abstract}
We present {\it Spitzer} Space Telescope observations of 11 regions 
southeast of the Bright Bar
in the Orion Nebula,
along a radial from 
the exciting star $\theta^1$~Ori~C, extending from 2.6 to 12.1$'$.
Our Cycle 5
programme obtained deep spectra with 
matching IRS short-high (SH)
and long-high (LH) aperture grid patterns.
Most previous IR missions observed only the inner few arcmin (the ``Huygens"
Region).  The extreme sensitivity of {\it Spitzer} in the 10-37~$\mu$m
spectral range permitted us to measure many lines of interest to
much larger distances from $\theta$$^1$~Ori~C.
Orion is the benchmark for studies of the interstellar medium,
particularly for elemental abundances.  {\it Spitzer} observations provide
a unique perspective on the neon and sulfur abundances
by virtue of observing the dominant ionization states of 
Ne (Ne$^+$, Ne$^{++}$) and S (S$^{++}$, S$^{3+}$) in Orion
and \ion{H}{2} regions in general.
The Ne/H abundance ratio is especially well determined, with a value of
$(1.01\pm{0.08})\times$10$^{-4}$ or
in terms of the conventional expression, 12~+ log~(Ne/H)~= 8.00$\pm$0.03.

We obtained corresponding new ground-based spectra at
Cerro Tololo Interamerican Observatory (CTIO). 
These optical data are used to estimate the electron temperature, 
electron density, optical extinction, and the S$^+$/S$^{++}$  
ionization ratio at each of our {\it Spitzer} positions.
That permits an adjustment for the total gas-phase sulfur
abundance because no S$^+$ line is observed by {\it Spitzer}.
The gas-phase S/H abundance ratio is 
$(7.68\pm0.30)\times10^{-6}$ or 12~+ log~(S/H)~= 6.89$\pm$0.02.
The Ne/S abundance ratio may be determined even when the weaker
hydrogen line, H(7--6) here, is not measured.
The mean value,  
adjusted for the optical S$^+$/S$^{++}$ ratio, is Ne/S = $13.0\pm{0.6}$.

 We derive the electron density ($N_e$) versus distance from 
$\theta$$^1$~Ori~C for [S~{\sc iii}] ({\it Spitzer}) and [S~{\sc ii}] (CTIO).
Both distributions are for the most part decreasing with increasing distance.
The values for $N_e$~[\ion{S}{2}] fall below those of $N_e$~[\ion{S}{3}] 
at a given distance except for the outermost position.
This general trend is consistent 
with the commonly accepted blister model for the Orion Nebula. 
The natural shape of such a blister is concave with an underlying
decrease in density with increasing distance from the source of
photoionization.

Our spectra are the deepest ever taken in these outer regions of Orion
over the 10-37~$\mu$m range.  Tracking the changes in ionization
structure via the line emission to larger distances provides much more
leverage for understanding the far less studied outer regions.
A dramatic find is the presence of high-ionization Ne$^{++}$ all the way
to the outer optical boundary $\sim$12$'$ from $\theta$$^1$~Ori~C.
This IR result is robust, whereas the optical evidence from observations 
of high-ionization species (e.g. O$^{++}$) at the outer optical boundary 
suffers uncertainty because of 
scattering of emission from the much brighter inner Huygens Region.
The {\it Spitzer} spectra are consistent with the Bright Bar being a
high-density `localized escarpment' in the larger Orion Nebula picture.
Hard ionizing photons reach most solid angles well SE of the Bright Bar.
The so-called Orion foreground `Veil', seen prominently in projection at
our outermost position 12$'$ from $\theta$$^1$~Ori~C, is likely 
an H~{\sc ii} region -- photo-dissociation region (PDR) interface.
The {\it Spitzer} spectra show very strong enhancements of PDR lines~-- 
[\ion{Si}{2}] 34.8~$\mu$m, 
[\ion{Fe}{2}] 26.0~$\mu$m,
and molecular hydrogen~-- at the outermost position.
\end{abstract}


\vspace{12pt}

\section{Introduction}

Most observational studies of the chemical evolution of the universe
rest on emission line objects.
\ion{H}{2} regions help elucidate the current mix of elemental abundances 
in the ISM.
They are laboratories for understanding physical 
processes in all emission-line sources and probes for stellar, 
galactic, and primordial nucleosynthesis. 
\ion{H}{2} regions are also among the best tracers of recent star formation.
The Orion Nebula (M42) is the benchmark for studies of the interstellar
medium (ISM), particularly as a gauge of elemental abundances.
In many ways this is similar to the role the Sun plays with respect to stars.
Because Orion is nearby and bright, it is one of the most observed nebulae.
Not surprisingly, most observations of Orion have been of the inner
bright region.
[Here we refer to this inner region as the classical ``Huygens Region".]
Detailed photoionization models, including our own (Baldwin~et~al.\ 1991;
Rubin~et~al.\ 1991a, b) as well as deep spectroscopic observations
interpreted via empirical analyses (Esteban~et~al.\ 2004; 
Baldwin~et~al.\ 2000) have concentrated on the Huygens Region.

The Bright Bar has been treated as the ``poster child"
H~{\sc ii}  region -- photo-dissociation region (PDR) interface.
The famous 3-colour image of the PDR 
(Tielens~et~al.\ 1993) demonstrated 
the  progressive separation of the 3.3~$\mu$m 
polycyclic aromatic hydrocarbon (PAH) feature (blue), 
H$_2$ 1-0 S(1) (green), and CO $J$~= 1-0 (red)
with increasing distance from $\theta^1$~Ori~C.
This was in good agreement with their theoretical model of
a plane-parallel slab for the Bright Bar.
Their result showed conclusively that the incident far-UV 
(non-ionizing) radiation field from  $\theta^1$~Ori~C
was responsible for this molecular structure in the Bright Bar.
Because their interest primarily concerned the structure, properties,
and observations of the Bright Bar PDR, they were not concerned with 
the emission that extends far beyond in the extended Orion
Nebula (obviously present, from any reasonably deep photograph).
With regard to the Huygens Region, one of our own papers 
derived a 3-dimensional model of the inner ionized region (Wen \& O'Dell 1995).
This work used detailed surface brightness images to delineate the 
3-dimensional position of the main ionization front
with increasing distance from the exciting  star $\theta^1$~Ori~C, 
and argued that the Bright Bar is almost perpendicular to the plane of the sky.

	With regard to the fainter extended outer nebula,
there has been progress in characterizing  the so-called
foreground ``Veil" with early {\it prima  facie} evidence
for its existence  stemming  from the
\ion{H}{1} 21-cm line absorption line work of van~der~Werf \& Goss (1989).
The Veil is seen in projection ($\sim$ edge-on) 
as the outer boundary of M42, the grayish colour extending from roughly
north counter clockwise to the southeast 
in the optical image shown here as Figure~5.
For a review of the structure of Orion, see O'Dell (2001) and references 
therein.
More recent studies of the Veil include  Abel~et~al.\ (2004 and 2006).

	Using the {\it Kuiper Airborne  Observatory (KAO)},
Simpson~et~al. (1986)
measured the [\ion{O}{3}] 51.8 and 88.4~$\mu$m lines at several positions
in Orion along a radial straight south from $\theta^1$~Ori~C,
extending as far as a position called P6 centred 3.75$'$
from $\theta^1$~Ori~C.
This did provide IR evidence of species as high ionization as
O$^{++}$ beyond the Bright Bar.
Except as noted, prior to {\it Spitzer}, high spectral resolution  
space-- or airborne--IR data have never 
extended to angular separations from $\theta^1$~Ori~C
that would place them  in the extended outer nebula.
To the best of our knowledge, the first such data 
exterior to the Bright Bar and the Huygens Region
were taken under the GTO~45 programme (PI: T.\ Roellig)
and the GO~1094 programme (PI: F.\ Kemper).
We did not examine the GTO~45  spectra, (``Orion Bar neutral"), 
a pair of short-high resolution (SH) and long-high resolution (LH)
aperture spectra centred close to and just SE of $\theta^2$~Ori~A,
taken in staring  mode.
Instead, we chose to examine a set of the GO~1094 paired SH and
LH aperture spectra  centred well SE of
the Bright Bar $\sim$3.4~arcmin from $\theta^1$~Ori~C.
These spectra were taken in staring mode with 
the minimum ramp (exposure) time of 6~s and total for each spectrum
of just 12~s.
As will be discussed later, the fields of view for the SH and LH were quite
different.
These spectra demonstrated that there were lines of 
high-ionization species ([\ion{Ne}{3}] and [\ion{S}{4}]) 
measurable with excellent signal-to-noise even beyond the PDR of
the Bright Bar.  We also determined that none of the emission lines
was saturated.
Those 24~s of data were an inspiration for us to propose using
{\it Spitzer} to probe even further from $\theta^1$~Ori~C.

{\it Spitzer} has a unique ability to address the
abundances of the elements neon and sulfur.
This is particularly true in the case of \ion{H}{2} regions,
where one can 
simultaneously observe four emission lines that
probe the dominant ionization states of Ne (Ne$^+$ and Ne$^{++}$)
and S (S$^{++}$ and S$^{3+}$).
The four lines, 
[\ion{Ne}{2}] 12.81,
[\ion{Ne}{3}] 15.56,
[\ion{S}{3}] 18.71, and [\ion{S}{4}] 10.51~$\mu$m
can be observed cospatially 
with the Infrared Spectrograph (IRS) on the 
{\it Spitzer}.
Because of the sensitivity of {\it Spitzer}, 
a special niche, relative to previous (and near-term foreseeable)
instruments, is for studies of fainter \ion{H}{2} regions.
Indeed many of the well-known Galactic \ion{H}{2} regions
would cause saturation problems if observed at their  brightest
positions. 
Because of this, prior to our Orion programme, we have used 
{\it Spitzer} to 
observe a number of
\ion{H}{2} regions in galaxies with 
various metallicities and other properties.
These studies were of the spiral galaxies  M83 (Rubin~et~al.\  2007,
hereafter R07),
M33 (Rubin~et~al.\  2008, hereafter R08),
and the dwarf irregular galaxy NGC~6822 (Rubin~et~al.\  2010,
hereafter R10).
To the extent that all the major forms of Ne and S
are observed, the true Ne/S abundance ratio could be inferred.
For Ne, this is a safe assumption, but for S, there is the possibility 
of non-negligible contributions due to S$^+$ as well as what could be 
tied up in dust.

	We have an ongoing interest to utilize this special capability
of {\it Spitzer} archival spectra to address the Ne/S abundance ratio.
Our current assessment of how much Ne/S may vary 
was discussed in Rubin~et~al.\ (2008), where we also included 
other {\it Spitzer} data, reanalyzed with a homogeneous atomic database.
In this paper, we make a careful assessment of the
Orion Nebula value for Ne/S.
This not only uses {\it Spitzer} measurements of the dominant
ionic species, but also new ground-based spectra that permit
an accounting for S$^+$, which {\it Spitzer} cannot do.
In the customary role of the Orion Nebula providing an important benchmark 
for the ISM, it is important to compare the 
Ne/S value with others, including
the uncertain and controversial
solar value  as well as what is
predicted  by nucleosynthesis, galactic chemical evolution (GCE) models.

The solar abundance, particularly of Ne,
remains the  subject of much controversy (e.g., Drake \& Testa 2005;
Bahcall, Serenelli, \& Basu 2006; and references therein).
The preponderance of evidence points to a Ne abundance
substantially higher in the solar neighborhood, and even in the Sun
itself, than the ``canonical" solar values,
Ne/S~$\sim$6.5 (Asplund~et~al. 2009).
While we cannot directly address the solar Ne value,
it is crucial to an understanding of nucleosynthesis and GCE
to have reliable benchmarks.
We made the  case that the solar Ne/S ratio is
`out of  line' with our {\it Spitzer}
\ion{H}{2}  region values (R07, R08, R10 and references therein).
{\it Note that the reason abundances are often derived as ratios
is to avoid absolute calibration problems.}
Previous to that,
Pottasch \& Bernard-Salas (2006)
discussed in their study of planetary nebulae  with 
the {\it Infrared Space Observatory (ISO)} that the solar neon abundance  
was likely too low.
They suggested that the planetary nebula neon abundance should be used instead.
Optical studies of planetary nebulae
and \ion{H}{2} regions have suggested an upward revision of the
solar  Ne/O ratio (Wang \& Liu 2008; Magrini, Stanghellini, \& Villaver 2009).
Recent observations of nearby B stars also suggest that the solar Ne/O
ratio should be higher (e.g., Morel \& Butler 2008).

With our new Orion data, we focus predominantly on neon, the fifth most 
abundant element in the Universe, and sulfur, one  of the top ten,
because of the specific capability that {\it Spitzer} provided.
Naturally, deriving abundances of other elements is
also important, but there  was no special ability to tackle
these with {\it Spitzer}.
Suffice it to say that to provide precision abundance measurements of
S and Ne is a major advance in basic data needed to understand
and test nucleosynthesis/GCE models.
While both S and Ne are `primary' $\alpha$-elements produced in
massive stars and released to the ISM in supernovae,
some differences in their  production and GCE may be expected.
$^{20}$Ne exists primarily in the C-burned shell of massive stars,
whereas
$^{32}$S arises during O-burning, probably explosively
(e.g., an interesting article with a useful cutaway schematic of the
fusion zones by  Clayton 2007).
According to figure~7 in the nucleosynthesis/GCE model of Woosley \& Heger (2007),
the Ne/S ratio is $\sim$8.6, when they start with the Lodders (2003)
solar abundances.

	We discuss the {\it Spitzer} observations in section 2. 
In section 3, our new ground-based spectra are presented.
In section 4, we discuss the variation 
of the electron density and three measures of the degree of ionization
with distance from the
exciting star.
Section 5 continues with the derivation of elemental abundance ratios:
Ne/S, Ne/H, S/H, and Fe/H.
In section 6, we present additional data in order to
characterize the Bright Bar and Outer Veil in 
the context of an overview of the entire Orion Nebula.
In section 7, there is additional discussion pertaining to the
major findings, including the Ne/S \& Ne/H ratios and the nature of
the Bright Bar and Outer Veil as an
\ion{H}{2} region -- PDR interface. 
Last, we provide a summary and conclusions in section~8.

\section{{\it Spitzer Space Telescope} Observations}

 We observed the outer Orion Nebula under our  
Cycle 5 {\it Spitzer Space Telescope} programme GO-50082.
The observations were  all southeast  of the famous bar, 
which we shall refer to as the Bright Bar (BB).
The fields chosen were centred along a radial 
outbound 
from
the exciting star $\theta$$^1$~Ori~C and approximately
orthogonal to  the BB (see Figure~1).
This radial coincides with our ``Slit~4", one of the slits defined
in our previous programme with {\it HST}/STIS long-slit spectra.
The SE tip passed through HH203  (Rubin et~al.\ 2003, colour fig.~1).
Our set of positions was selected to examine the far side of the Bright Bar.
There are 11 locations that 
start at 2.6$'$ and extend to 12.1$'$
from $\theta^1$~Ori~C (see Figure~1). 
In order of increasing distance (D) from $\theta^1$~Ori~C,
the positions
are called ``inner" (I4, I3, I2,  I1),
``middle" (M1, M2, M3, M4), 
and ``veil" (V1, V2, V3).
Table 1 lists the coordinates for the centres of the areas mapped and
the projected angular distance D.
We note that for the inner positions, the time-sequence  order 
of observations was indeed I1,  I2, I3, and I4. 
We chose that just in case the brightest I4 region might suffer
some saturation effect, which might then cause a latency problem
with the subsequent observation position.  Fortunately, we experienced no
saturation issues.

We obtained deep spectra with both the 
{\it Spitzer} Infrared Spectrograph (IRS) 
short wavelength, high dispersion (spectral resolution, R $\sim$ 600) 
configuration, called the short-high (SH) module
and the long wavelength, high dispersion (R $\sim$ 600) 
configuration, called the long-high (LH) module
(e.g.,  Houck~et~al.\ 2004).
These cover respectively the wavelength range 
from 9.9~-- 19.6~$\mu$m and from $\sim$19~-- $\sim$36~$\mu$m.
The SH slit size is 
4.7$''$ $\times$ 11.3$''$, while the LH is
11.1$''$ $\times$ 22.3$''$. 
The SH observations permit cospatial observations of 
five important emission lines:
 [\ion{S}{4}] 10.51,
hydrogen H(7--6) (Hu$\alpha$)  12.37,
[\ion{Ne}{2}] 12.81,
[\ion{Ne}{3}] 15.56,
and [\ion{S}{3}] 18.71~\micron.
The LH observations permit cospatial observations of 
several more important emission lines: 
[\ion{Fe}{3}] 22.93, 
[\ion{Fe}{2}] 25.99, 
[\ion{S}{3}] 33.48, 
[\ion{Si}{2}] 34.82~\micron. 
In order that we could use {\bf all} the emission lines observed
with both modules,
we made a concerted effort to match the field of view (FOV) for
the SH and LH modules.
However, a perfect match is not possible because the
SH and LH rectangular apertures are not exactly orthogonal
(84.8$^{\rm o}$).
With the ``mapping mode" for the IRS, we had the ability
to overlap apertures by offsetting in either the parallel direction
(along  the long-axis of the rectangular aperture) or
the perpendicular direction
(along  the short-axis of the aperture).
By selecting the following scheme, the resulting
SH and LH aperture grid patterns 
(henceforth {\it `chex'}, after the breakfast cereal) 
very closely match the same area in the nebula:
with SH, one displacement of 5$''$ parallel and 9 
displacements of 2.3$''$ perpendicular;
with LH, one displacement of 4.5$''$ parallel and one 
displacement of 4.5$''$ perpendicular.
We used the {\it Spitzer} software {\sc SPOT} to measure our chex size.
The SH is 25.4$''$$\times$16.3$''$ (area 414.0~arcsec$^2$) 
and the LH is 26.8$''$$\times$15.5$''$ (area 415.4~arcsec$^2$),
indeed a good match (see Figure~1). 
Another very important purpose of overlapping the apertures 
is that most spatial positions will be covered in several locations 
on the array, minimizing the effects of bad pixels.

To save overhead, we clustered our 11 positions into 5 
on-source
{\it Spitzer} Astronomical Observing Requests (AORs).
Because much more integration time was necessary to observe
the fainter veil positions (V1, V2, V3~-- all three included
in the {\it same} AOR), we needed to split these 
into 3 separate AORs, that were designated veil1, veil2, and veil3.
The other two AORs clustered all of the inner positions in one
and all  the middle positions in the other.
We did not control the scheduling of the AORs which were actually in 
the following time sequence (with the data set number, and total time in min.):
veil3 (25381120, 261.94);
middle (25381362, 276.97);
inner (25381376, 198.80);
veil2 (25380864, 261.96); and
veil1 (25380608, 326.37).
For various reasons, we {\bf changed the nomenclature} herein
for the three veil data sets~-- veil1, veil2, and veil3 refer to 
respective data sets 25380864, 25380608, and 25381120.
Throughout this paper Vx-y means chex x and AOR y.
For example, V3-1 means chex V3 and veil AOR 1 (data set 25380864).

The entire programme was executed between 2008 November 14 and
November 21 (UT), thereby causing very little sky-rotation of
the FOV.  Immediately adjacent in time to each on-source AOR, 
a background off-source AOR was taken.
These were all done at the same position~--
$\alpha$, $\delta$ = $5^{\rm h}32^{\rm m}36\fs5$, 
$-5^{\rm o}$17\arcmin47$''$ (J2000)~--
in ``staring mode", which utilizes a single aperture with
a shift along the long-slit axis  (parallel direction) of 1/3 the 
aperture dimension.
More time was used to observe those background observations
associated with the fainter regions.
Our choice of ramp (exposure) times and number of mapping 
cycles was as follows:
inner chex, SH 6~s, 8 cycles and LH 6~s, 8 cycles;
middle chex, SH 6~s, 12 cycles and LH 14~s, 6 cycles;
for all veil chex, SH 30~s and LH 14~s.
For both AORs veil2 and veil3, there were 5 and 11 cycles 
respectively for the SH and LH,
while more time was used in veil1 with 6 and 17 cycles 
respectively to fill up our {\it Spitzer} allotment.

	Our data were processed and calibrated 
with version S18.5 of the standard IRS pipeline at the 
{\it  Spitzer} Science Center.
To build our post-BCD (basic  calibrated data) data products,
we use {\sc cubism}, the CUbe Builder for IRS Spectral Mapping, 
(version 1.6) 
(Smith  et~al.\ (2007a, b and references therein).
{\sc cubism} was used to build maps, which account for aperture overlaps,
and to deal effectively with bad pixels.  
   From the IRS mapping observations, it can combine these data
into a single 3-dimensional cube with two spatial and one spectral dimension.
For each of our regions,
we constructed a data cube.
Global bad pixels (those occurring at the same pixel in every BCD) were
removed manually.  
Record level bad pixels (those occurring only within individual BCDs)~-- 
that deviated by 5~$\sigma$ from the median pixel value
and occurred within at least 10~per~cent of the BCDs~-- were removed 
automatically in 
{\sc cubism} with the ``Auto Bad Pixels" function.  
In reducing our data, we were careful to ensure that the 
``Auto Bad Pixels" function did not incorrectly flag any of the pixels 
on our programme spectral lines as bad.
Our Orion chex are in a fairly ``smooth" area, and as such,
it is more appropriate to reduce 
our data 
assuming each region is uniformly extended within the SH and LH apertures.
This is the default option and the one we used with {\sc cubism}.

   The fully processed background-subtracted  spectra that we use 
are presented in a colour montage showing all 11 chex in Figure 2 for
the SH and Figure 3 for the LH. For the veil chex, we show
only the longest-exposure spectra (the one formed from
data set 25380608) in order
not to clutter the figures.  These figures provide a useful overview of
the changes that occur at the varying distances from  the
exciting star.  The changes to the continuum levels and the
PAH features can also be seen.  
For instance, it is apparent that  the continuum 
intensity decreases 
with increasing distance from $\theta^1$~Ori~C from I4 through V1, 
but then increases from V1 to V3.
All of the spectral lines that we discuss in the paper are labeled in Figures 2 and 3.
There are some features that we do not measure or discuss that are also labeled.
These include the PAH bands and weaker lines such as H(8-7).
In addition, the very recently identified C$_{60}$ feature near 18.9~$\mu$m
(Cami~et~al.\ 2010)
is also marked.  They found this in the young planetary nebula Tc~1 and as they discuss,
a minor fraction  of this emission feature is due to  C$_{70}$ also.

	Our further analysis of these spectra used
the line-fitting routines in  the IRS Spectroscopy Modeling Analysis 
and Reduction Tool ({\sc smart}, Higdon et~al.\ 2004).
The emission lines were measured with {\sc smart} using a Gaussian line fit.
The continuum baseline was fit with a linear or
quadratic function.
Figures~4 (a)--(d) show the data and fits for several 
lines at chex V3 for one of the three veil AORs, the
one we call veil1 (using data set 25380864).
Most of our line measurements have higher signal-to-noise (S/N) than these.
We display this set to illustrate that lines from species as highly 
ionized as Ne$^{++}$ are clearly measurable all the way to the outer 
extended optical boundary.

A line is deemed to be detected if the intensity is  at least
as  large as the 3~$\sigma$ uncertainty.
We measure the uncertainty by the product of the full-width-half-maximum
(FWHM) and the
root-mean-square variations in the adjacent, line-free continuum;
it does not include systematic  effects.
The possible uncertainty in the absolute
flux calibration of the spectroscopic products delivered by the pipeline
is likely confined to between 5 and 10~per~cent
(see discussion on p.~1411 of R07).  
Any uncertainty in the flux due to pointing errors is probably small
and in the worst case should not exceed 10~per~cent.  
	For the brighter lines
the systematic uncertainty far exceeds the measured (statistical) uncertainty.
Even for the fainter lines, we estimate that the systematic uncertainty
exceeds the measured uncertainty.
In addition to the line intensity, the measured FWHM and
heliocentric radial velocities (V$_{helio}$)
are listed in Table~2.
Both the FWHM and V$_{helio}$ are useful in judging the
reliability of the line measurements. 
The FWHM is expected to be the instrumental width for all our lines.  
With a resolving power for the SH
and LH  modules of $\sim$600, our lines should have a
FWHM of roughly 500~km~s$^{-1}$.
The values for 
V$_{helio}$ 
should straddle the heliocentric systemic radial
velocity for M42.
For the Huygens Region, heliocentric velocities of the 
higher ionization lines are $\sim$+18~km~s$^{-1}$, 
those for the lower-ionization species near the main ionization
front are $\sim$+25~km~s$^{-1}$, 
while those for the PDR lines 
are $\sim$+28~km~s$^{-1}$
(O'Dell 2001).
Subject to the coarse spectral resolution with {\it Spitzer},
most of our measurements are in agreement with these expectations.

\section{Ground-based Observations}

The ground-based spectroscopy was performed with the Boller \&\ Chivens spectrograph
mounted on the 1.5 m telescope at the Cerro Tololo Interamerican Observatory 
on the nights of 
2008 November 18, 19, 22, 24  and 2009 December 9, 10, 13 (UT). 
Observations were made with a long
slit crossing at or near most of the positions measured with {\it Spitzer}.
The illuminated
portion of the 2.6\arcsec\ wide slit was 429\arcsec\ long in the 2008 observations and 345\arcsec\ during the
2009 observations. The slit was opened to greater than 5\arcsec\ width during observations of the photometric
reference stars Feige 15, Feige 25, and Hiltner 600, which was wide enough to include all of the wavelengths
measured over the limited range of zenith distances (25\arcdeg\ to 51\arcdeg) employed and the astronomical
seeing image size of no more than 1.0\arcsec. Feige 15 observations were made early each night at multiple
zenith distances in 2008 and multiple reference stars were observed once each night in 2009.
Photometrically clear conditions applied during all observations of the reference stars and the nebula.

All observations were made such that the first order of the grating was
employed with a chopping filter
(GG 385 in 2008 and GG 395 in 2009) that permitted measurement of the red 
end of the spectrum 
without contamination by signal from the overlapping second order.  
Each pixel of the 
Loral 1K CCD subtended 1.30\arcsec\ along the slit.
For the  400 lines/mm (blaze 8000~\AA) grating 58 observations on the first
three nights in 2008 (November 18, 19, 22), 
each pixel along the dispersion was about 2.2~\AA\  and the FWHM 
of the emission lines was about 6.7~\AA.  
The 300 lines/mm, blaze 4000 \AA\  grating 09  used on the night of 
2008 November 24 and for the 2009 observations gave a slightly higher 
wavelength range, had a scale of 2.9~\AA\  per pixel, and
FWHM~= 6.8~\AA.  
A position angle
(PA) of 134.6\arcdeg\ was used for observations centring the star JW~831
(Jones \&\ Walker 1988) and PA=59.9\arcdeg\  used for JW~873. On the third night  in 2008 the PA=90\arcdeg\
slit was placed 11.7\arcsec\ south of JW 887, while on the fourth night of 2008 the PA=90\arcdeg\ slit was
carefully displaced to the south from the brightest Trapezium star \thC\ 
distances of 120\arcsec,
150\arcsec, and 180\arcsec. 
During the 2009 observations, JW 887 was used for displacements to positions
V1 and M4,
and JW 975 was used for the displacement to V3. 
The location of the slits are shown in Figure 5.

Sky observations were made at
two locations selected to be well removed from nebular emission, these being
identified from wide field of view \Ha +[N~II] images of the region.  
The sky positions were
$\alpha$, $\delta$ = $5^{\rm h}26^{\rm m}03^{\rm s}$, 
$-0^{\rm o}$25\arcmin42$''$ 
and 
$5^{\rm h}28^{\rm m}19^{\rm s}$, 
$-7^{\rm o}$08\arcmin36$''$ 
(J2000)
and the measurements were indistinguishable from one another.
In 2008 on the first night of the JW 831 observation of a bright portion 
of the nebula,
sky observations
totaling 3600 seconds were made.
On the second night of the JW 873  observations,
sky observations totaling
2700 seconds were made.
On the third night of the JW 887 observations,
four sky observations totaling
3600 seconds were made, and on the fourth night of the observations 
displaced from $\theta$$^1$~Ori~C,
frequent observation sets of 2400 seconds were interleaved with the 
observations of the nebula. 
In 2009, 3600 seconds of sky observations were made on 
December 9 and 7200 seconds of sky observations on each of December 10 and 13. 
Observations of the twilight sky were made and used to determine the
illumination correction along the slit.

Where necessary, a series of exposure times were used since the strongest emission-lines entered the
non-linear portion of the CCD detector during the long exposures.  In all cases the exposures were made
in pairs, which were then used for correction of cosmic-ray tracks. For the JW~831 observations, twin
exposures of 60, 300, 600, and 1200 seconds were made.  For the JW~873 observations, twin exposures of
600 seconds and two twin exposures of 1800 seconds were made. For the JW~887 observations twin exposures
of 900 seconds were made. 
For the fourth night observations displaced south from 
$\theta$$^1$~Ori~C, 
exposure times
were 60 seconds for 120\arcsec, 120 seconds for 150\arcsec, and 150 seconds for 180\arcsec.
The total signal per pixel along the slit in the \Hb\ reference line ranged from 2200 to 7200
analog-digital-units (ADU) at a gain of 0.7 ADU per electron event for the shortest exposures in the
faintest to brightest regions sampled.  In the case of the V1, 
V3, and M4 observations in 2009, total
exposure times of 3900 seconds, 5700 seconds, and 3900 seconds were used.  
{\sc iraf}\footnote{{\sc iraf} is distributed by the 
National Optical Astronomy Observatories, which is operated by the
Association of Universities for Research in Astronomy, Inc.\ under cooperative agreement with the
National Science foundation.} tasks
were used to process and spectro-photometrically calibrate the observations.

Samples from along the slits that correspond to different {\it Spitzer}
observations were taken. The location 
of the sampled regions are also shown in Figure 5.  
The total intensity in each emission-line was measured 
by fitting each line 
with a Lorentzian line profile 
using the task `splot'.  
Features that were identified as a blend of emission from 
two or more ions, using the high spectral resolution results of  
Esteban et~al.\ (2004) as a guide, were 
not measured. All the 
measured line intensities were then normalized to \Hb. 
A representative spectrum is 
shown in Figure 6. 
Because of the wide range of intensities, this 
M4 position spectrum is shown as a logarithm of the intensity.

The effects of  interstellar extinction were removed by comparing the 
observed \Ha/\Hb\ flux ratio with the value of 2.89 
expected from recombination theory assuming
case~B, electron density ($N_e$)~= 1000~cm$^{-3}$, and 
electron temperature ($T_e$)~= 8500~K (Storey \& Hummer 1995),
and employing the 
recently determined reddening curve derived by Blagrave~ et~al.\ (2007) from 
the nebular \ion{He}{1} lines.  
Note that the predicted \Ha/\Hb\ flux ratio changes little with
$N_e$ and $T_e$ over our range of interest.
The results are expressed as the commonly used 
logarithmic extinction at \Hb\ (\cHb) and are given in Table 3. 
This table also gives the extinction 
corrected surface brightness of the sample in the \Hb\ line.  
Tables 4~-- 7 present  the observed (F$_{\lambda}$) and
extinction corrected (I$_{\lambda}$) line intensities relative
to \Hb\ for the 16 different spectral samples.
In the case of the southwest-most samples, the observed \Ha/\Hb\ ratios 
were less than theoretically expected. 
The theoretical \Ha/\Hb\ ratios vary only slowly with $T_e$ and matching 
the observations would require temperatures twice as high as those 
derived from heavy ion line ratios.
The dominance of higher temperatures in the \Ha\ and \Hb\ emitting 
regions is probably not the correct 
interpretation of these data because hydrogen recombination emission 
increases  with decreasing $T_e$.
Thus this emission should selectively come from any lower 
$T_e$ regions along the line of sight.

The explanation of these anomalously low \Ha/\Hb\ ratios probably 
lies with the fact that these 
regions have important components of the emission 
illuminated
from the much brighter part of the 
nebula that are being scattered by material along these outer lines of 
sight.  One knows from high spectral resolution studies (O'Dell 1992, 
Henney 1994, Henney 1998, O'Dell 2001) that even in the inner 
nebula, the dust component of the PDR 
beyond the main ionization front scatters 
several tens of per~cent of the emission and that the nebular 
continuum (Baldwin et~al. 1991) is much 
stronger than expected for an atomic continuum because of scattered 
light from the Trapezium stars.  
The anomalously low line ratio would indicate that the bluer \Hb\ line 
is scattered more efficiently 
than the \Ha\ line.  Since the effects of such scattering have not been 
modeled and there is a pattern 
of decreasing extinction in the direction of the anomalous line ratios,  
we have assumed that there is 
no extinction in those four samples.  This assumption and the uncertainties 
of the role of the scattered 
emission-line radiation probably introduce an uncertainty of the 
derived line ratios of about 10~per~cent.

Electron temperatures were determined 
from line ratios using the 
{\sc iraf-stsdas} task {\sc temden}
from the [\ion{N}{2}] ratio [I(6548) + I(6583)]/I(5755) 
and the [\ion{O}{3}] ratio [I(4959) + I(5007)]/I(4363). 
Electron densities 
were determined using the [\ion{S}{2}] I(6716)/I(6731) ratios
but {\it updating} the atomic data as discussed in the next section.
These combinations give the particularly useful advantage of sampling 
different regions along the line of sight. [\ion{S}{2}]  
emission will arise essentially at the main ionization front, [\ion{N}{2}]  
emission comes from a zone where hydrogen is ionized and helium is neutral, 
and the [\ion{O}{3}] emission comes from a zone where 
H is ionized and He is singly ionized (O'Dell 1998).  
The results of the calculations are presented in Table~8.  

\section{Variations with Distance from the Exciting Star}

\subsection{Variations in Electron Density}

The {\it Spitzer} data provide an excellent diagnostic  of electron
density ($N_e$) in the S$^{++}$ region
from the line flux ratio [\ion{S}{3}] 18.7/33.5~$\mu$m.
Likewise, the ground-based observations 
provide an excellent diagnostic of $N_e$ in the S$^+$ region
from the line flux ratio [\ion{S}{2}] 6716/6731~\AA.
Both of these diagnostic tools are very insensitive to $T_e$ 
(e.g., Rubin 1989).
For our analyses, we will use $T_e$~= 8000~K.
The optical spectra discussed in the last section permit an
assessment of $T_e$ [\ion{N}{2}] and $T_e$ [\ion{O}{3}] values
(see Table 8) from classical forbidden line ratios.
While these values for  $T_e$ are somewhat higher than the
8000~K adopted, we point  out a well-known bias.
That is,  both $T_e$[\ion{O}{3}] and $T_e$[\ion{N}{2}] derived from
the ratio of fluxes of `auroral' to `nebular' lines
are systematically higher than the so-called `$T_0$',
which is the ($N_e$$\times$$N_i$$\times$$T_e$)--weighted average, where  
$N_i$  is  the ion  density of  interest.
The amount of  this bias depends on the degree of $T_e$
variations in the observed volume
(see Peimbert 1967, and many forward references).
In our analyses, for $N_e$ now, and in later sections using the set 
of IR lines, it is more 
appropriate to be using a $T_e$ that is similar to $T_0$.
Because  of the insensitivity of the volume emissivities
to $T_e$, particularly when working with ratios for these IR lines,
our results depend very little on this $T_e$ choice. 

Figure~7 shows 
$N_e$~[\ion{S}{3}] and $N_e$~[\ion{S}{2}] versus D (the projected distance
in arcmin from $\theta^1$~Ori~C to the centre of the chex or optical
sample).
For [\ion{S}{3}], we use the  effective collision strengths
from Tayal \& Gupta (1999) and the transition probabilities 
(A-values) from the recent compilation 
``Critically Evaluated Atomic Transition Probabilities for Sulfur 
\ion{S}{1}~--  \ion{S}{15}"  
(Podobedova, Kelleher \& Wiese 2009).
The original source they cite is Froese Fischer, Tachiev \& Irimia (2006).
For [\ion{S}{2}], we use the  effective collision strengths
from Ramsbottom, Bell \& Stafford (1996) and the A-values 
from Podobedova et~al.\ (2009) with  the original source 
Irimia \& Froese Fischer (2005).

These two $N_e$ distributions provide a unique perspective
for the extended outer Orion Nebula.
Clearly the values for 
$N_e$~[\ion{S}{2}] fall below those of $N_e$~[\ion{S}{3}] 
at a given D except for the outermost regions, including V3.
For any given {\it Spitzer} chex or optical sample, we view a column 
along the line of sight with a rectangular cross section.  
Due to ionization stratification, S$^{++}$/S$^+$ will be  
selectively highest in the column near the minimal projected distance from 
$\theta^1$~Ori~C. 
Along this line-of-sight,
at distances on either side of  
the minimum impact parameter,
S$^{++}$/S$^+$ will be expected to be decreasing because the 
actual  3-D distance to $\theta^1$~Ori~C is larger.  
In this picture, there would not be a plane-parallel density profile 
but one that had a degree of concavity with respect to 
$\theta^1$~Ori~C and an approximately monotonically
decreasing density with increasing D from the exciting star.

	There are several other considerations.
A blister is not only the commonly accepted model for the Orion  
Nebula, it is also a natural configuration once a nebula enters the  
champagne-phase (e.g., Tenorio-Tagle 1979). 
Ionizing radiation leads to the creation of a dense  
PDR and an ionization stratified layer facing the dominant ionizing  
source ($\theta^1$~Ori~C). 
The natural shape of such a blister is concave,  
thus explaining the general form of the Huygens Region (Wen \& O'Dell 1995).
The factors that produce the concavity in the  
Huygens Region will also be at play further away as one gets beyond  
the perturbation of the Bright Bar.
In quasi-steady state, there would be a gas density drop going away from  
the PDR into the ionized layer.

When viewing [\ion{S}{2}] emission, we are seeing material that is 
for the most part very close to the H$^+$--H$^0$ ionization front.
Just interior to this H-ionization front is where sulfur  
transitions from S$^{++}$ to S$^+$.  
There is then the possibility that the bulk of the [\ion{S}{2}] emission 
arises from a region where there is only partial ionization of hydrogen.
Hence $N_e$ as measured by $N_e$~[\ion{S}{2}] 
would be lower than that obtained from 
$N_e$~[\ion{S}{3}] even though 
the {\it total} gas density could be higher (as the PDR is approached)
than the total gas density nearby, but closer to $\theta^1$~Ori~C. 

In order to explain why $N_e$~[\ion{S}{2}] exceeds $N_e$~[\ion{S}{3}]
at the outermost position V3, we offer the following.
As one views far enough away from $\theta^1$~Ori~C,
scattered light becomes more important. 
By comparing \Hb\ and the radio  continuum, O'Dell \& Goss (2009)
showed that in the outer Orion regions the dust in the PDR is not 
only scattering Trapezium optical starlight, but also scattering 
nebular emission line radiation produced in the much brighter Huygens Region.  
While this can be important for the [\ion{S}{2}] emission,
the infrared [\ion{S}{3}] emission will be far less affected by scattering.
The optical spectrum at V3 has a strong continuum, indicating 
substantial scattered optical light. 
This is likely why $N_e$~[\ion{S}{2}] is larger than 
$N_e$~[\ion{S}{3}] because the [\ion{S}{2}] 
flux is a mix of local (low $N_e$) emission and scattered light  
from the higher $N_e$ Huygens Region.

\subsection{Variations in Degree of Ionization}

     From the measured infrared intensities, we are able to estimate ionic 
abundance ratios for three elements in adjacent
ionic states: Ne$^{++}$/Ne$^+$, S$^{3+}$/S$^{++}$ and 
Fe$^{++}$/Fe$^+$.
Important advantages compared with optical studies 
of various other ionic ratios are: 
(1) the IR lines have a weak and similar $T_e$ 
dependence, while the collisionally-excited
optical lines vary exponentially with $T_e$
(e.g., Osterbrock \& Ferland 2006), and 
(2) the IR lines suffer far less from interstellar extinction and
scattering.
Indeed for our purposes,  the differential extinction correction
is negligible as the lines are relatively close in wavelength.
In our analysis, we deal with ionic abundance ratios
and therefore line intensity ratios.
In order to derive the 
ionic abundance ratios, we perform the usual semiempirical
analysis assuming a constant $T_e$ and $N_e$
to obtain the volume emissivities for the pertinent transitions.
We use the  atomic data described in Simpson et~al.\ (2004) and 
Simpson et~al.\ (2007) except
for the A-values for the  sulfur ionic species.
Earlier we discussed [\ion{S}{3}] and  [\ion{S}{2}].
We also use the A-values in Podobedova et~al.\ 2009 for
[\ion{S}{4}]. 
The original source they cite is `Froese Fischer 2002a, downloaded from
http://atoms.vuse.vanderbilt.edu/ on 2005 December 21'.
In addition, we use a different effective collision strength for 
the [\ion{Ne}{2}] line, as detailed in the next paragraph.

\subsubsection{Ne$^{++}$/Ne$^+$}

	We present both the variation of the observed flux ratio 
F(15.6)/F(12.8) and Ne$^{++}$/Ne$^+$ with D in Figure~8 using the values 
from Table~2 and Table~10, respectively.
Here and throughout, the error values represent the propagated intensity
measurement uncertainties and do not include the systematic uncertainties.
In this paper, we commence
to use the effective collision strengths 
for [\ion{Ne}{2}] of Griffin~et al.\ (2001).\footnote{The value at 8000~K 
is 0.310 from the more complete set of effective collision strengths, 
available on the Controlled Fusion Atomic Data Center Web Site at ORNL,
www-cfadc.phy.ornl.gov/data\_and\_codes.}
In our previous papers (R07, R08, and R10), we had used the values
from Saraph \& Tully (1994).
Compared to those, the Griffin~et al.\ values are approximately 
10~per~cent higher at the $T_e$'s  characteristic of \ion{H}{2} regions. 
The Griffin~et al.\ (2001) values appear to be the best available now
(as also judged by Witthoeft~et~al.\ 2007).
We continue to use the same effective collision strengths 
for  [\ion{Ne}{3}] (McLaughlin \& Bell 2000).

In our empirical derivation of ion ratios, as already discussed,
we use the derived $N_e$~[\ion{S}{3}] and $T_e$~= 8000~K throughout.
The F(15.6) decreases monotonically with D by almost
a factor of 700 from I4 to V3.
We note that F(12.8) is a monotonically decreasing relation as well
except for a rise at V2 of $\sim$30~per~cent compared with V1.
Even though Ne$^+$ is the dominant neon ion beyond the 
Bright Bar, the [\ion{Ne}{3}] 15.6 line is clearly  present all the 
way to the outer boundary (see Figure~4).
In fact, there is a very dramatic increase in the Ne$^{++}$/Ne$^+$ 
ratio for all three V3 observations by a factor of $\sim$4.8
over the three V2 observations.
The main reason for this jump is likely due to the large drop in
$N_e$~[\ion{S}{3}] by a factor of 3 from V2 to V3. 
Ionization equilibrium dictates that 
Ne$^{++}$/Ne$^+$~$\propto$ $N_e$$^{-1}$ all other things being equal.
Whether the rest of the decrease in the  neon ionization equilibrium
(factor of $\sim$4.8) is necessary to attribute to other causes
is difficult to determine.
We could speculate that there might be another source
of hard ionizing photons besides  $\theta^1$~Ori~C at this
outer boundary, perhaps even external to the Orion Nebula.

\subsubsection{S$^{3+}$/S$^{++}$}

	As for neon, we present both the variation with D of the 
observed flux ratio\break
F(10.5)/F(18.7) as well as the derived ionic ratio 
S$^{3+}$/S$^{++}$ (Figure~9).
Both [\ion{S}{4}] 10.5 and [\ion{S}{3}] 18.7 intensities decrease 
monotonically with D.
Clearly F(10.5) decreases more steeply than F(18.7) with increasing D.
The [\ion{S}{4}] 10.5 line was detected in just one of the three
V3 observations, V3-2.
As for the Ne$^{++}$/Ne$^+$ ratio,
the analysis shows that there is a similar dramatic increase in the 
S$^{3+}$/S$^{++}$ ratio for V3-2 by more than a factor of 5 over the 
V2 observations.
The reasons provided in the last subsection would have a bearing
for this ionic ratio as well.
Following Table~2, we show non-detections in the plots  as 3~$\sigma$
upper limits.

\subsubsection{Fe$^{++}$/Fe$^+$}

	By virtue of the simultaneous measurement 
of both [\ion{Fe}{3}] 22.9 and [\ion{Fe}{2}] 26.0 lines 
with the LH module, the line flux ratio covers
exactly the same sky area (as did ratios involving lines 
observed with the SH module). 
Here we present both the variation with D of the 
observed flux ratio F(22.9)/F(26.0) and the derived ionic ratio 
Fe$^{++}$/Fe$^+$ (Figure~10).
Both [\ion{Fe}{3}] 22.9 and [\ion{Fe}{2}] 26.0 intensities decrease 
with increasing D except that there is a dramatic increase in
F(26.0) at V2 by a factor of 2.2 compared to the intensity at V1.
An increase was also noted above for the [\ion{Ne}{2}] 12.8 line intensity. 
The [\ion{Fe}{3}] 22.9  line was not detected in any of the three
V3 observations and is treated as a 3~$\sigma$ upper limit  in the plot.
In Figure~10, the observed ratio F(22.9)/F(26.0) follows a
very different pattern with D than those seen in Figures 8 and 9
with the higher ionization line in the numerator and the lower
ionization line in the denominator.
The primary reason for this is that the 
[\ion{Fe}{2}] 26.0 line has a very substantial PDR contribution
(Kaufman~et~al.\ 2006),
because it arises from the second energy level
just 385~cm$^{-1}$ above ground 
(e.g., see  discussion on p.\ 1126 of Simpson et~al.\  2007).
Our analysis of the Fe$^{++}$/Fe$^+$ ratio {\it does not account}
for the PDR contribution to the [\ion{Fe}{2}] 26.0 line intensity.
We derive Fe$^+$ by assuming the 26.0 line intensity is excited 
by electron collisions only.
Even for this excitation route, we have not accounted for the 
PDR contribution, which occurs at the lower $T_e$~$\sim$500~K
for the upper (second) energy level.
Thus the Fe$^{++}$/Fe$^+$ ratios derived using our measured
[\ion{Fe}{2}] 26.0 line intensity must be {\bf lower limits}.

There is another [\ion{Fe}{2}] line 
$^4F_{7/2}$--$^4F_{9/2}$ at 17.936~$\mu$m
that has a purer \ion{H}{2} region origin.
This arises from a level  2430~cm$^{-1}$ above ground 
(characteristic temperature $\sim$3500~K).
Unfortunately, this is a weak line and at the SH spectral
resolution, blended with [\ion{P}{3}] 
$^2P_{3/2}$--$^2P_{1/2}$ at 17.885~$\mu$m (see Figure 2).
We are able to measure this [\ion{Fe}{2}] line only
at chex V1 and V2.
At V1 the [\ion{P}{3}] line is the brighter while at
V2 the [\ion{Fe}{2}] line becomes the brighter.
The Fe$^{++}$/Fe$^+$ ratio derived using this weak line
is also shown in Figure~10 as the star symbol
(red in the colour version). 
As expected, these few Fe$^{++}$/Fe$^+$ values are much higher than those
inferred using the 26.0~$\mu$m line and should be considered the
truer estimate of the Fe$^{++}$/Fe$^+$ ratio.

Figure~10 may hold some important clues about the behaviour
of the outer Orion regions.  Notable compared with the neon and
sulfur plots is the increase in both F(22.9)/F(26.0) and 
Fe$^{++}$/Fe$^+$ beginning from I2 to I1 
(between D~= 3.7~-- 4.4$'$).
While {\it both} F(22.9) and F(26.0) are decreasing with D
for all the inner and middle chex, between I2 and I1, the drop
in F(26.0) is much larger (factor  of 2.27) than that for F(22.9)
(factor of 1.25).
The lower F(22.9)/F(26.0) ratios at I4,  I3 and I2 may be due
to some residual influence of the Bright Bar contributing
significantly to F(26.0), although I2 is well removed from the BB.
Another factor that may contribute to the `inversion' in
F(22.9)/F(26.0) with D is the decrease in $N_e$.   
Again, ionization equilibrium would require that 
Fe$^{++}$/Fe$^+$~$\propto$ $N_e$$^{-1}$, all other things being equal.
Finally, another possibility that might contribute to the 
increased F(22.9)/F(26.0) ratio between I2 to I1
is the presence of [\ion{Fe}{4}].
In fact,  [\ion{Fe}{4}] is believed to be the most abundant
ion in the Orion Nebula  according  to detailed photoionization
models (Rubin et~al.\ 1991a, 1991b; Baldwin et~al.\ 1991).
The discovery of the [\ion{Fe}{4}] 2837~\AA\ line in Orion
(Rubin et~al.\ 1997) 
was used to estimate the iron abundance.
A more recent discussion may be found  in Rodr\'\i guez \& Rubin (2005).
If the transition from Fe$^{3+}$ to Fe$^{++}$ is occurring between
chex I2 and I1, this would help to explain the `inversion'.

\section{Determination of Elemental Abundance Ratios}

   In this section we derive several ratios of elemental abundances 
that may be addressed with our {\it Spitzer} data.
As stated earlier, we have been particularly interested in the Ne/S ratio
and have undertaken several studies to utilize the special
ability of {\it Spitzer} spectroscopy in this regard (R07, R08,  R10).
In this section, we first cover Ne/S.
Then we derive and discuss three measures of metallicity:
Ne/H, S/H and Fe/H.

\subsection{Neon to Sulfur abundance ratio}

	For \ion{H}{2} regions, using {\it Spitzer} data only,
the gas-phase Ne/S ratio may be approximated as 
(Ne$^+$ + Ne$^{++}$)/(S$^{++}$ + S$^{3+}$).
This includes the dominant ionization states of these two elements.
However this relation does not account for S$^+$,
which should be present at some level.
We may safely ignore the negligible contributions
of neutral Ne and S in the ionized region.
Figure~11 shows our approximation for Ne/S versus D.

Our ground-based observations, which cover 
[\ion{S}{2}] 6716, 6731~\AA\ and
[\ion{S}{3}] 6312~\AA\ cospatially, allow for a correction
to the {\it Spitzer}-data-only measurements.
In order to estimate the downward corrections that 
apply to the individual chex, we derive S$^+$/S$^{++}$ from
the above optical lines.
Because the position of the spectral long-slit sample extractions are
usually not the same as the chex and always a much smaller
area on the sky, we use the {\it optical sample closest to the
various chex}.  The volume emissivities used
in conjunction with the extinction-corrected intensities
for the [\ion{S}{2}] 6716, 6731 and [\ion{S}{3}] 6312 lines 
are those for $N_e$~[\ion{S}{2}] and 
$N_e$~[\ion{S}{3}] respectively; we continue to use $T_e$~= 8000~K
for both.
With these S$^+$/S$^{++}$ values, we correct the
{\it Spitzer}-data-only estimate to
obtain Ne/S~= (Ne$^+$ + Ne$^{++}$)/(S$^+$ + S$^{++}$ + S$^{3+}$).

The derived S$^+$/S$^{++}$ ratio is always less than 0.19
for any of the inner or middle chex.  
For the three sets of observations of the veil chex,
it is no higher than 0.44.
Thus S$^{++}$ remains the dominant S ion even in the outermost regions.
While we find a fairly constant Ne/S for the 8 chex comprising I4 -- M4,
Figure~11 indicates a steep increase in Ne/S with D in the veil positions.
We surmise that this may be due to a significant and increasing
amount of S
being tied up in dust grains.  It is a safe assumption that 
there will be negligible Ne in grains.
Thus  while the  gas-phase Ne/S ratio may indeed be larger
for these veil positions, the values presented in
Figure~11 must be considered {\it upper limits for the {\bf total}}
Ne/S abundance ratio.
Because of the likelihood that not all forms of a significant
amount of sulfur are accounted for in the veil positions,
our best estimate of the true Ne/S abundance ratio for the
Orion Nebula is obtained from the eight values,
corrected for S$^+$,  for the I4~-- M4 chex.  
The median value is 12.8. 
From the internal scatter amongst these 8 values,
we obtain a sample mean and variance of 13.01$\pm$0.64.
The uncorrected median for these same 8 chex is 15.0.

\subsection{Ne/H and S/H} 

By virtue of measuring the H(7--6) line in the same SH spectra
as the two neon and two sulfur lines, we are able to derive the
Ne/H and S/H abundances.
The H(7--6) line provides a measure of H$^+$ from recombination theory
(Storey \& Hummer 1995). 
There is a bit of a complication here because at Spitzer's spectral 
resolution, the H(7-6) line is blended with the H(11-8) line.
Their respective $\lambda$(vac)~= 12.371898 and 12.387168~$\mu$m.
In order to correct for the contribution of
the H(11-8) line, we use the relative intensity of H(11-8)/H(7-6)
from recombination theory (Storey \& Hummer 1995) assuming
case~B and $N_e$~= 500~cm$^{-3}$.
The  ratio H(11-8)/H(7-6)~= 0.122 and holds over 
our range of interest $N_e$~= 100~-- 1000~cm$^{-3}$ and $T_e$~= 8000~K.
Indeed, it is appropriate for $T_e$~= 10000~K and for case~A as well.

There is also the possible blending with the H(7-6) line by He(7-6),
that we do not account for in this paper, but now discuss
with regard to how this would affect our analysis of metallicity.
In an {\it ISO} short wavelength spectrometer (SWS)
IR spectrum of the inner Orion Nebula (within the Huygens Region),
the spectral resolution (R~$\sim$2000) permitted a separation of the H(5-4) 
from the strongest He(5-4) components (Rubin et~al.\ 1998).
They were then able to derive a robust He$^+$/H$^+$ ratio of 0.085$\pm$0.003 from
those H and He Br$\alpha$ transitions. 
In the present case, all the strongest fine-structure components of the He(7-6) 
transition remain blended with the H(7-6) line at the {\it Spitzer} spectral
resolution.
We have used the photoionization code {\sc cloudy} to predict the 
intensities of the He(7-6) lines relative to the H(7-6) line.
This has incorporated the physics described in  
Porter et al.\ (2005).
The estimate is made using a $T_e$ of 8500~K and $N_e$ of 1000~cm$^{-3}$
consistent with those used in this paper and case B recombination theory.
The strongest He(7-6) component is the combined triplet and singlet multiplet
$7i~^3I$ $\rightarrow$ $6h~^3H^o$ and $7i~^1I$ $\rightarrow$ $6h~^1H^o$
at 12.366519~$\mu$m.
Next strongest is the combined triplet and singlet multiplet
$7h~^3H^o$ $\rightarrow$ $6g~^3G$ and  $7h~^1H^o$ $\rightarrow$ $6g~^1G$
at 12.3657~$\mu$m.
This is followed by the combined triplet and singlet multiplet
$7g~^3G$ $\rightarrow$ $6f~^3F^o$ and $7g~^1G$ $\rightarrow$ $6f~^1F^o$ 
at 12.3618~$\mu$m.
Other multiplets that would also blend are weaker and not used for this
estimate.
If the appropriate He$^+$/H$^+$ value were  0.085 at the location of our chex,
then summing the above transitions for He(7-6) would result in an
expected flux ratio He(7-6)/H(7-6)~= 0.065.
In terms of the contribution of the He(7-6) components to the {\it entire}
observed blend [H(7-6)~+ H(11-8)~+ He(7-6)], it would be 0.055.
However, it is very unlikely that at our chex locations SE of the Bright Bar,
that He$^+$/H$^+$ is that large.
Because we are unable to estimate how much smaller the ratio
might be, we do not apply any correction to values for Ne/H and S/H
derived herein.
We may safely conclude that any upward adjustment to these metallicities
would be {\it no larger than a factor of 1.055} and likely only a few percent.
We note that all three He(7-6) components are on the
blue side of H(7-6) while H(11-8) is on the red side.
At the limited {\it Spitzer} spectral resolution, we see no systematic velocity shift
or increase in the H(7-6) FWHM with respect to the other lines measured in  
Table~2.

Figure~12 shows the Ne/H values.  These are the sum of the 
Ne$^+$/H$^+$ and Ne$^{++}$/H$^+$ ratios listed in Table~10 along
with the propagated uncertainties.
There appears to be little variation with position for all chex.
The H(7-6) line was not detected at V3, thus there are only lower
limits at this outermost position.
Following the same method as for the Ne/S ratio, 
utilizing just the innermost 8 chex, 
the median  value 
Ne/H~= 1.01$\times$10$^{-4}$;
the sample mean and variance yields
(0.99$\pm$0.07)$\times$10$^{-4}$.
If we also include the 6 independent measurements at V1 and V2,
the median becomes Ne/H~= 1.03$\times$10$^{-4}$,
while the sample of 14 mean and variance is
(1.01$\pm$0.08)$\times$10$^{-4}$.
In terms of the conventional expression,  
this is 12~+ log~(Ne/H)~= 8.00$\pm$0.03.

Figure~13 shows the S/H estimates from the {\it Spitzer} data.  
These are the sum of the S$^{++}$/H$^+$ and S$^{3+}$/H$^+$ ratios 
in Table~10 along with the propagated uncertainties.
There appears to be little variation with position until 
reaching the V2 position.
Once again we use the mean for the innermost 8 chex as the best value 
S/H~= 6.58$\times$10$^{-6}$.
The drop in the estimated S/H as indicated by all three independent
measurements at V2 is likely due to the onset of more sulfur being
tied up in grains.
For these 8 innermost chex, we again make a 
correction for S$^+$, unseen by {\it Spitzer}, by
using the S$^+$/S$^{++}$ ratios derived from the optical data here.
The best {\it corrected} 
S/H~= $(7.68\pm0.30)\times10^{-6}$ or
12~+ log~(S/H)~= 6.89$\pm$0.02.

Esteban et~al.\ (2004) made deep optical echelle spectra 
within the inner Huygens Region.
They used empirical methods to derive gas-phase elemental abundances.
According to their table~14, for collisionally-excited lines (CELs), 
they range from
12~+ log(Ne/H)~=  $7.78\pm0.07$ to $8.05\pm0.07$
(Ne/H~= 6.03$\times$10$^{-5}$ to 1.12$\times$10$^{-4}$)
depending on various ionization correction factors and
whether they assume no $T_e$ variations or a mean-square
$T_e$ variation factor, $t^2$  (Peimbert 1967) of 0.022,
respectively.
Similarly for sulfur, they found
12~ + log(S/H)~=  7.06$\pm$0.04 to 7.22$\pm$0.04
(S/H~= 1.15$\times$10$^{-5}$ to 1.66$\times$10$^{-5}$).

\subsection{Fe/H}

The discussion in section 4.2.3 is very relevant to
our derivation of the Fe/H abundances.
Figure~14 plots the Fe/H estimates from the {\it Spitzer} data.  
These are the sum of the Fe$^+$/H$^+$ and Fe$^{++}$/H$^+$ ratios 
in Table~10 along with the propagated uncertainties.
There appears to be little variation with position except
for the V3 position.
We stress that the Fe$^+$/H$^+$ ratios are derived from the
[\ion{Fe}{2}] 26~$\mu$m line, which as discussed no doubt has an
unknown significant PDR contribution.  Because of this, the
Fe$^+$/H$^+$ ratios are overestimated, causing the
Fe/H estimates for chex I4~-- V2  in Figure~14 to be deemed an 
{\it upper limit}.
While the surface brightness of the 
[\ion{Fe}{2}] 26~$\mu$m line is somewhat smaller at V3 compared
with V2, the derived Fe$^+$/H$^+$ ratios are much higher
because the H(7-6) line is not detected at V3.
The three separate V3 points are plotted as lower limits
because we use the 3~$\sigma$ upper limit for  the H(7-6) line.
Nevertheless, the same caveat applies here too, that is, we
have not accounted for any PDR contribution to the
26~$\mu$m line.  Hence, it is incorrect to conclude that
the gas-phase Fe/H abundance at V3 is as high as these
3 points indicate.
Subject to all the uncertainty, we follow the same method of 
using the median for the innermost 8 chex to estimate 
an upper limit for
the gas-phase (Fe$^+$~+ Fe$^{++}$)/H$^+$~= 1.39$\times$10$^{-6}$.
However, Fe$^{3+}$ has not been accounted for
and that would necessitate an {\it increase} in 
the estimate above for an assessment of the 
{\it total gas-phase} Fe/H.

Indeed, there is little that can be contributed in this paper
to the determination of the total or even the gas-phase Fe/H abundance.
As mentioned in section 4.2.3, there is  the uncertainty
of how much Fe$^{3+}$ there might be, which could be
particularly important for the inner chex positions.
Furthermore there have been a number of studies that
conclude iron must be substantially tied up in dust grains
even within the \ion{H}{2} region (e.g., Rodr{\'{\i}}guez 2002
and references therein).

From their deep optical echelle spectra within the inner Huygens Region,
Esteban et~al.\ (2004) used empirical methods to also derive 
the gas-phase Fe/H abundance ratio.
According to their table~14, they range from
12 + log(Fe/H)~=  5.86$\pm$0.10 to 6.23$\pm$0.08
(Fe/H~= 7.24$\times$10$^{-7}$ to 1.70$\times$10$^{-6}$)
depending on various ionization correction factors and
whether they assume no $T_e$ variations or a mean-square
$T_e$ variation factor, $t^2$  (Peimbert 1967) of 0.022,
respectively.

\section{Characterization of the Bright Bar 
and Outer Veil as an \ion{H}{2} region -- PDR interface}

While {\it  Spitzer} is an admirable machine for measuring both Ne and S
abundances in \ion{H}{2} regions, the neon abundances are determined
more reliably.
As previously mentioned, this is because with {\it Spitzer} observations alone,
we are neither accounting for S$^+$ nor  S  that  may  be
tied up  in dust.
Thus it is preferable here to ratio silicon (and other heavy elements) 
to neon because
neon is so well determined with both the 12.8 and 15.6~$\mu$m
lines well measured all the way to the extended Orion outer boundary at V3. 
We list the Si$^+$/(Ne$^+$ + Ne$^{++}$) ratio in Table~10
and show it  versus D in Figure~15. 
Our derivation of  the Si$^+$ abundance assumes that {\it all}
the [\ion{Si}{2}] 34.8~$\mu$m line emission arises within 
the ionized region and does not include the very significant
PDR contribution at much lower characteristic temperatures
(e.g., Kaufman~et~al.\ 2006).
This caveat is similar to what was discussed for 
the [\ion{Fe}{2}] 26~\micron ~line (see section 4.2.3).
Thus, the Si$^+$/Ne values here must be considered 
{\it upper limits}.
Figure~15 shows at first a monotonic decrease in this ratio moving 
outward from the Bright Bar from I4 to M1  
(D~=  2.6~-- 5.1$'$).
The ratio then increases with distance from V1 to V3 
(D~=  8.8~-- 12.1$'$)
with excellent
repeatability amongst  the 3 independent observations.
There is a dramatic increase at V3.

It is well established that the [\ion{Si}{2}] 34.8~$\mu$m line
in Orion predominantly arises in the PDR but also is 
produced in the ionized region
(Rubin, Dufour \& Walter 1993). 
It is possible that the drop in the estimated Si$^+$/Ne ratio from 
I4 to M1 
(D~=  2.6~-- 5.1$'$)
is due to a residual influence of the Bright Bar contributing
significantly to F(34.8), although this is a stretch for I1 and M1
given that they are far from the BB.
Nevertheless, there is a robust conclusion that we may draw here;
that the dramatic rise at V3 must be due to a very substantial
PDR 34.8~$\mu$m contribution.
This is a strong piece of evidence that V3 is viewing an 
\ion{H}{2} region -- PDR interface.
This picture is consistent with many of the other Figures
indicating a large change at V3.
In a manner similar to Figure~15, we also have plotted 
the (Fe$^+$ + Fe$^{++}$)/(Ne$^+$ + Ne$^{++}$) ratio versus D
(not included in this paper).
This shows a giant leap up at V3 even when we take Fe$^{++}$
as zero (recall it was not detected at V3).
We attribute this rise due to a very substantial
PDR 26.0~$\mu$m contribution.

   The set of measured hydrogen lines may also prove particularly
useful to disentangle emission arising in the  ionized 
\ion{H}{2} region and the PDR.
Figure~16 displays in four panels
the flux ratio versus D of the H(7--6) line, 
which arises in the \ion{H}{2} region, along with the three H$_2$ lines~--
H$_2$~S(2) 12.28, H$_2$~S(1) 17.04, and H$_2$~S(0) 28.22~$\mu$m~--
which arise in the PDR. 
Here we discuss panel~(a) only~-- the flux ratio
of the adjacent lines H(7--6) 12.4/H$_2$~S(2) 12.3.
The intensity of the H(7--6) line falls monotonically with increasing D,
except for an increase at V2 compared with V1.
For all three observations at V3, H(7--6) was not detected
(see the upper limits in Table~2), which indicate that it is
faintest by far at V3.
The\break
H(7--6)/H$_2$~S(2) flux ratio shows an increase at I1, M2, and M3
compared to adjacent chex.
This is somewhat reminiscent of the behaviour of the F(22.9)/F(26.0)
ratio (see Fig.~10),
where we raised the possibility that  the lower F(22.9)/F(26.0) ratios 
at I4,  I3 and I2 might be due 
to some residual influence of the Bright Bar contributing
significant PDR F(26.0) emission.
In the case of Figure~16~(a),
the H(7--6)/H$_2$~S(2) flux ratio would be lower because
of the Bright Bar PDR still enhancing  the H$_2$ lines.
However, the ratio at M1 does not fit the pattern.
More definitively, the upper limit to the flux
ratio at V3 does comport with the other evidence
that there is a very substantial PDR line contribution at V3.
Indeed the H$_2$~S(2) and H$_2$~S(1) lines have become brighter
with increasing D from V1 to V3, and at V3 are brighter than at I1
and almost as bright as at M1 (see Table~2).
This is yet another strong piece of evidence that V3 is 
indeed sampling an \ion{H}{2} region -- PDR interface.
While it is beyond the scope of this paper,
we do note that this set of 
{\it Spitzer} data should provide a  means to 
compare, test, and interpret with a detailed photoionization modeling  
effort that treats both the \ion{H}{2} region and the PDR.

\section{Discussion}

    After a 2009 conference talk on the subject of this paper,
one of the leading experts on PDR modeling, and the Orion Nebula 
Bright Bar (BB) specifically, told RR that he/she
was surprised to 
hear that there were lines of high-ionization species beyond (SE of)
the BB.  This individual thought that the BB quenched all 
ionizing radiation.
After all, there is a definite transition from the
ionized \ion{H}{2} region to the PDR at the BB~--
per the famous 3-colour image of the PDR by Tielens~et~al.\ (1993) 
mentioned earlier.
We posit that the reconciliation of that view with the 
observations/analysis/results here supplies important
information regarding the BB.
As generally believed, the BB may be treated as a $\sim$~plane-parallel
slab, viewed nearly edge-on to the line of sight.
This slab is at much higher density than the adjacent material 
within the \ion{H}{2} region (NW of the BB, that is, the side
closer to $\theta$$^1$~Ori~C).
The amount of matter at these higher densities within the slab is sufficient
to soak up all the ionizing ($\geq$13.6~eV) photons, causing the PDR.
Our {\it Spitzer} results demand a scenario in which copious ionizing 
photons penetrate to {\bf much larger distances} SE of the BB.
A simple and reasonable explanation is that the slab representing the BB
is a \underbar{{\it localized escarpment}} within the confines of the 
larger Orion Nebula picture.

In this picture, the BB slab will quench the ionizing photons
emanating from $\theta$$^1$~Ori~C over a very limited solid angle.
There will then be foreground and background emission along
sight lines to the BB that is not produced in the BB.  
Because of the high density within the slab,
the contribution to the emission measure through the (edge-on) length 
of the BB will be by far the majority of the 
emission measure integrated over the entire line-of-sight column.
Hence this foreground and background emission, including spectral lines
of higher-ionization species,  not generated within the BB will 
be dwarfed by the emission produced within the BB.
Once the line of sight is clear of the dominating influence of
the BB, the character of this harder spectrum can be seen SE of the bar.
It would be expected that the BB will create a shadow-zone volume
that is devoid of direct ionizing photons from $\theta$$^1$~Ori~C,
but again over a limited solid angle.
There is the possibility that the BB is clumpy and/or has holes, allowing
radiation to penetrate to the `shadowed' side.
However this appears to be ruled out by the observations and modeling of the
BB (Tauber~et~al.\ 1994; Tielens~et~al.\ 1993).

	There was previous IR evidence of species as high ionization as
O$^{++}$ beyond the BB from {\it KAO} observations (Simpson~et~al.\ 1986).
Without question, there is abundant evidence from optical observations
beyond the BB
of line emission from O$^{++}$, as well as other ionic species
found in \ion{H}{2} regions.
Indeed, one need look no further than the optical spectra
presented here in Tables 4--7.
Even at the most distant position V3, lines are measured from the 
following higher-ionization species (along
with the ionization potential to create the ion):
\ion{He}{1} (24.6~eV), 
[\ion{Ar}{3}] (27.6~eV), 
[\ion{O}{3}] (35.1~eV), 
and [\ion{Ne}{3}] (41.0~eV).
The problem with interpreting these optical observations
is due to the fact that much of the emission may be 
photons scattered from the much brighter inner Huygens Region
(O'Dell 2001; O'Dell \& Goss 2009).
Because scattering is wavelength dependent, it is unknown how much 
of the observed optical line emission is produced in situ and how much
is the scattered component.

The mid-IR {\it Spitzer} lines suffer far less from
scattering than do the optical lines,
providing another inherent advantage when interpreting
them in terms of nebular properties, including abundances.
As mentioned in \S4.2, the other advantages, compared with the optical, 
are that they are far less sensitive to $T_e$ and 
fluctuations in $T_e$ ($t^2$) and suffer far less from extinction.
Because of these important advantages, together with the 
ability of {\it Spitzer} to measure all the pertinent neon species
along with the H(7--6) line in the same spectra, and the
fact that Ne will not be incorporated in grains and molecules,
the Orion Nebula Ne/H~= (1.01$\pm$0.08)$\times$10$^{-4}$ 
(12~+ log~(Ne/H)~= 8.00$\pm$0.03) is one of the most robust determinations
of {\it total} metallicity for any element in any \ion{H}{2} region.
It is somewhat ironic that while Ne/H is the
poorest determined amongst the most abundant elements in the Sun,
it is (arguably) the best determined heavy element abundance ratio
in Orion~-- a worthy benchmark standard.

There have been more estimates of the gas-phase Ne/S abundance ratio
using {\it Spitzer} data than Ne/H due to the weakness of the
H(7--6) line relative the Ne and S lines used.
We reviewed the situation with regard to Ne/S in R08 (see figures
11 and 12 in that paper).
The value we determine here $13.0\pm{0.6}$
is in reasonable accord with 
those found in R08 for the higher ionization regions.
However, all of the results in R08 used a different 
effective collision strength for [\ion{Ne}{2}] 
as discussed earlier.
Our transition to using Griffin~et al.\ (2001)
instead of Saraph \& Tully (1994) values
will result in a downward revision to Ne/S in the R08 estimates
by as much as 10~per~cent for the lower ionization \ion{H}{2} regions,
but a smaller change for those at higher ionization.
We defer a reanalysis of the results in R08 to a later paper
in which we will also present our {\it Spitzer} observations of a 
number of \ion{H}{2} regions in the dwarf irregular galaxy NGC~6822.

\section{Summary and conclusions}

   We obtained {\it Spitzer} IRS observations at 11 positions in
the Orion Nebula all southeast of the Bright Bar and extending 
in a straight line to more than 12$'$ from the exciting star 
$\theta$$^1$~Ori~C.
These spectra were taken with both the short-high (SH)
and long-high (LH) modules using an aperture grid patterns chosen to
very closely match the same area in the nebula. 
In addition, we have made new ground-based, long-slit spectra 
that correspond closely with the 11 regions observed with {\it Spitzer}.
Orion is the benchmark for studies of the interstellar medium,
particularly for elemental abundances.  
With these data, we focus predominantly on neon, the fifth most 
abundant element in the Universe, and sulfur, one  of the top ten,
because of the specific capability that {\it Spitzer} provided.
Our major points are enumerated below.

\noindent
(1) The Ne/H abundance ratio is especially well determined, with a value of
$(1.01\pm{0.08})\times$10$^{-4}.$
In terms of the conventional expression,  
this is 12~+ log~(Ne/H)~= 8.00$\pm$0.03.
This may well be the {\it gold standard} for a determination
of metallicity in an \ion{H}{2} region.

\noindent
(2) We estimate the Ne/S gas-phase abundance ratio by
observing the dominant ionization states of 
Ne (Ne$^+$, Ne$^{++}$) and S (S$^{++}$, S$^{3+}$) with {\it Spitzer}.
The optical data are used
to correct our {\it Spitzer}-derived Ne/S ratio for S$^+$, 
which is not observed with {\it Spitzer}.
Excluding all three outermost `Veil' positions,
we find the median value adjusted for the optical 
S$^+$/S$^{++}$ ratio is Ne/S = 12.8.
From the internal scatter amongst these 8 values,
we obtain a sample mean and variance of 13.01$\pm$0.64.

\noindent
(3) A dramatic find is the presence of species as high-ionization 
as Ne$^{++}$ all the way to the outer optical boundary $\sim$12$'$ 
from $\theta$$^1$~Ori~C.
At these locations beyond the Bright Bar, where the transition from ionized
to photo-dissociation region lines is purported to be complete,
it was somewhat surprising to find the high ionization lines
of [\ion{S}{4}] 10.51 and [\ion{Ne}{3}] 15.56~$\mu$m
present with excellent signal-to-noise (S/N) ratios.
A likely possibility is that the Bright Bar is an escarpment
that is quenching the ionizing radiation from
$\theta^1$~Ori~C over a {\it localized solid angle}.
As usually characterized, the Bright Bar is seen nearly edge-on.
The depth along the line of sight is not known.
Thus there can be copious ionizing radiation
in the foreground (and the background)
that does not encounter the Bar at all.
Such a scenario very much modifies a common viewpoint of the Nebula
in the SE quadrant.
This picture of the ionized \ion{H}{2} region continuing
SE of the Bar is further supported by our long-slit
spectra that sample all the chex.
From these we infer $T_e$ values at least as high as
8300~K from the familiar diagnostic
line intensity ratios, [\ion{N}{2}] 6584/5755~\AA\ and
[\ion{O}{3}] 5007/4363~\AA\ --
values that are typical for the ionized \ion{H}{2} region, not PDRs.
Likewise, our estimate for the fractional ionic abundance for
S$^+$ is significantly smaller than that for S$^{++}$.

This IR result is robust, whereas the optical evidence from 
observation of high-ionization (e.g. O$^{++}$) at the outer optical 
boundary suffers uncertainty because of the possible scattering of 
emission from the much brighter inner Huygens Region.
The {\it Spitzer} spectra are consistent with the Bright Bar being a
high-density {\bf `localized escarpment'} in the larger Orion Nebula picture.
Hard ionizing photons reach most solid angles well SE of the Bright Bar.

\noindent
(4) The {\it Spitzer} data provide an excellent diagnostic of electron
density in the S$^{++}$ region from the line flux ratio 
[\ion{S}{3}] 18.7/33.5~$\mu$m.
Likewise, the ground-based observations 
provide an excellent diagnostic of $N_e$ in the S$^+$ region
from the line flux ratio [\ion{S}{2}] 6716/6731~\AA.
From these, we derive the electron density versus distance from 
$\theta$$^1$~Ori~C (see Figure 7).
These two $N_e$ distributions provide a unique perspective
for the extended outer Orion Nebula, with the values for 
$N_e$~[\ion{S}{2}]~$<$ $N_e$~[\ion{S}{3}] 
at a given distance except for the outermost region V3.
The fact that $N_e$~[\ion{S}{2}] is lower than $N_e$~[\ion{S}{3}] 
for the most part is expected, as explained in \S4.1,
where reasons for the behaviour in the outermost region
are also offered.

\noindent
(5)  The {\it Spitzer} data provide substantial evidence
that at chex V3, the observations are 
sampling an \ion{H}{2} region -- PDR interface.
This should not be unexpected since visually this
appears to be the outer boundary of the Orion Nebula in this
direction.
As mentioned in the introduction,
it is also the position of the ``Veil" seen 
in projection (essentially edge-on) along our observed
radial from $\theta$$^1$~Ori~C. 
As described in O'Dell (2001), 
early evidence for this foreground ``Veil" 
stemmed from \ion{H}{1} 21-cm line absorption 
line observations (van~der~Werf \& Goss 1989).
The Veil is seen in projection 
as the outer boundary of M42, the grayish colour extending from roughly
north counter clockwise to the southeast (see Figure~5).
In a very recent paper (O'Dell \& Harris 2010), the case is made that
the more likely picture is the following.
Instead of the foreground Veil curving back away from the observer to
be seen edge-on near V3, it is the background \ion{H}{2} region~ -- PDR boundary
that is curving up toward the observer.
In this view, they suggest the word ``Rim" to define this
feature.  As such, our position V3 is then sampling the ``Rim  wall" in this
particular radial direction from $\theta$$^1$~Ori~C.
This difference in perception and nomenclature does
not alter the conclusions of the present paper.
The following {\it Spitzer} data support the inference that
at V3, we are indeed sampling an \ion{H}{2} region -- PDR interface.

From the plot of the Si$^+$/Ne versus D (Figure~15), derived using
the [\ion{Si}{2}] 34.8~$\mu$m line, 
there is a dramatic increase in this ratio at the outermost V3 position.
As detailed in \S6, our estimate of Si$^+$/Ne assumes {\it all}
of the 34.8~$\mu$m emission arises in the ionized region 
and does not account for an unknown PDR contribution.
The large increase at the outermost V3 position is strong evidence
that the bulk of the [\ion{Si}{2}] 34.8 emission arises in a PDR
at this \ion{H}{2} region~ -- PDR boundary.
In a manner similar to Figure~15, we also plotted 
the (Fe$^+$ + Fe$^{++}$)/(Ne$^+$ + Ne$^{++}$) ratio versus D
(not included in this paper).
This shows a giant leap up at V3 even when we take Fe$^{++}$
as zero (recall it was not detected at V3).
We attribute this rise due to a very substantial
PDR 26.0~$\mu$m contribution.
For all three observations at V3, H(7--6) was not detected,
an indication that by far it is faintest at V3.
On the other hand, the H$_2$~S(2) and H$_2$~S(1) lines,
with an origin only in the PDR, have become brighter
with increasing D from V1~-- V3 and at V3 are brighter than at I1
and almost as bright as at M1.

\begin{acknowledgments}
This work is based on observations made with the 
{\it Spitzer Space Telescope},
which is operated by the Jet Propulsion Laboratory, California Institute
of Technology under NASA contract 1407.  Support for this work was provided
by NASA for this {\it Spitzer} programme identification 50082.
In addition to the {\it Spitzer} support, CRO was supported in part by 
{\it HST} grant AR 10967. 
GJF gratefully acknowledges support by NSF (0607028 and 0908877) and NASA (07-ATFP07-0124).
We thank Don Clayton and Stan Woosley for providing information 
on the Ne/S ratio from a nucleosynthesis,
galactic chemical evolution perspective.
We are grateful for the help of our students~-- David Ng, Tim Craven, 
Savannah Lodge-Scharff, Evan Gitterman, Chris Lo, and Atish Agarwala~-- 
with various stages of this work.
We thank the referee for valuable comments.
\end{acknowledgments}

\begin{figure}
\centering
\resizebox{15.8cm}{!}{\includegraphics{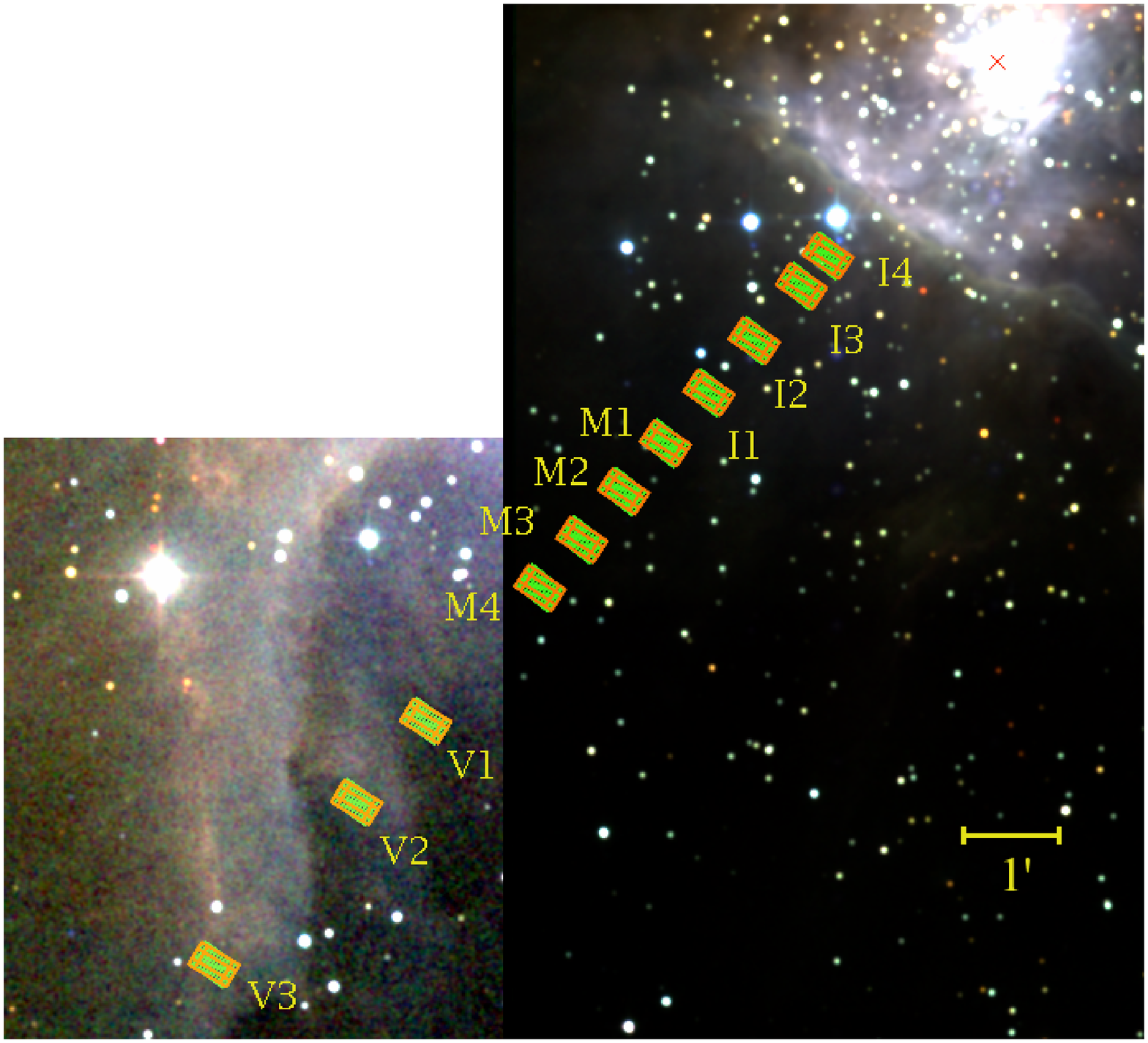}}
\vskip0.1truein
\caption{This shows our 11 observed {\it Spitzer} positions
for the Orion Nebula (M42, NGC~1976)
overlaid on a composite 2MASS image with
H-~(blue), J-~(green), and K-band (red).
The aperture mapping (or grid) patterns 
(that we call ``chex", after the breakfast cereal)
for the short high (SH)
and long high (LH) modules are shown in green and orange respectively.
These are labelled as defined in Table~1.
The SH and LH  {\it individual}
aperture sizes are respectively  4.7$''$ $\times$ 11.3$''$
and 11.1$''$ $\times$ 22.3$''$
with the orientations roughly orthogonal.
The Trapezium is at the top right with the dominant
ionizing star $\theta^1$~Ori~C marked with a red X.
For reference, the star $\theta^2$~Ori~A is just N of our NW-most
aperture cluster.
N is up, E to the left.}
\end{figure}

\begin{figure}
\centering
\resizebox{14.0cm}{!}{\includegraphics{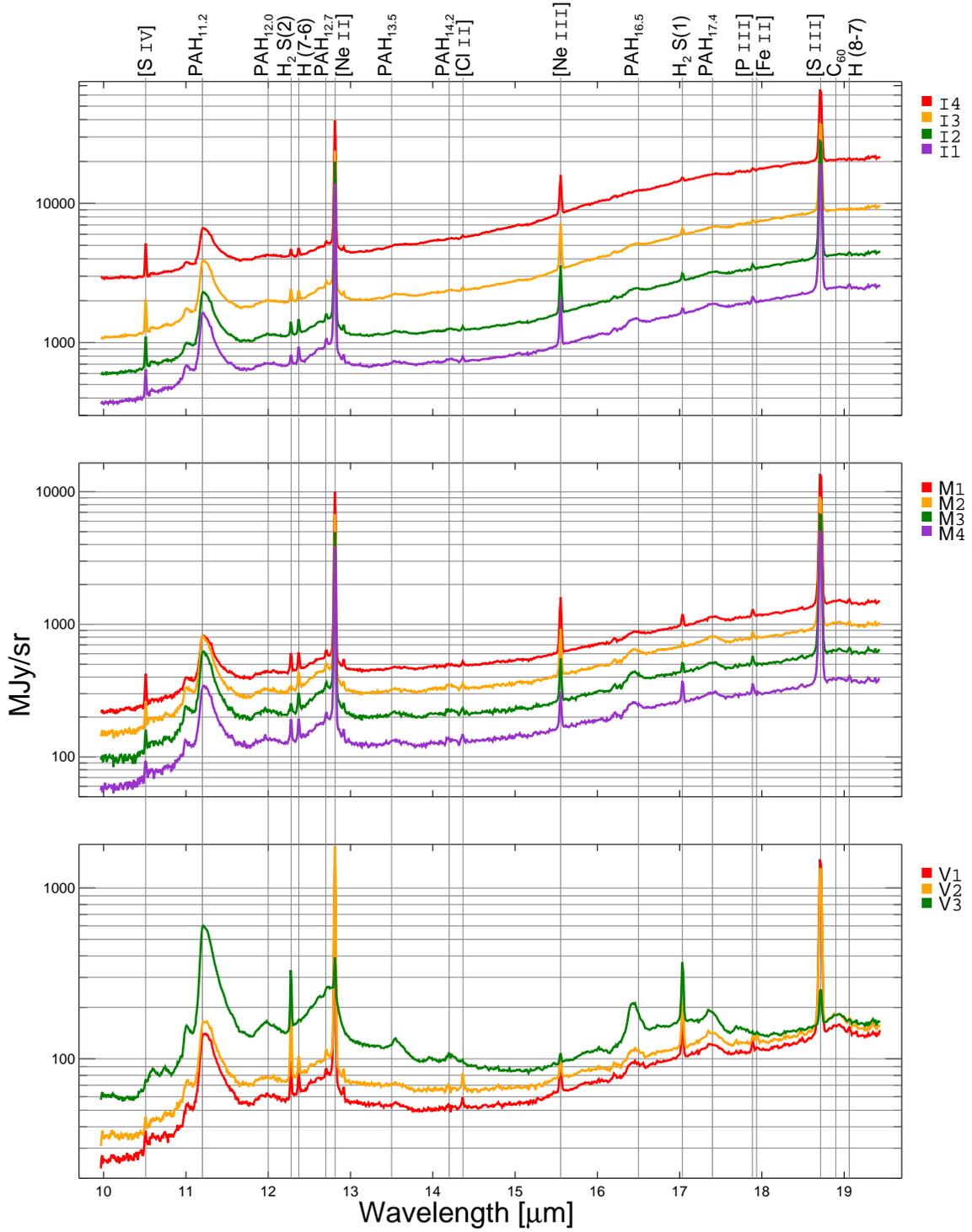}}
\vskip0.1truein
\caption{{\it  Spitzer} short-high (SH) full spectra of all 11 chex.
This composite of surface brightness (intensity) versus wavelength
has labelled the dominant features. Vertical fiducial lines guide how
the features vary with distance from the exciting star $\theta^1$~Ori~C.}
\end{figure}

\begin{figure}
\centering
\resizebox{14.0cm}{!}{\includegraphics{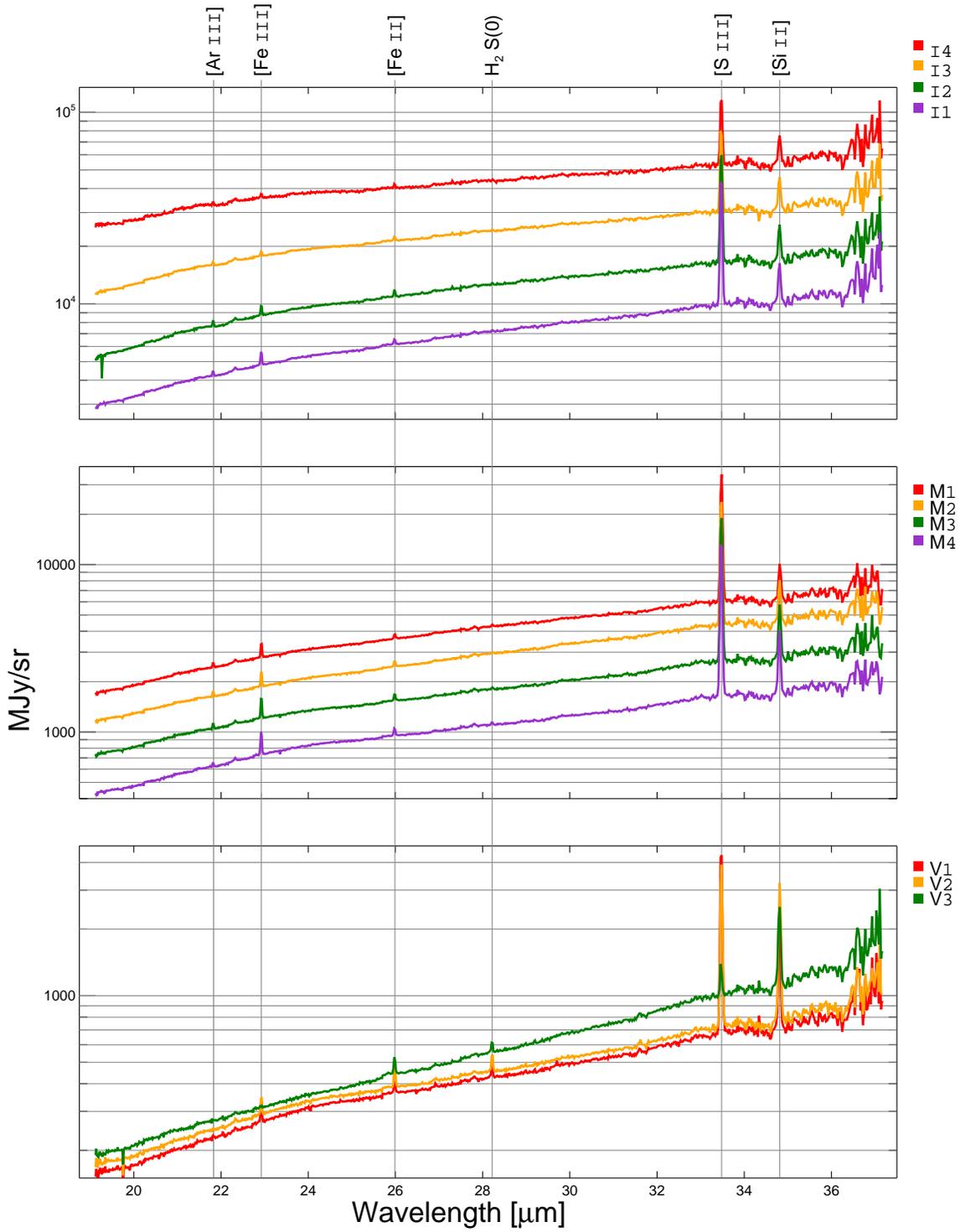}}
\vskip0.1truein
\caption{The same as Figure 2 for the set of
long-high (LH) full spectra of all 11 chex.}
\end{figure}

\begin{figure}
  \begin{center}
\vskip-0.25truein
    \begin{tabular}{cc}
\resizebox{70mm}{!}{\includegraphics{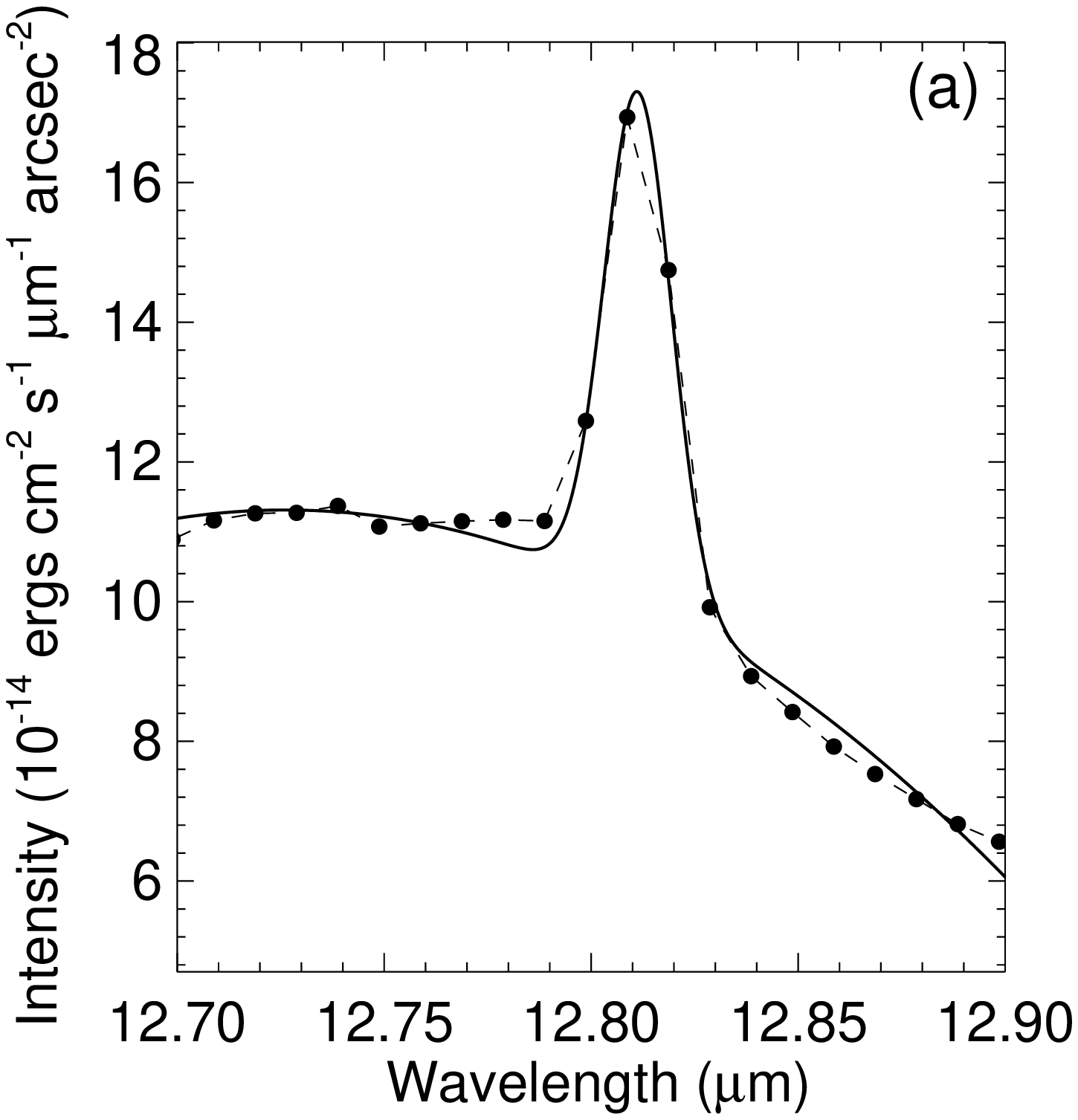}} &
\resizebox{70mm}{!}{\includegraphics{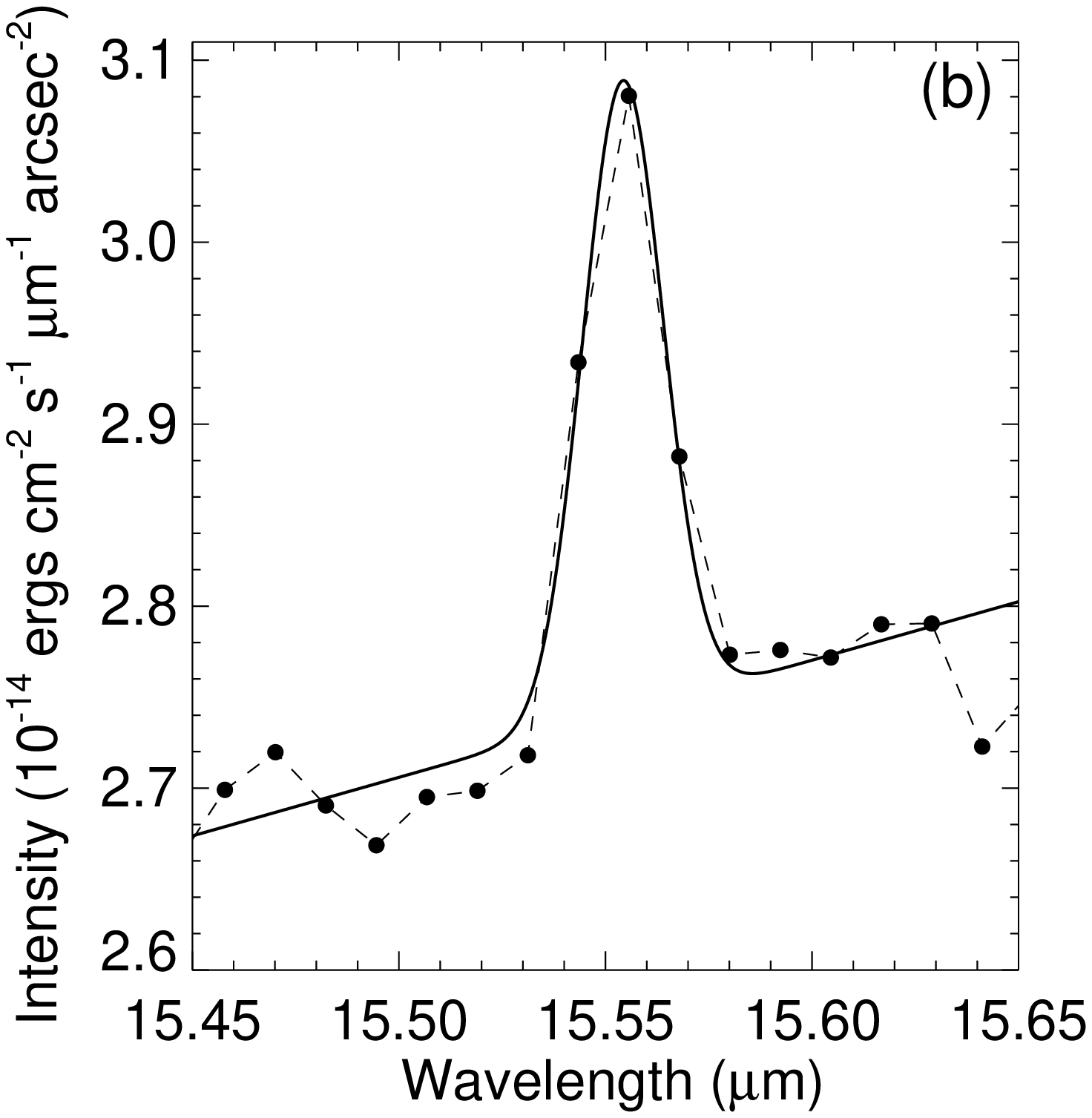}} \\
\resizebox{70mm}{!}{\includegraphics{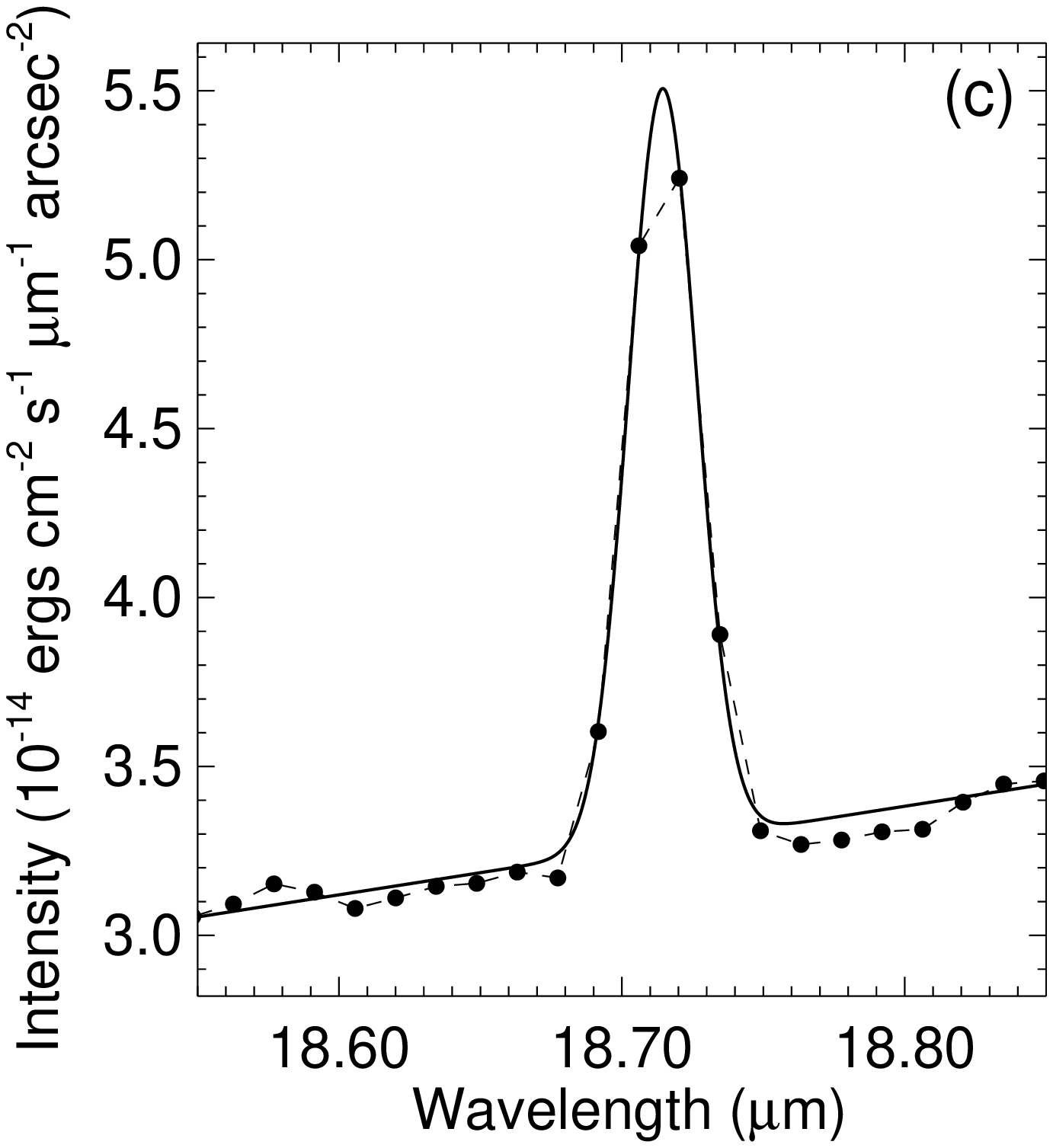}} &
\resizebox{70mm}{!}{\includegraphics{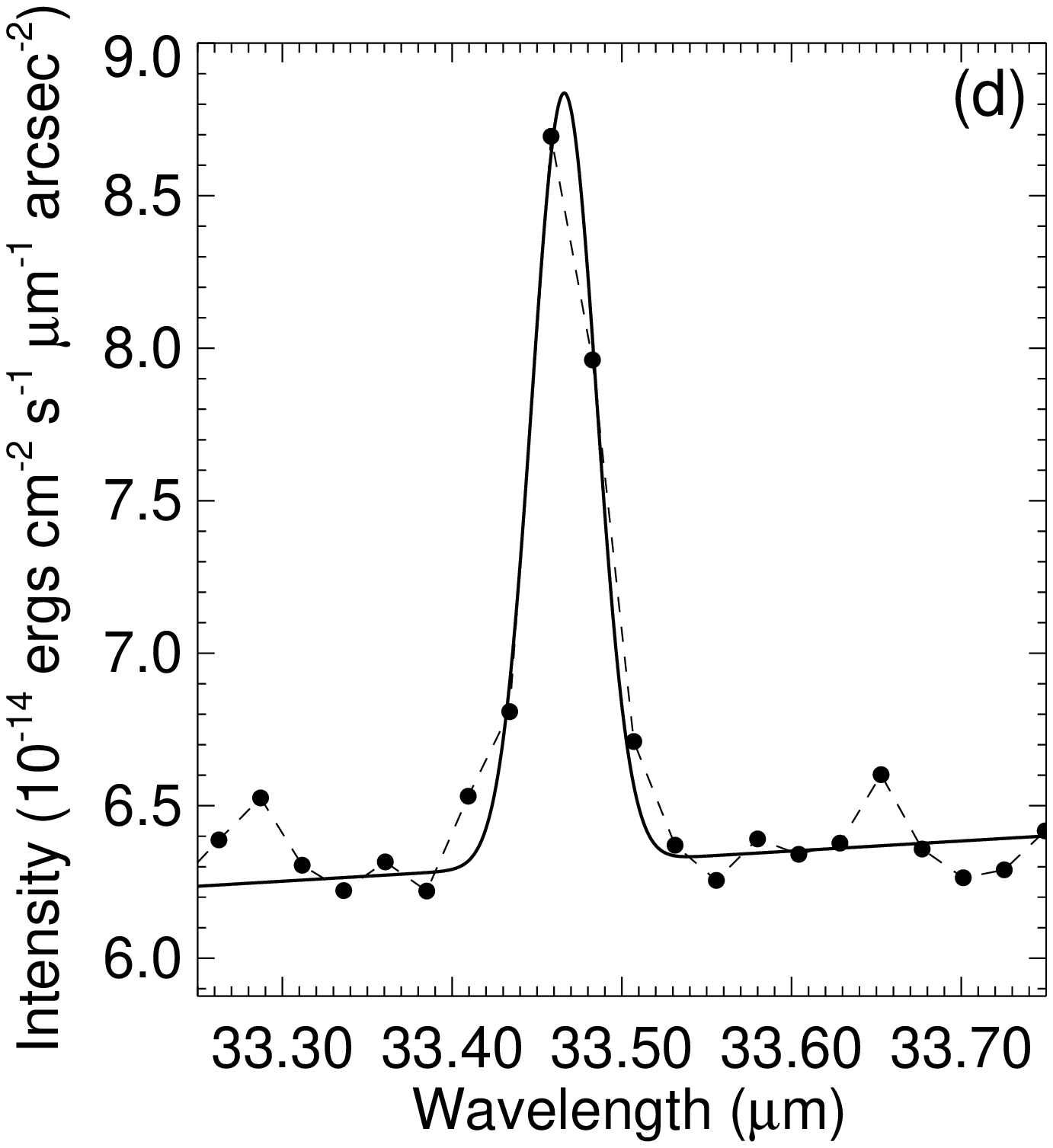}} \\
    \end{tabular}
\vskip0.1truein
\caption{Measurements of four emission lines in chex V3, the outermost 
one (see Fig.~1) and for just one (data set 25380864)
of the three AORs  
where all three of the veil chex were observed:
{\bf (a)} [\ion{Ne}{2}] 12.8~$\mu$m;
       {\bf (b)} [\ion{Ne}{3}] 15.6~$\mu$m; 
       {\bf (c)} [\ion{S}{3}] 18.7~$\mu$m; and 
       {\bf (d)} [\ion{S}{3}] 33.5~$\mu$m.
       The data points are the filled circles.
       The fits to the continuum and Gaussian profiles are the solid lines.
       These are among the lines listed  in Table~2 for  V3-1, that is
       chex V3 and veil AOR 1.
       Such measurements provide the set of line intensities 
       for further analysis.}
    \label{test4}
  \end{center}
\end{figure}

\begin{figure}
\epsscale{1.0}
\plotone{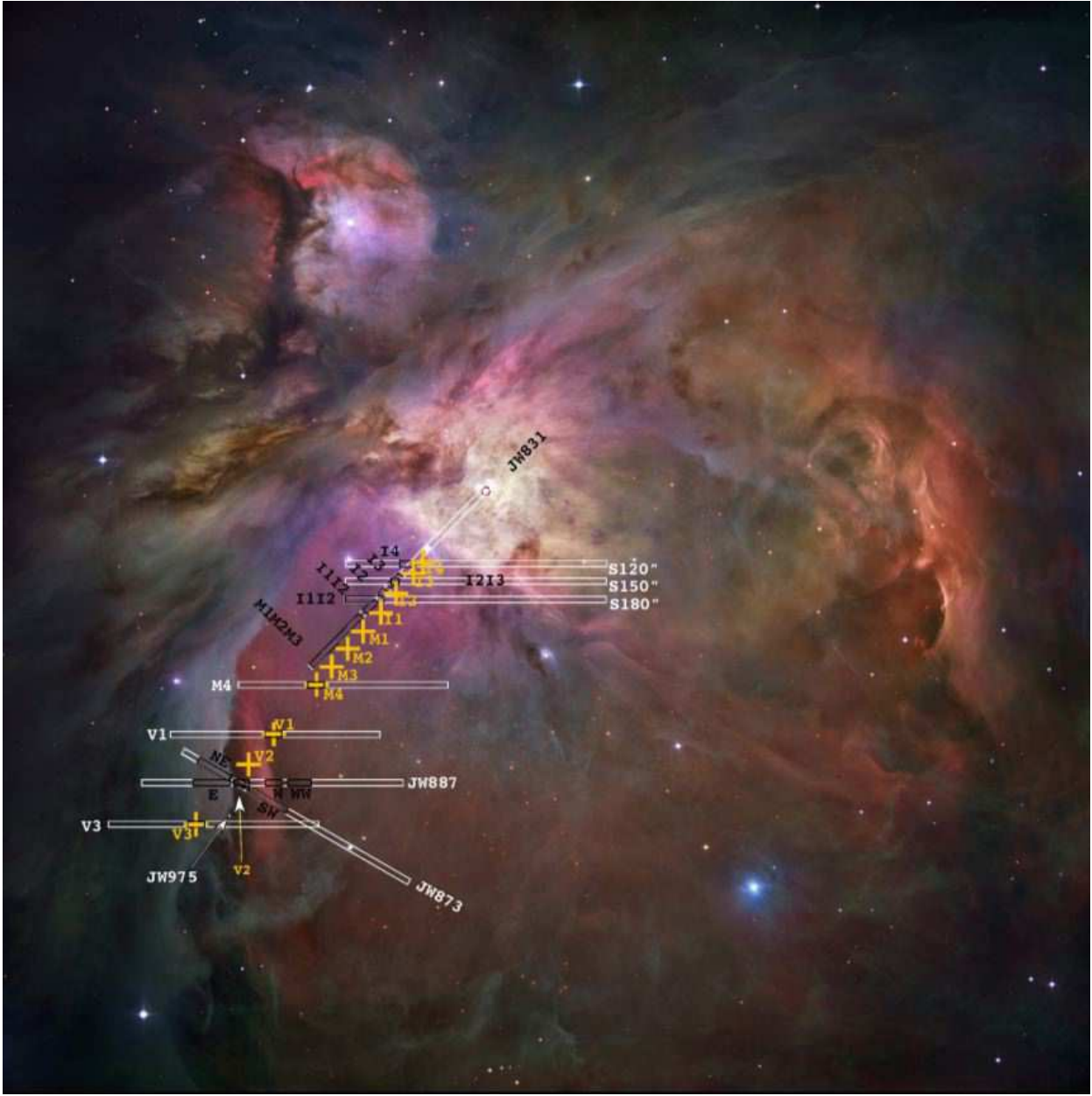}
\caption{
This 1200\arcsec\  x 1200\arcsec\ image of the Orion Nebula taken from Henney et~al. (2007) shows the 
regions sampled in our spectroscopy. The white boxes show the slit positions and are labelled with the 
reference star used, this being $\theta \rm ^{1}$ Ori C for the east-west slits displaced south of that 
star, or with the name of the {\it Spitzer}  chex. 
For clarity the slits are shown as 10\arcsec\ wide 
even though they were actually 2.6\arcsec\ wide. 
The yellow crosses indicate the centre of regions observed 
with {\it Spitzer}, while the dark boxes and labels indicate 
the ground-based optical spectroscopy samples.
\label{samples}}
\end{figure}
\clearpage

\begin{figure}
\epsscale{1.0}
\plotone{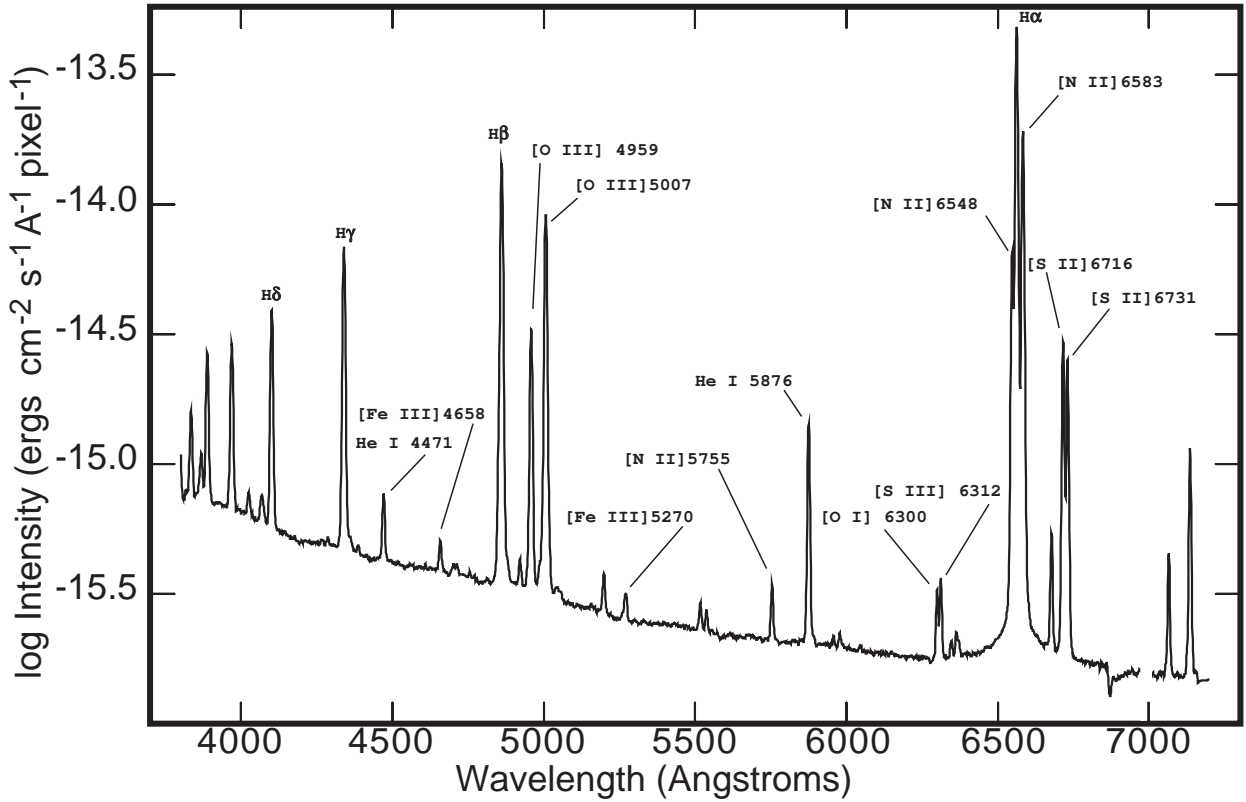}
\caption{This is a logarithmic presentation of a representative optical 
spectrum. It is the result of 3900 seconds of integration over five exposures 
of the M4 extraction along 21 pixels of the M4 long slit (see Figure~5).
The gap in data  near 7000~\AA\ is due
to a column defect in the CCD detector.
\label{spectrum}}
\end{figure}
\clearpage

\begin{figure}
\centering
\resizebox{14.0cm}{!}{\includegraphics{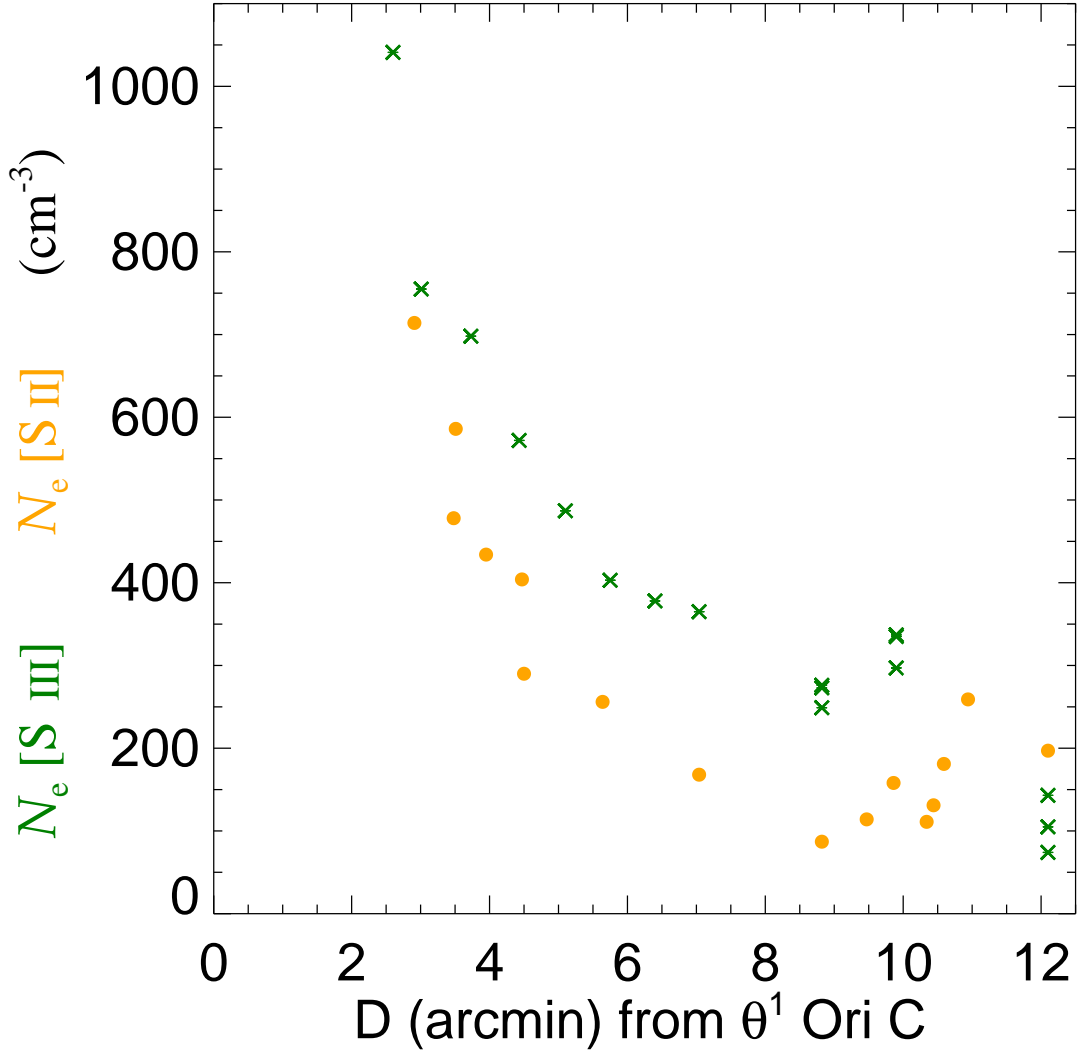} }
\vskip0.1truein
\caption[]{Plot of the electron density
$N_e$~[\ion{S}{3}] (dark or green in colour) x's 
and $N_e$~[\ion{S}{2}] (gray or yellow in colour) circles versus D (the distance
in arcmin from $\theta^1$~Ori~C to the centre of the
{\it Spitzer} chex or ground-based optical sample).}

\end{figure}

\begin{figure}
\centering
\resizebox{14.0cm}{!}
{\includegraphics{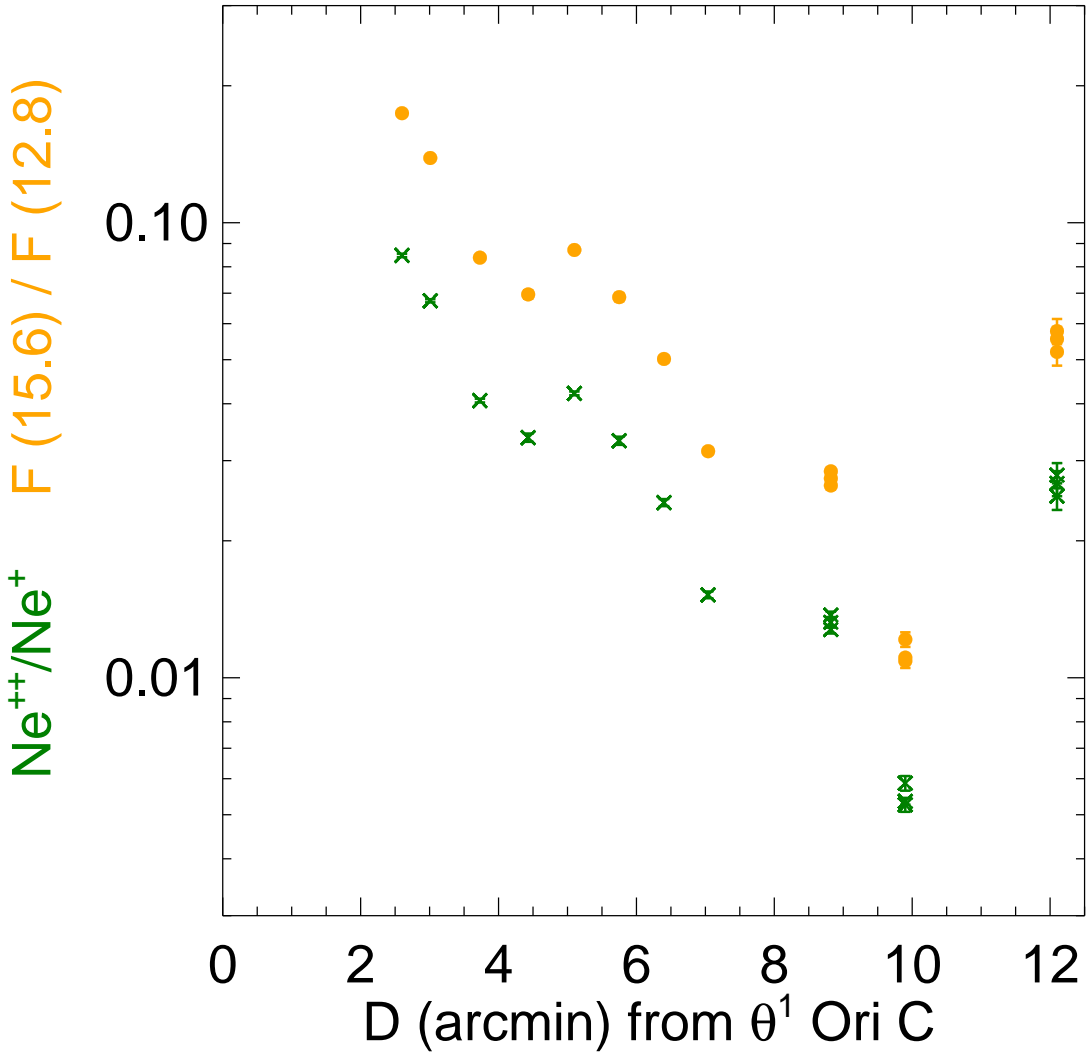} }
\vskip0.1truein
\caption[]{Plot of the line flux ratio 
[\ion{Ne}{3}] 15.6/[\ion{Ne}{2}] 12.8 (gray or yellow in colour)
and the derived Ne$^{++}$/Ne$^+$ (black or green in colour)
versus D.
Error bars here and in Figs 9--16 are for the propagated measurement
uncertainties and do not include the systematic uncertainties (see text).}

\end{figure}

\begin{figure}
\centering
\resizebox{14.0cm}{!}{\includegraphics{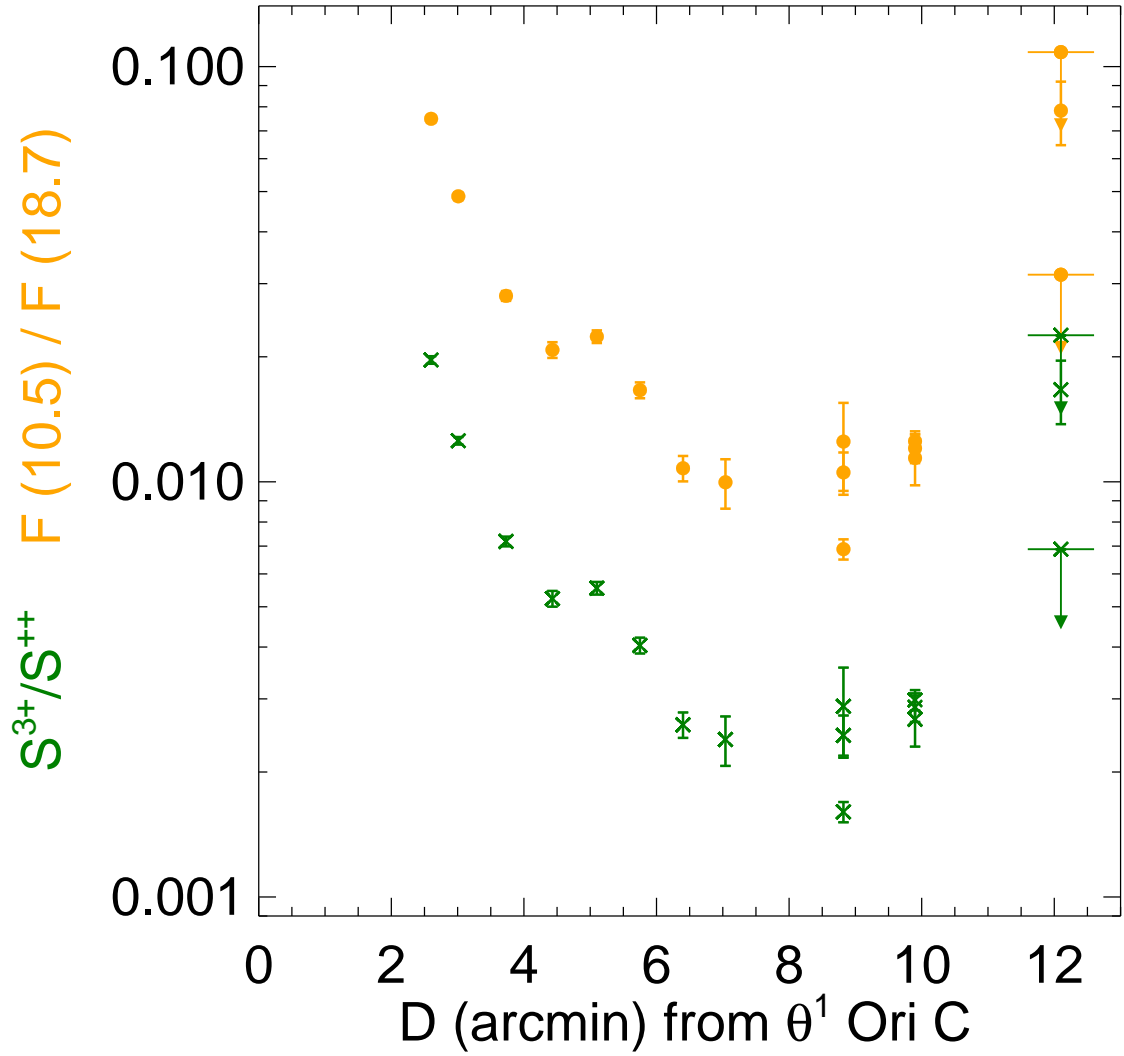} }
\vskip0.1truein
\caption[]{Plot of the line flux ratio 
[\ion{S}{4}] 10.5/[\ion{S}{3}] 18.7 (gray or yellow in colour)
and the derived S$^{3+}$/S$^{++}$ (black or green in colour)
versus D.}

\end{figure}

\begin{figure}
\centering
\resizebox{14.0cm}{!}{\includegraphics{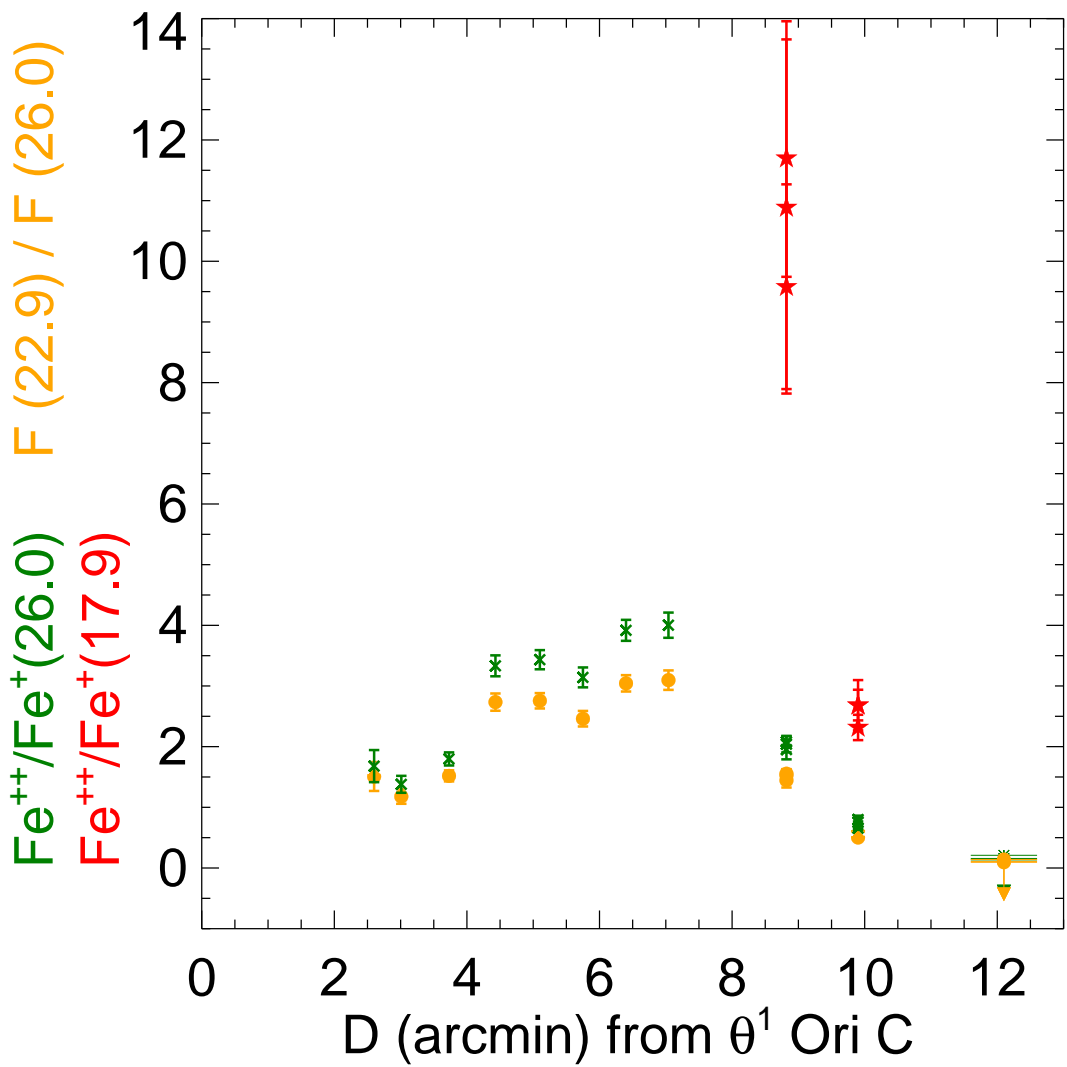} }
\vskip0.1truein
\caption[]{Plot of the line flux ratio 
[\ion{Fe}{3}] 22.9/[\ion{Fe}{2}] 26.0 (gray or yellow circles in colour)
and the derived Fe$^{++}$/Fe$^+$ (black or green x in colour)
versus D.  In addition, the [\ion{Fe}{2}] 17.9~$\mu$m line
was measured at V1 and V2 only.  
The Fe$^{++}$/Fe$^+$ derived from the
[\ion{Fe}{3}] 22.9/[\ion{Fe}{2}] 17.9 ratio is shown (black or red 
stars in colour).
These higher values are a more
accurate measure of Fe$^{++}$/Fe$^+$  than those using the 
[\ion{Fe}{2}] 26.0 line, which are {\bf lower limits} (see text).}

\end{figure}

\begin{figure}
\centering
\resizebox{14.0cm}{!}{\includegraphics{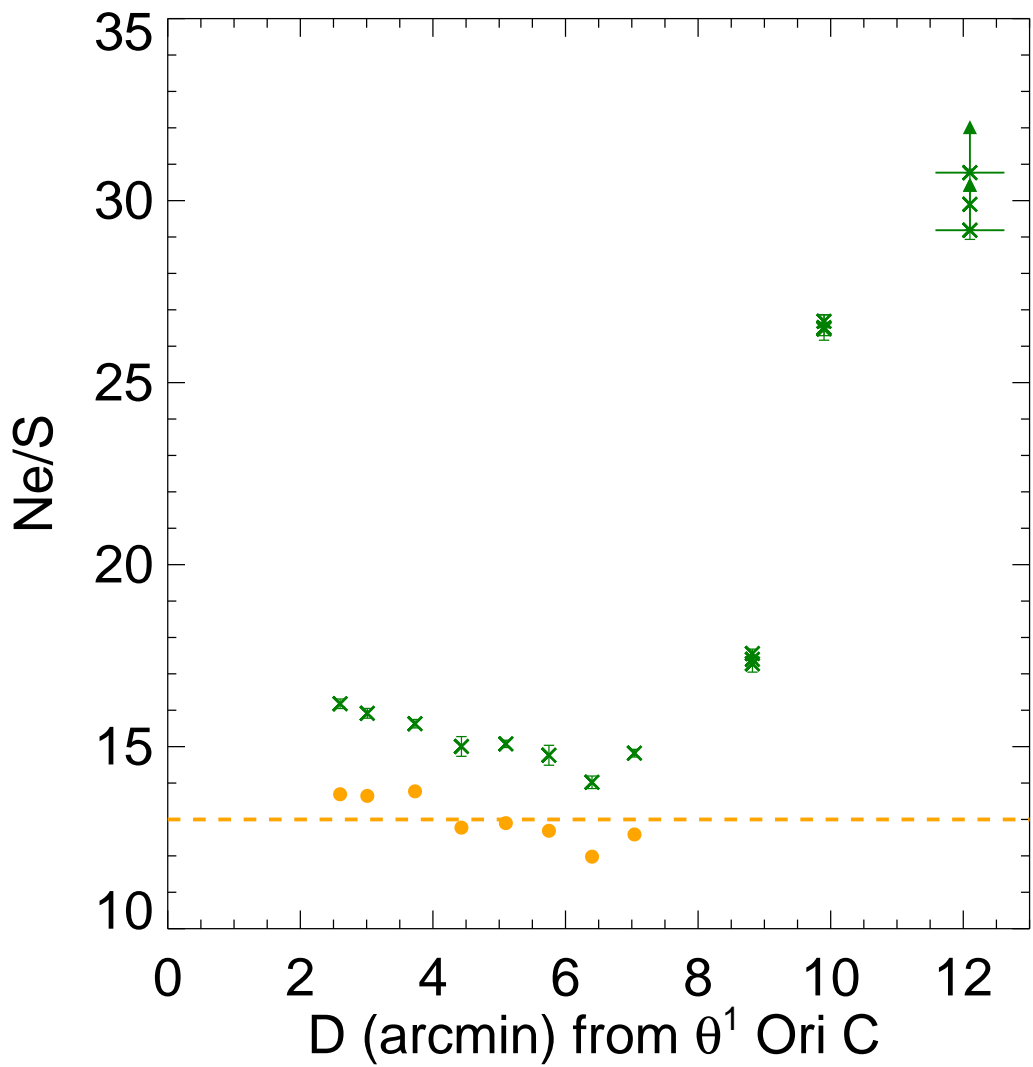} }
\vskip0.1truein
\caption[]{Plot of Ne/S versus D.
The dark or green x in colour represent the
the gas-phase Ne/S ratio as approximated by the
(Ne$^+$ + Ne$^{++}$)/(S$^{++}$ + S$^{3+}$) ratio
derived from the {\it Spitzer} data only.
These should be considered as upper limits
to the Ne/S ratio because sulfur in S$^+$ has not been accounted for.
The gray or yellow circles in colour show the 
(Ne$^+$ + Ne$^{++}$)/(S$^+$ + S$^{++}$ + S$^{3+}$) ratio
after adjusting the 
{\it Spitzer}--only data
by the optically-determined S$^+$/S$^{++}$ ratios
for the inner 8 chex (see text).
The dashed horizontal line is the mean value for these
8 chex and represents our best estimate for the gas-phase
Ne/S~= $13.0\pm0.6$.}

\end{figure}

\begin{figure}
\centering
\resizebox{14.0cm}{!}{\includegraphics{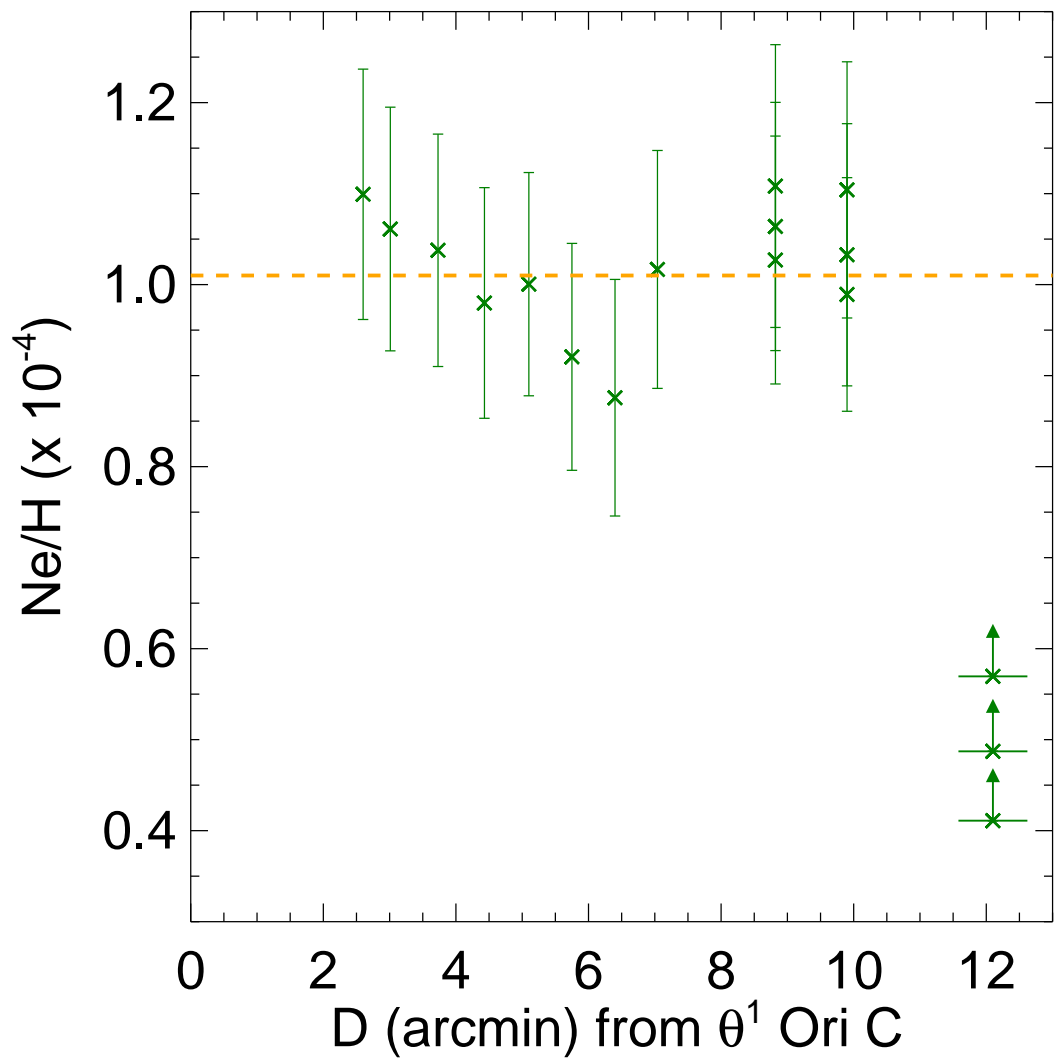} }
\vskip0.1truein
\caption[]{Plot of Ne/H versus D.
Except for chex V3 where the H(7--6) line was not detected, 
the ratios vary little.
We include the 6 independent measurements at V1 and V2
and take the mean for the 10 innermost chex as the best value,
Ne/H~= $(1.01\pm0.08)\times10^{-4}$.
In terms of the conventional expression,  
this is 12~+ log~(Ne/H)~= 8.00$\pm$0.03.
This may well be the {\it gold standard} for a determination
of metallicity in an \ion{H}{2} region (see text)}

\end{figure}

\begin{figure}
\centering
\resizebox{14.0cm}{!}{\includegraphics{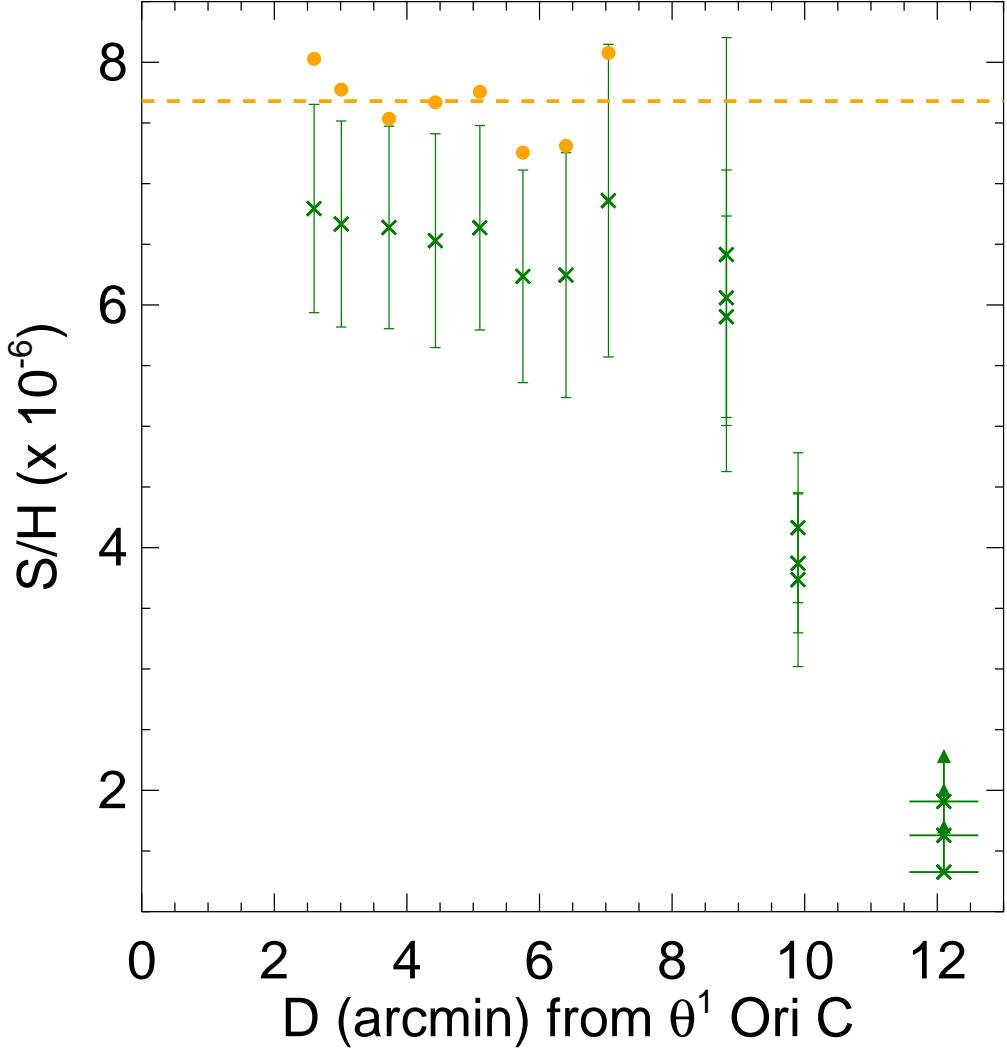} }
\vskip0.1truein
\caption[]{This figure shows the S/H estimates versus D.
Here we plot the sum of the S$^{++}$/H$^+$ and S$^{3+}$/H$^+$ ratios 
using  the {\it  Spitzer} data only.
As for Ne/S, we use the mean for the innermost 8 chex as the best value 
S/H~= $(6.58\pm0.23)\times10^{-6}$.
For these 8 innermost chex, we again make a 
correction for S$^+$, unseen by {\it Spitzer}, by
using the S$^+$/S$^{++}$ ratios derived from our optical data.
These points are the filled circles (yellow in the colour version).
The best {\it corrected} 
S/H~= $(7.68\pm0.30)\times10^{-6}$ or
12~+ log~(S/H)~= 6.89$\pm$0.02.}

\end{figure}

\begin{figure}
\centering
\resizebox{14.0cm}{!}{\includegraphics{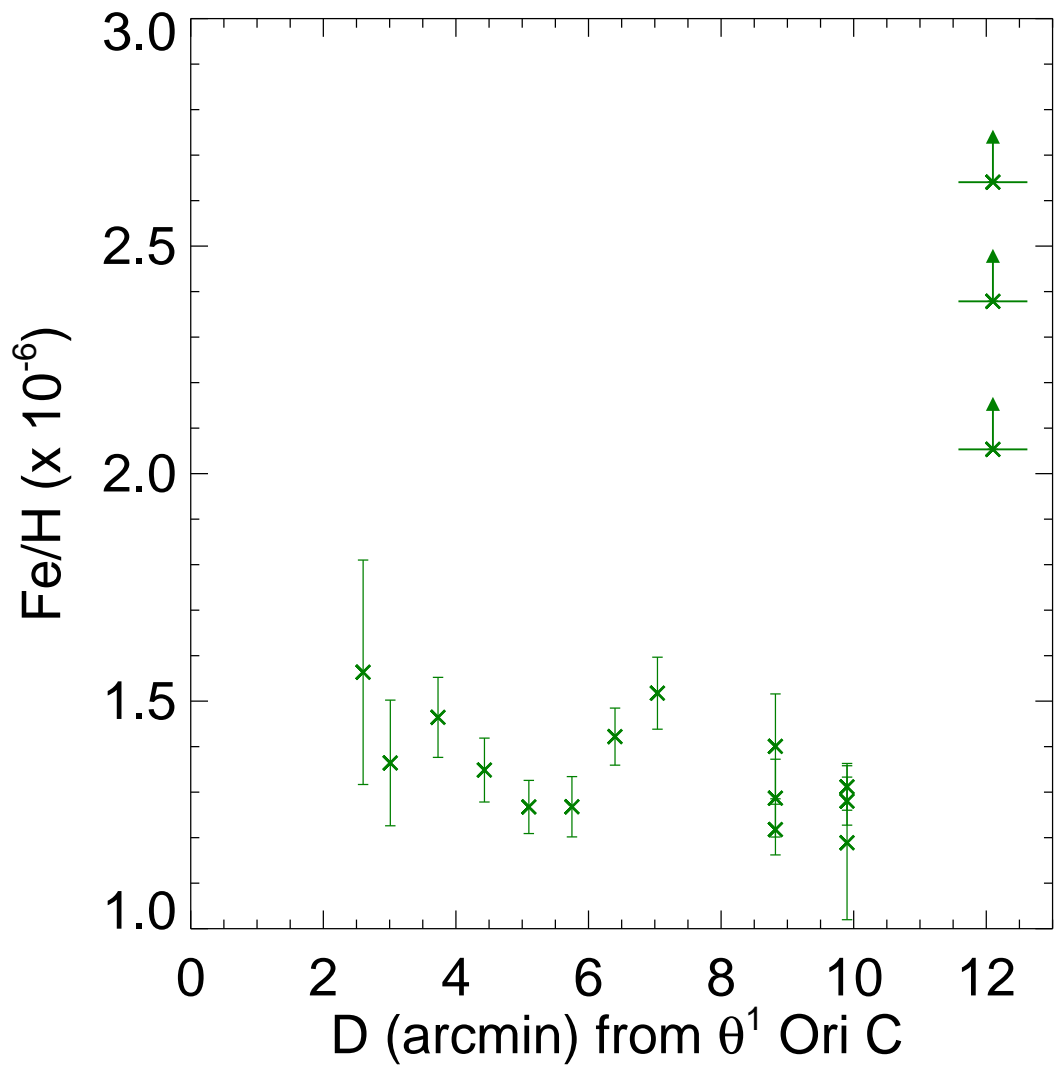} }
\vskip0.1truein
\caption[]{Plot of gas-phase Fe/H estimated from the 
the sum of the Fe$^+$/H$^+$ and Fe$^{++}$/H$^+$ ratios 
(see Table~10).
The Fe$^+$/H$^+$ ratios are derived using the
[\ion{Fe}{2}] 26~$\mu$m line, but we do not account for an unknown, 
significant PDR contribution (see text).  Because of this, the
Fe$^+$/H$^+$ ratios are upper limits.
This causes the Fe/H ratio here to be {\it overestimated}.
However, since Fe$^{3+}$ has not been accounted for,
that would {\it increase} an assessment of gas-phase Fe/H (see text).}

\end{figure}

\begin{figure}
\centering
\resizebox{14.0cm}{!}{\includegraphics{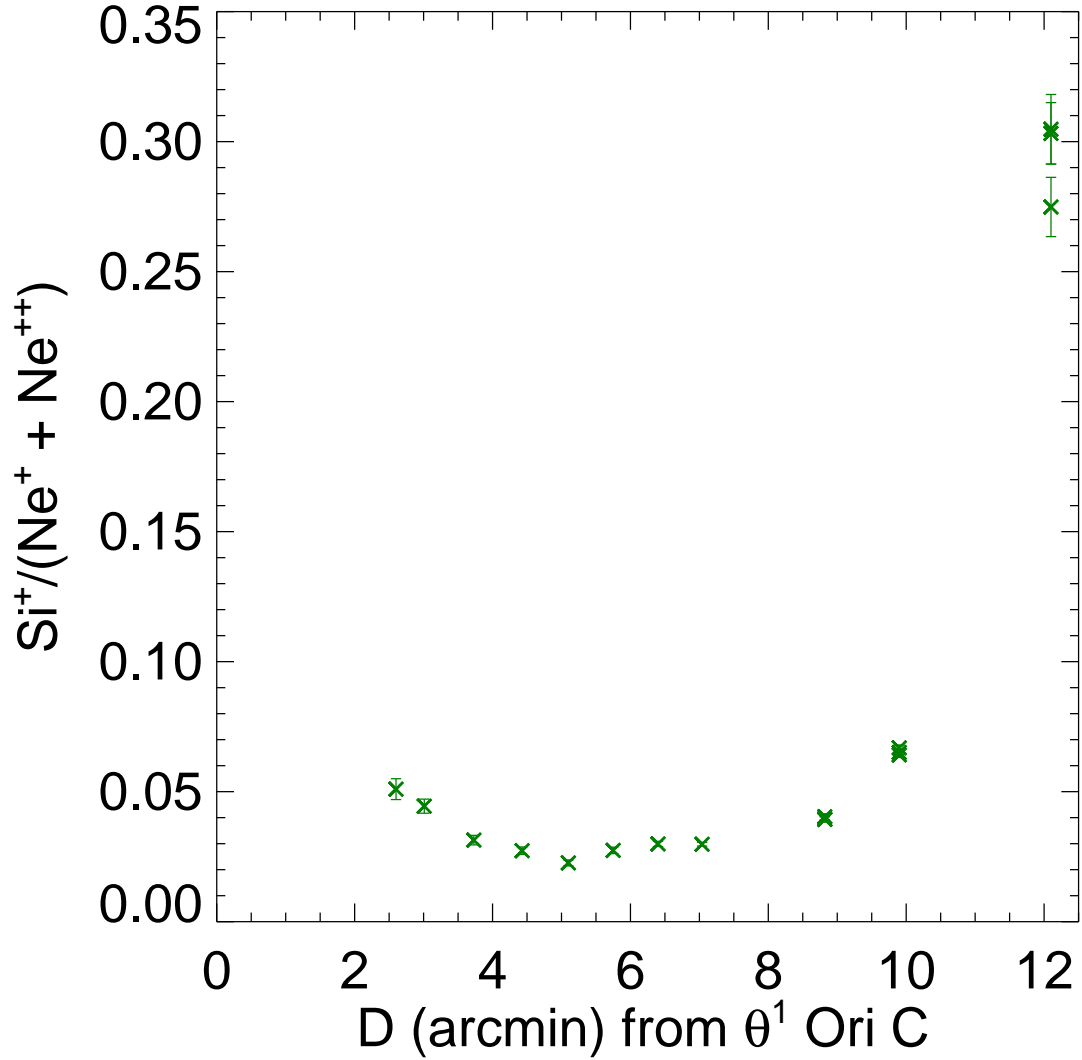} }
\vskip0.1truein
\caption[]{Plot of Si$^+$/Ne.
This ratio is derived using 
the [\ion{Si}{2}] 34.8~$\mu$m line assuming that it is produced
in the ionized region {\it only} and does not account for an unknown, 
significant PDR contribution (see text).
The large increase at the outermost V3 position is strong evidence
that the bulk of the [\ion{Si}{2}] 34.8 emission arises in a PDR
at this \ion{H}{2} region -- PDR interface.}

\end{figure}

\begin{figure}
  \begin{center}
\vskip-0.25truein
    \begin{tabular}{cc}
\resizebox{70mm}{!}{\includegraphics{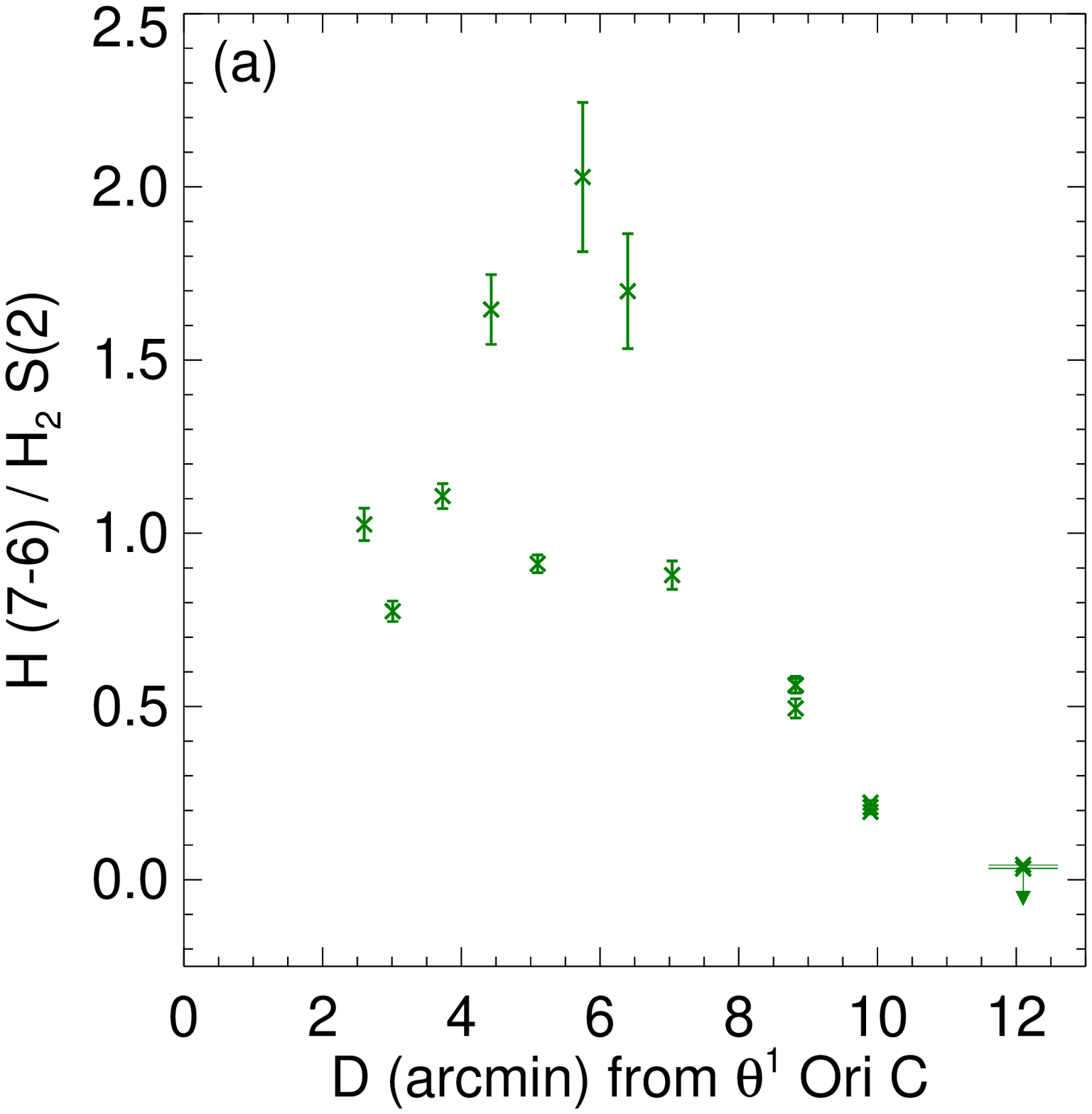}} &
\resizebox{70mm}{!}{\includegraphics{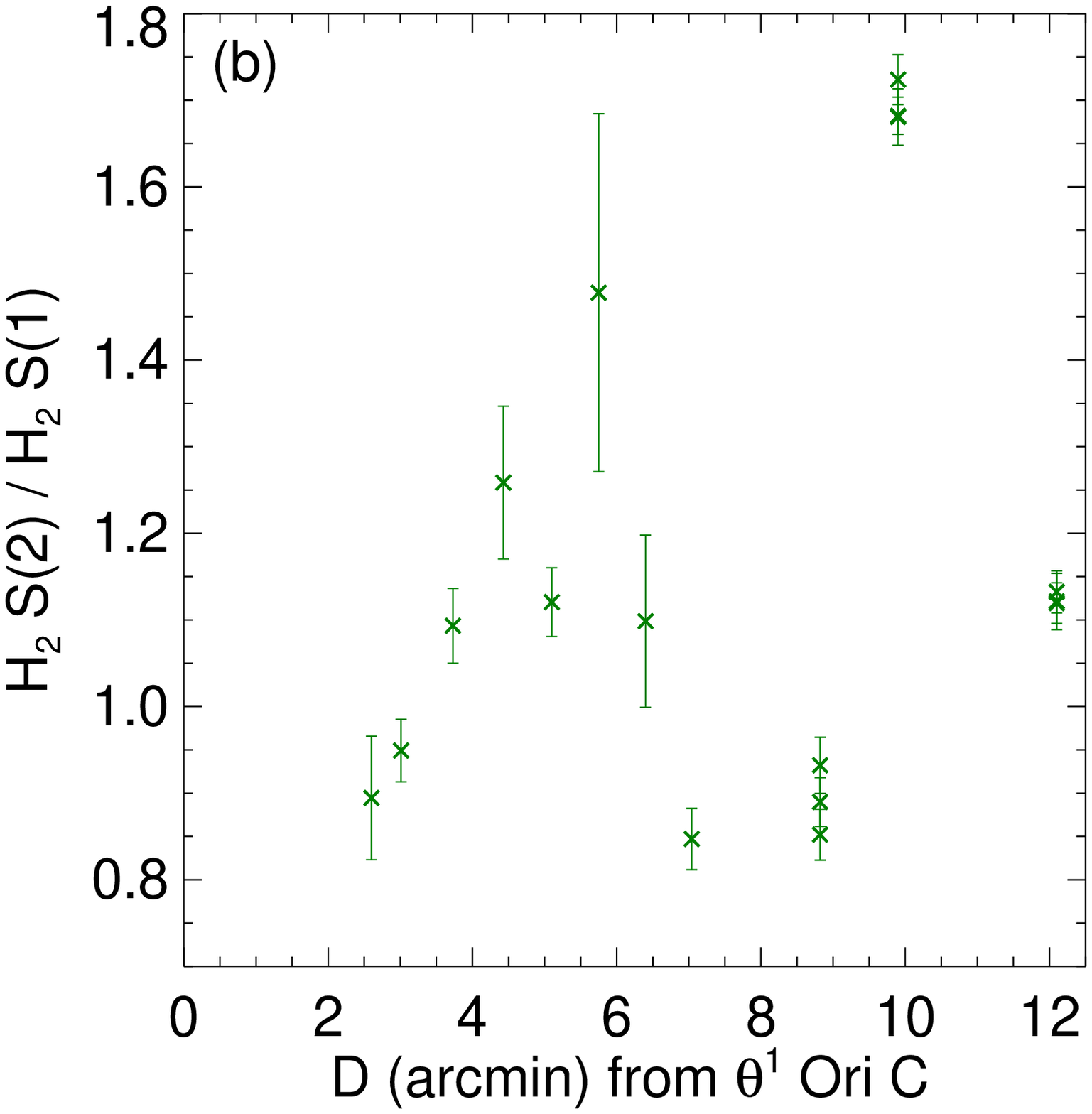}} \\
\resizebox{70mm}{!}{\includegraphics{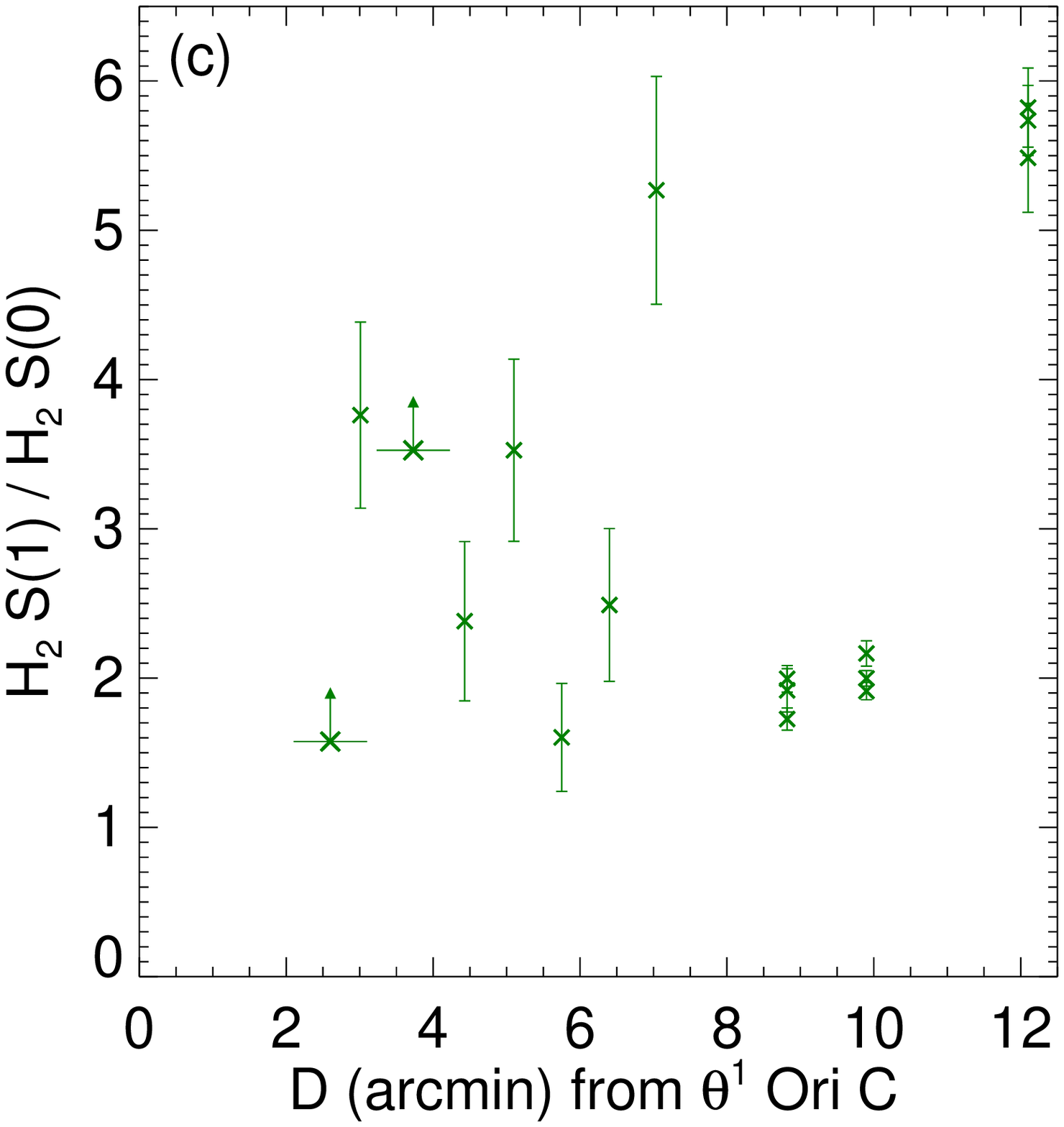}} &
\resizebox{70mm}{!}{\includegraphics{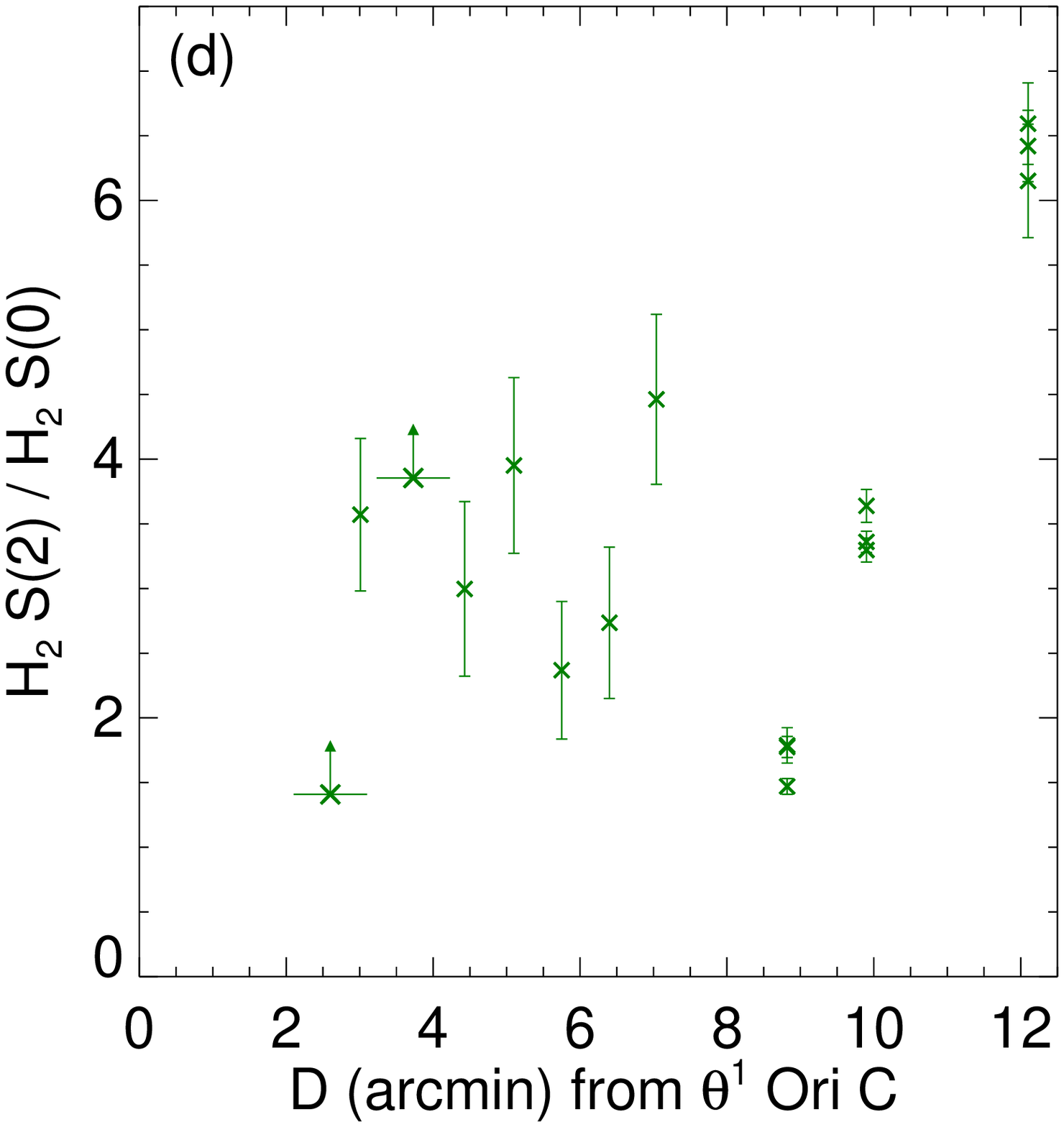}} \\
    \end{tabular}
\caption{This plots the flux ratio versus D of the H(7--6) line, 
which arises in the \ion{H}{2} region, to the three H$_2$ lines,
which arise in the PDR. 
       {\bf (a)} H(7--6)/H$_2$~S(2);
       {\bf (b)} H$_2$~S(2)/H$_2$~S(1);
       {\bf (c)} H$_2$~S(1)/H$_2$~S(0);and 
       {\bf (d)} H$_2$~S(2)/H$_2$~S(0).}
    \label{ }
  \end{center}
\end{figure}

\begin{deluxetable}{cccc}
\tabletypesize{\scriptsize}
\tablecaption{Regions Observed in M42}
\tablewidth{0pt}
\tablenum{1}
\tablehead{
\colhead{Chex} & \colhead{Distance} & \colhead{RA} & \colhead{DEC}\\
                             & \colhead{(arcmin)}  & \multicolumn{2}{c}{{(J2000)}}
}
\startdata

I4 & 2.60 & 5 35 23.3 & -5 25 20.7 \\

I3 & 3.01 & 5 35 24.4 & -5 25 39.1 \\

I2 & 3.73 & 5 35 26.3 & -5 26 12.1 \\

I1 & 4.43 & 5 35 28.1 & -5 26 43.8 \\

M1 & 5.10 & 5 35 29.9 & -5 27 14.4 \\

M2 & 5.75 & 5 35 31.6 & -5 27 43.6 \\

M3 & 6.40 & 5 35 33.3 & -5 28 12.9 \\

M4 & 7.04 & 5 35 35.0 & -5 28 42.1 \\

V1 & 8.82 & 5 35 39.7 & -5 30~~2.7 \\

V2 & 9.90 & 5 35 42.5 & -5 30 51.9 \\

V3 & 12.08 & 5 35 48.3 & -5 32 30.3 \\

\enddata

\end{deluxetable}

\begin{deluxetable}{ccccccr}    
\tabletypesize{\scriptsize}
\tablecaption{M42 {\it Spitzer} Line Measurements}
\tablewidth{0pt}
\tablenum{2}
\tablehead{
 \colhead {Chex}
& \colhead {Species}
& \colhead {Line}
& \colhead{Intensity}
& \colhead{1$\sigma$ error}
& \colhead{FWHM}
& \colhead{V$_{helio}$} \\

& 
& \colhead{($\mu$m)}
& \multicolumn{2}{c}{{(ergs cm$^{-2}$ s$^{-1}$ arcsec$^{-2}$)}}

& \colhead{(km s$^{-1}$)}
& \colhead{(km s$^{-1}$)}

}

\startdata

I4 & [S \sc{iv}] & 10.5 & 2.48E-14 & 4.84E-16 & 483 & 11 \\
   & H$_2$ S(2) & 12.3 & 5.42E-15 & 1.98E-16 & 464 & -44 \\
   & H \sc{i} 7-6 & 12.4 & 5.56E-15 & 1.52E-16 & 474 & -38 \\
   & [Ne \sc{ii}] & 12.8 & 3.18E-13 & 1.27E-15 & 449 & -60 \\
   & [Ne \sc{iii}] & 15.6 & 5.53E-14 & 4.31E-16 & 451 & -33 \\
   & H$_2$ S(1) & 17.0 & 6.06E-15 & 4.29E-16 & 491 & -22 \\
   & [S \sc{iii}] & 18.7 & 3.31E-13 & 2.40E-15 & 485 & -11 \\
   & [Fe \sc{iii}] & 22.9 & 1.12E-14 & 1.50E-15 & 539 & -32 \\
   & [Fe \sc{ii}] & 26.0 & 7.43E-15 & 6.21E-16 & 306 & -81 \\
   & H$_2$ S(0) & 28.2 & 3.85E-15$^a$ & --- & ---  & ---  \\
   & [S \sc{iii}] & 33.5 & 2.39E-13 & 4.05E-15 & 448 & -75 \\
   & [Si \sc{ii}] & 34.8 & 8.92E-14 & 7.00E-15 & 554 & -38 \\
   
\\

I3 & [S \sc{iv}] & 10.5 & 1.03E-14 & 2.03E-16 & 507 & 9 \\
   & H$_2$ S(2) & 12.3 & 4.66E-15 & 1.15E-16 & 467 & -42 \\
   & H \sc{i} 7-6 & 12.4 & 3.61E-15 & 1.05E-16 & 490 & -25 \\
   & [Ne \sc{ii}] & 12.8 & 2.02E-13 & 8.54E-16 & 449 & -49 \\
   & [Ne \sc{iii}] & 15.6 & 2.80E-14 & 1.92E-16 & 451 & -27 \\
   & H$_2$ S(1) & 17.0 & 4.91E-15 & 1.41E-16 & 462 & 11 \\
   & [S \sc{iii}] & 18.7 & 2.12E-13 & 1.60E-15 & 486 & 1 \\
   & [Fe \sc{iii}] & 22.9 & 5.95E-15 & 5.02E-16 & 524 & -50 \\
   & [Fe \sc{ii}] & 26.0 & 5.05E-15 & 2.83E-16 & 473 & -93 \\
   & H$_2$ S(0) & 28.2 & 1.30E-15 & 2.13E-16 & 332 & 0 \\
   & [S \sc{iii}] & 33.5 & 1.87E-13 & 2.04E-15 & 450 & -101 \\
   & [Si \sc{ii}] & 34.8 & 5.93E-14 & 3.62E-15 & 536 & -63 \\

\\

I2 & [S \sc{iv}] & 10.5 & 5.02E-15 & 1.31E-16 & 484 & 2 \\
   & H$_2$ S(2) & 12.3 & 2.76E-15 & 6.88E-17 & 459 & -46 \\
   & H \sc{i} 7-6 & 12.4 & 3.05E-15 & 6.40E-17 & 493 & -30 \\
   & [Ne \sc{ii}] & 12.8 & 1.71E-13 & 6.26E-16 & 446 & -54 \\
   & [Ne \sc{iii}] & 15.6 & 1.44E-14 & 1.27E-16 & 453 & -32 \\
   & H$_2$ S(1) & 17.0 & 2.52E-15 & 7.72E-17 & 454 & 19 \\
   & [S \sc{iii}] & 18.7 & 1.79E-13 & 1.21E-15 & 485 & -1 \\
   & [Fe \sc{iii}] & 22.9 & 6.01E-15 & 2.36E-16 & 438 & -9 \\
   & [Fe \sc{ii}] & 26.0 & 3.96E-15 & 1.80E-16 & 395 & -99 \\
   & H$_2$ S(0) & 28.2 & 7.15E-16$^a$ & --- &---   &---   \\
   & [S \sc{iii}] & 33.5 & 1.65E-13 & 1.65E-15 & 452 & -75 \\
   & [Si \sc{ii}] & 34.8 & 3.62E-14 & 2.12E-15 & 536 & -54 \\

\\

I1 & [S \sc{iv}] & 10.5 & 2.61E-15 & 1.12E-16 & 482 & 0 \\
   & H$_2$ S(2) & 12.3 & 1.33E-15 & 6.95E-17 & 495 & -34 \\
   & H \sc{i} 7-6 & 12.4 & 2.20E-15 & 7.06E-17 & 503 & -15 \\
   & [Ne \sc{ii}] & 12.8 & 1.17E-13 & 2.01E-15 & 439 & -58 \\
   & [Ne \sc{iii}] & 15.6 & 8.14E-15 & 9.86E-17 & 450 & -35 \\
   & H$_2$ S(1) & 17.0 & 2.52E-15 & 7.72E-17 & 454 & 19 \\
   & [S \sc{iii}] & 18.7 & 1.25E-13 & 8.92E-16 & 488 & 0 \\
   & [Fe \sc{iii}] & 22.9 & 4.80E-15 & 1.23E-16 & 483 & -36 \\
   & [Fe \sc{ii}] & 26.0 & 1.75E-15 & 7.92E-17 & 466 & -96 \\
   & H$_2$ S(0) & 28.2 & 4.45E-16 & 9.75E-17 & 412 & 37 \\
   & [S \sc{iii}] & 33.5 & 1.29E-13 & 1.25E-15 & 450 & -93 \\
   & [Si \sc{ii}] & 34.8 & 2.37E-14 & 1.20E-15 & 521 & -60 \\

\\

M1 & [S \sc{iv}] & 10.5 & 2.00E-15 & 6.90E-17 & 480 & 8 \\
   & H$_2$ S(2) & 12.3 & 1.70E-15 & 3.60E-17 & 474 & -41 \\
   & H \sc{i} 7-6 & 12.4 & 1.55E-15 & 2.89E-17 & 477 & -20 \\
   & [Ne \sc{ii}] & 12.8 & 8.37E-14 & 2.69E-16 & 441 & -67 \\
   & [Ne \sc{iii}] & 15.6 & 7.30E-15 & 6.59E-17 & 450 & -42 \\
   & H$_2$ S(1) & 17.0 & 1.52E-15 & 4.33E-17 & 433 & 28 \\
   & [S \sc{iii}] & 18.7 & 8.92E-14 & 5.41E-16 & 482 & -8 \\
   & [Fe \sc{iii}] & 22.9 & 3.23E-15 & 5.32E-17 & 426 & 7 \\
   & [Fe \sc{ii}] & 26.0 & 1.17E-15 & 5.03E-17 & 486 & -40 \\
   & H$_2$ S(0) & 28.2 & 4.31E-16 & 7.36E-17 & 400 & 13 \\
   & [S \sc{iii}] & 33.5 & 1.01E-13 & 4.75E-16 & 453 & -32 \\
   & [Si \sc{ii}] & 34.8 & 1.53E-14 & 7.81E-16 & 529 & 11 \\

\\

M2 & [S \sc{iv}] & 10.5 & 9.80E-16 & 4.28E-17 & 474 & 1 \\
   & H$_2$ S(2) & 12.3 & 5.47E-16 & 5.32E-17 & 771 & 11 \\
   & H \sc{i} 7-6 & 12.4 & 1.11E-15 & 4.73E-17 & 508 & -27 \\
   & [Ne \sc{ii}] & 12.8 & 5.62E-14 & 2.08E-16 & 442 & -71 \\
   & [Ne \sc{iii}] & 15.6 & 3.80E-15 & 3.72E-17 & 452 & -44 \\
   & H$_2$ S(1) & 17.0 & 3.70E-16 & 3.72E-17 & 500 & 39 \\
   & [S \sc{iii}] & 18.7 & 5.90E-14 & 3.52E-16 & 482 & -13 \\
   & [Fe \sc{iii}] & 22.9 & 2.27E-15 & 5.30E-17 & 432 & -3 \\
   & [Fe \sc{ii}] & 26.0 & 9.21E-16 & 4.29E-17 & 454 & -68 \\
   & H$_2$ S(0) & 28.2 & 2.31E-16 & 4.68E-17 & 431 & 25 \\
   & [S \sc{iii}] & 33.5 & 7.24E-14 & 7.39E-16 & 439 & -75 \\
   & [Si \sc{ii}] & 34.8 & 1.33E-14 & 5.53E-16 & 512 & -35 \\

\\

M3 & [S \sc{iv}] & 10.5 & 5.00E-16 & 3.52E-17 & 477 & 21 \\
   & H$_2$ S(2) & 12.3 & 5.16E-16 & 3.94E-17 & 557 & -22 \\
   & H \sc{i} 7-6 & 12.4 & 8.76E-16 & 5.33E-17 & 540 & -34 \\
   & [Ne \sc{ii}] & 12.8 & 4.20E-14 & 5.18E-16 & 443 & -60 \\
   & [Ne \sc{iii}] & 15.6 & 2.11E-15 & 2.75E-17 & 459 & -29 \\
   & H$_2$ S(1) & 17.0 & 4.70E-16 & 2.28E-17 & 333 & 56 \\
   & [S \sc{iii}] & 18.7 & 4.64E-14 & 1.03E-16 & 488 & 1 \\
   & [Fe \sc{iii}] & 22.9 & 2.11E-15 & 3.54E-17 & 425 & 0 \\
   & [Fe \sc{ii}] & 26.0 & 6.93E-16 & 2.84E-17 & 443 & -62 \\
   & H$_2$ S(0) & 28.2 & 1.89E-16 & 3.77E-17 & 448 & 36 \\
   & [S \sc{iii}] & 33.5 & 5.87E-14 & 6.50E-16 & 443 & -55 \\
   & [Si \sc{ii}] & 34.8 & 1.12E-14 & 3.53E-16 & 509 & -18 \\

\\

M4 & [S \sc{iv}] & 10.5 & 3.44E-16 & 4.68E-17 & 601 & -34 \\
   & H$_2$ S(2) & 12.3 & 6.76E-16 & 2.34E-17 & 473 & -25 \\
   & H \sc{i} 7-6 & 12.4 & 5.94E-16 & 1.85E-17 & 486 & -23 \\
   & [Ne \sc{ii}] & 12.8 & 3.33E-14 & 1.18E-16 & 442 & -64 \\
   & [Ne \sc{iii}] & 15.6 & 1.05E-15 & 1.75E-17 & 453 & -35 \\
   & H$_2$ S(1) & 17.0 & 7.98E-16 & 1.86E-17 & 425 & 35 \\
   & [S \sc{iii}] & 18.7 & 3.44E-14 & 2.25E-16 & 485 & -1 \\
   & [Fe \sc{iii}] & 22.9 & 1.53E-15 & 2.95E-17 & 475 & -37 \\
   & [Fe \sc{ii}] & 26.0 & 4.95E-16 & 2.39E-17 & 483 & -96 \\
   & H$_2$ S(0) & 28.2 & 1.51E-16 & 2.17E-17 & 426 & -35 \\
   & [S \sc{iii}] & 33.5 & 4.42E-14 & 4.04E-16 & 448 & -96 \\
   & [Si \sc{ii}] & 34.8 & 8.88E-15 & 2.18E-16 & 503 & -67 \\

\\

V1-1 & [S \sc{iv}] & 10.5 & 6.54E-17 & 3.66E-18 & 331 & -63 \\
   & H$_2$ S(2) & 12.3 & 3.47E-16 & 8.14E-18 & 442 & -40 \\
   & H \sc{i} 7-6 & 12.4 & 1.96E-16 & 6.93E-18 & 503 & 5 \\
   & [Ne \sc{ii}] & 12.8 & 1.11E-14 & 3.83E-17 & 444 & -65 \\
   & [Ne \sc{iii}] & 15.6 & 3.03E-16 & 8.03E-18 & 464 & -22 \\
   & H$_2$ S(1) & 17.0 & 3.90E-16 & 8.26E-18 & 464 & 25 \\
	&	[Fe \sc{ii}]	&	17.9	&	3.00E-17	&	4.89E-18	&	487	&	103	\\   
   & [S \sc{iii}] & 18.7 & 9.51E-15 & 3.78E-17 & 487 & -1 \\
   & [Fe \sc{iii}] & 22.9 & 3.61E-16 & 1.28E-17 & 460 & -15 \\
   & [Fe \sc{ii}] & 26.0 & 2.36E-16 & 1.33E-17 & 521 & -88 \\
   & H$_2$ S(0) & 28.2 & 1.96E-16 & 7.70E-18 & 406 & -50 \\
   & [S \sc{iii}] & 33.5 & 1.37E-14 & 1.18E-16 & 440 & -84 \\
   & [Si \sc{ii}] & 34.8 & 4.31E-15 & 1.17E-16 & 499 & -52 \\

\\

V1-2 & [S \sc{iv}] & 10.5 & 1.01E-16 & 1.18E-17 & 503 & -77 \\
   & H$_2$ S(2) & 12.3 & 3.44E-16 & 7.99E-18 & 425 & -52 \\
   & H \sc{i} 7-6 & 12.4 & 1.93E-16 & 5.68E-18 & 493 & -19 \\
   & [Ne \sc{ii}] & 12.8 & 1.13E-14 & 4.22E-17 & 448 & -80 \\
   & [Ne \sc{iii}] & 15.6 & 3.21E-16 & 5.69E-18 & 461 & -53 \\
   & H$_2$ S(1) & 17.0 & 4.04E-16 & 1.03E-17 & 454 & 13 \\
   		&	[Fe \sc{ii}]	&	17.9	&	3.42E-17	&	5.90E-18	&	416	&	90	\\
   & [S \sc{iii}] & 18.7 & 9.59E-15 & 5.86E-17 & 488 & -18 \\
   & [Fe \sc{iii}] & 22.9 & 3.38E-16 & 1.16E-17 & 453 & -4 \\
   & [Fe \sc{ii}] & 26.0 & 2.18E-16 & 6.52E-18 & 449 & -79 \\
   & H$_2$ S(0) & 28.2 & 2.34E-16 & 8.03E-18 & 414 & -18 \\
   & [S \sc{iii}] & 33.5 & 1.39E-14 & 1.33E-16 & 436 & -84 \\
   & [Si \sc{ii}] & 34.8 & 4.41E-15 & 1.09E-16 & 495 & -53 \\
\\

V1-3 & [S \sc{iv}] & 10.5 & 1.19E-16 & 2.84E-17 & 722 & -9 \\
   & H$_2$ S(2) & 12.3 & 3.68E-16 & 9.68E-18 & 474 & -39 \\
   & H \sc{i} 7-6 & 12.4 & 1.82E-16 & 8.96E-18 & 461 & -17 \\
   & [Ne \sc{ii}] & 12.8 & 1.11E-14 & 1.33E-16 & 440 & -70 \\
   & [Ne \sc{iii}] & 15.6 & 2.94E-16 & 4.92E-18 & 454 & -41 \\
   & H$_2$ S(1) & 17.0 & 3.95E-16 & 8.95E-18 & 446 & 25 \\
   		&	[Fe \sc{ii}]	&	17.9	&	3.12E-17	&	8.61E-18	&	429	&	124	\\
   & [S \sc{iii}] & 18.7 & 9.51E-15 & 5.80E-17 & 483 & -7 \\
   & [Fe \sc{iii}] & 22.9 & 3.60E-16 & 2.06E-17 & 460 & 12 \\
   & [Fe \sc{ii}] & 26.0 & 2.49E-16 & 1.48E-17 & 466 & -39 \\
   & H$_2$ S(0) & 28.2 & 2.06E-16 & 1.48E-17 & 395 & -32 \\
   & [S \sc{iii}] & 33.5 & 1.42E-14 & 1.69E-16 & 445 & -54 \\
   & [Si \sc{ii}] & 34.8 & 4.59E-15 & 1.20E-16 & 504 & -18 \\
   
\\

V2-1 & [S \sc{iv}] & 10.5 & 1.03E-16 & 5.76E-18 & 485 & 72 \\
   & H$_2$ S(2) & 12.3 & 1.21E-15 & 7.38E-18 & 442 & -46 \\
   & H \sc{i} 7-6 & 12.4 & 2.52E-16 & 1.09E-17 & 508 & 8 \\
   & [Ne \sc{ii}] & 12.8 & 1.45E-14 & 5.00E-17 & 444 & -68 \\
   & [Ne \sc{iii}] & 15.6 & 1.76E-16 & 6.46E-18 & 473 & -36 \\
   & H$_2$ S(1) & 17.0 & 7.18E-16 & 1.32E-17 & 461 & 24 \\
   		&	[Fe \sc{ii}]	&	17.9	&	1.21E-16	&	9.98E-18	&	545	&	59	\\
   & [S \sc{iii}] & 18.7 & 8.19E-15 & 4.91E-17 & 487 & -8 \\
   & [Fe \sc{iii}] & 22.9 & 2.73E-16 & 9.90E-18 & 458 & 14 \\
   & [Fe \sc{ii}] & 26.0 & 5.42E-16 & 1.06E-17 & 470 & -30 \\
   & H$_2$ S(0) & 28.2 & 3.32E-16 & 1.14E-17 & 416 & -11 \\
   & [S \sc{iii}] & 33.5 & 1.09E-14 & 5.13E-17 & 450 & -31 \\
   & [Si \sc{ii}] & 34.8 & 8.87E-15 & 1.44E-16 & 506 & 11 \\
\\

V2-2 & [S \sc{iv}] & 10.5 & 1.02E-16 & 7.95E-18 & 534 & -55 \\
   & H$_2$ S(2) & 12.3 & 1.24E-15 & 7.91E-18 & 433 & -38 \\
   & H \sc{i} 7-6 & 12.4 & 2.43E-16 & 6.97E-18 & 477 & 6 \\
   & [Ne \sc{ii}] & 12.8 & 1.49E-14 & 4.58E-17 & 440 & -62 \\
   & [Ne \sc{iii}] & 15.6 & 1.65E-16 & 2.69E-18 & 424 & -54 \\
   & H$_2$ S(1) & 17.0 & 7.36E-16 & 8.13E-18 & 451 & 39 \\
   		&	[Fe \sc{ii}]	&	17.9	&	1.15E-16	&	9.91E-18	&	541	&	47	\\
   & [S \sc{iii}] & 18.7 & 8.47E-15 & 1.13E-16 & 482 & 1 \\
   & [Fe \sc{iii}] & 22.9 & 3.01E-16 & 1.13E-17 & 421 & 2 \\
   & [Fe \sc{ii}] & 26.0 & 4.92E-16 & 5.37E-18 & 423 & -47 \\
   & H$_2$ S(0) & 28.2 & 3.69E-16 & 8.66E-18 & 412 & -8 \\
   & [S \sc{iii}] & 33.5 & 1.13E-14 & 1.21E-16 & 445 & -46 \\
   & [Si \sc{ii}] & 34.8 & 8.89E-15 & 1.17E-16 & 501 & -5 \\
   
\\

V2-3 & [S \sc{iv}] & 10.5 & 9.37E-17 & 1.32E-17 & 692 & -84 \\
   & H$_2$ S(2) & 12.3 & 1.19E-15 & 1.05E-17 & 444 & -45 \\
   & H \sc{i} 7-6 & 12.4 & 2.65E-16 & 6.69E-18 & 546 & -38 \\
   & [Ne \sc{ii}] & 12.8 & 1.46E-14 & 5.81E-17 & 443 & -67 \\
   & [Ne \sc{iii}] & 15.6 & 1.59E-16 & 5.51E-18 & 434 & -54 \\
   & H$_2$ S(1) & 17.0 & 6.92E-16 & 9.84E-18 & 464 & 24 \\
   	&	[Fe \sc{ii}]	&	17.9	&	1.07E-16	&	6.82E-18	&	483	&	81	\\
   & [S \sc{iii}] & 18.7 & 8.21E-15 & 4.92E-17 & 486 & -7 \\
   & [Fe \sc{iii}] & 22.9 & 2.90E-16 & 4.09E-17 & 462 & -20 \\
   & [Fe \sc{ii}] & 26.0 & 5.08E-16 & 9.48E-18 & 431 & -70 \\
   & H$_2$ S(0) & 28.2 & 3.62E-16 & 9.67E-18 & 418 & -63 \\
   & [S \sc{iii}] & 33.5 & 1.15E-14 & 9.75E-17 & 450 & -91 \\
   & [Si \sc{ii}] & 34.8 & 8.96E-15 & 1.35E-16 & 500 & -56 \\

\\

V3-1 & [S \sc{iv}] & 10.5 & 2.15E-17$^a$  & --- & --- & --- \\
 & H$_2$ S(2) & 12.3 & 1.55E-15 & 2.84E-17 & 421 & -50 \\
   & H \sc{i} 7-6 & 12.4 & 6.63E-17$^a$   & --- & --- & --- \\
   & [Ne \sc{ii}] & 12.8 & 1.47E-15 & 4.40E-17 & 446 & -52 \\
   & [Ne \sc{iii}] & 15.6 & 8.50E-17 & 4.78E-18 & 442 & -18 \\
   & H$_2$ S(1) & 17.0 & 1.37E-15 & 1.52E-17 & 461 & 20 \\
   & [S \sc{iii}] & 18.7 & 6.81E-16 & 1.32E-17 & 458 & 23 \\
   & [Fe \sc{iii}] & 22.9 & 2.48E-17$^a$  & --- & --- & --- \\
   & [Fe \sc{ii}] & 26.0 & 4.15E-16 & 1.82E-17 & 478 & -100 \\
   & H$_2$ S(0) & 28.2 & 2.35E-16 & 1.03E-17 & 396 & -55 \\   
   & [S \sc{iii}] & 33.5 & 1.20E-15 & 2.43E-17 & 401 & -134 \\
   & [Si \sc{ii}] & 34.8 & 5.37E-15 & 1.76E-16 & 513 & -66 \\

\\

V3-2 & [S \sc{iv}] & 10.5 & 5.12E-17 & 8.90E-18 & 384 & -34 \\
   & H$_2$ S(2) & 12.3 & 1.63E-15 & 2.84E-17 & 422 & -53 \\
   & H \sc{i} 7-6 & 12.4 & 5.40E-17$^a$  & --- & --- & --- \\
   & [Ne \sc{ii}] & 12.8 & 1.42E-15 & 4.19E-17 & 445 & -58 \\
   & [Ne \sc{iii}] & 15.6 & 7.88E-17 & 4.40E-18 & 345 & -71 \\
   & H$_2$ S(1) & 17.0 & 1.45E-15 & 1.68E-17 & 461 & 17 \\
   & [S \sc{iii}] & 18.7 & 6.54E-16 & 9.78E-18 & 460 & 14 \\
    & [Fe \sc{iii}] & 22.9 & 4.11E-17$^a$  & --- & --- & --- \\
   & [Fe \sc{ii}] & 26.0 & 4.01E-16 & 1.86E-17 & 465 & -98 \\
   & H$_2$ S(0) & 28.2 & 2.53E-16 & 9.94E-18 & 411 & -41 \\
   & [S \sc{iii}] & 33.5 & 1.24E-15 & 2.31E-17 & 396 & -130 \\
   & [Si \sc{ii}] & 34.8 & 5.44E-15 & 1.43E-16 & 515 & -65 \\

\\

V3-3 & [S \sc{iv}] & 10.5 & 7.54E-17$^a$  & --- & --- & --- \\
   & H$_2$ S(2) & 12.3 & 1.54E-15 & 4.22E-17 & 419 & -54 \\
   & H \sc{i} 7-6 & 12.4 & 4.94E-17$^a$  & --- & --- & --- \\
   & [Ne \sc{ii}] & 12.8 & 1.51E-15 & 4.84E-17 & 449 & -54 \\
   & [Ne \sc{iii}] & 15.6 & 7.88E-17 & 4.66E-18 & 352 & -71 \\
   & H$_2$ S(1) & 17.0 & 1.38E-15 & 1.31E-17 & 465 & 17 \\
   & [S \sc{iii}] & 18.7 & 6.95E-16 & 1.02E-17 & 465 & 21 \\
    & [Fe \sc{iii}] & 22.9 & 5.80E-17$^a$  & --- & --- & --- \\
   & [Fe \sc{ii}] & 26.0 & 4.15E-16 & 2.06E-17 & 470 & -87 \\
   & H$_2$ S(0) & 28.2 & 2.51E-16 & 1.65E-17 & 437 & -57 \\
   & [S \sc{iii}] & 33.5 & 1.40E-15 & 3.69E-17 & 493 & -119 \\
   & [Si \sc{ii}] & 34.8 & 5.51E-15 & 1.51E-16 & 521 & -69 \\

\enddata

 $^a$ Intensity less than 3$\sigma$, considered an upper limit

\end{deluxetable}

\begin{deluxetable}{lcccl}
\tabletypesize{\scriptsize}
\tablecaption{Surface Brightness and Extinction Values}
\tablewidth{0pt}
\tablenum{3}
\tablehead{
\colhead{Sample Name} &
\colhead{Distance} &
\colhead{\cHb}&
\colhead{\sB(corrected)}&
\colhead{Comments} \\

& \colhead{(arcmin)}
& & \colhead{(ergs \cms\  \pers\ \perarc)}

}
\startdata
I4-S120             & 2.91     & 0.19  & 2.97x10$^{-13}$ & --- \\
I3-JW831            & 3.51    & 0.21 &  2.52x10$^{-13}$ & --- \\
I2I3-S150          & 3.48      & 0.14 & 2.09x10$^{-13}$ & --- \\
I2-JW831          & 3.95     & 0.20 &  2.16x10$^{-13}$ & --- \\
I1I2-S180      & 4.50   & 0.10 & 1.39x10$^{-13}$ & --- \\
I1I2-JW831        & 4.47      & 0.23 & 2.05x10$^{-13}$ & --- \\
M1M2M3-JW831    & 5.64        & 0.15 & 8.99x10$^{-14}$ & --- \\
M4                   & 7.04     &  0.05 & 4.80x10$^{-14}$ & --- \\
V1                  & 8.82    & 0.03 & 1.88x10$^{-14}$ & --- \\
V2-JW873-NE   & 10.59  & 0.11 & 2.30x10$^{-14}$ & ---\\
V2-JW887-E    & 10.94  & 0.13 & 2.07x10$^{-14}$ & ---\\
V2              & 10.44       & 0       & 1.47x10$^{-14}$ &  \Ha /\Hb=2.82\\
V2-JW887-W      & 9.86  & 0      & 1.93x10$^{-14}$ &\Ha /\Hb=2.82\\
V2-JW887-WW    & 9.47 & 0      &  1.17x10$^{-14}$ & \Ha /\Hb=2.73\\
V2-JW873-SW    & 10.34  & 0      & 1.73x10$^{-14}$ & \Ha /\Hb=2.78\\
V3                & 12.10       & 0.11 & 1.52x10$^{-14}$ & --- \\
\enddata
\tablecomments{\cHb\ is derived from the Blagrave~et~al.\ (2007) 
extinction curve, the observed \Ha /\Hb\ ratio and an assumed intrinsic 
ratio of 2.89, appropriate for the range of electron temperatures and 
densities in this paper.}
\end{deluxetable}

\begin{deluxetable}{lccccc}
\tabletypesize{\scriptsize}
\tablecaption{Observed and Extinction Corrected Line Ratios-1
\label{ratios1}}
\tablewidth{0pt}
\tablenum{4}
\tablehead{
\colhead{Region} &
\colhead{} & 
\colhead{I4-S120}  &
\colhead{I3-JW831} &
\colhead{I213-S150} &
\colhead{I2-JW831}\\
\colhead{$\lambda$~(\AA)} &
\colhead{Ion} &
\colhead{F$_{\lambda}$~~~~~~ I$_{\lambda}$} &
\colhead{F$_{\lambda}$~~~~~~ I$_{\lambda}$} &
\colhead{F$_{\lambda}$~~~~~~ I$_{\lambda}$} &
\colhead{F$_{\lambda}$~~~~~~ I$_{\lambda}$}}
\startdata
3869 & [Ne~III] & 0.0337 0.0359 &  & 0.0268 0.0278 &\\
4070 & [S~II]    & 0.0123 0.0130 &  & 0.0145 0.0151&\\
4102 & H I         & 0.215 0.226 & & 0.225 0.234&\\
4340 & H I         & 0.437 0.454 & 0.456 0.478 & 0.450 0.463& 0.438 0.456\\
4363 & [O~III]   & 0.0035 0.0036 & 0.0068 0.0071 & & \\
4471 & He I       & 0.0272 0.0280 & 0.0284 0.0293 & 0.0322 0.0329 & 0.0342 0.0352\\
4658 & [Fe~III] & 0.0077 0.0078 & 0.0088 0.0089 & 0.0108 0.0109 & 0.0113 0.0115\\
4861 & H I          & 1.000 1.000& 1.000 1.000& 1.000 1.000 & 1.000 1.000\\
4922 & He I       & 0.0068 0.0068 & 0.0053 0.0053 & 0.0083 0.0083 & 0.0085 0.0085\\
4959 & [O~III]   & 0.326 0.324 & 0.355 0.352 & 0.316 0.315& 0.283 0.281\\
5007 & [O~III]   & 0.997 0.987 & 1.066 1.055 & 0.960 0.953& 0.939 0.930\\
5048 & He I       &  & 0.0056 0.0055 & & \\
5056 & Si II       &  & 0.0036 0.0035 & 0.0027 0.0027 & 0.0023 0.0023\\
5199 & [N~I]    & 0.0073 0.0072 & 0.0069 0.0067 & 0.0082 0.0081& 0.0066 0.0064\\
5270 & [Fe~III] & 0.0052 0.0051 & 0.0048 0.0047 & 0.0048 0.0047 & 0.0063 0.0061\\
5518 & [Cl~III] & 0.0037 0.0035 & 0.0037 0.0035 & 0.0046 0.0045 & 0.0043 0.0041\\
5538 & [Cl~III] & 0.0025 0.0024 & 0.0030 0.0029 & 0.0048 0.0047& 0.0044 0.0042\\
5755 & [N~II]   & 0.0076 0.0072 & 0.0072 0.0068 & 0.0065 0.0062 & 0.0062 0.0059\\
5876 & He I      & 0.0951 0.0895 & 0.0943 0.0882 & 0.102 0.0976 & 0.106 0.0998\\
5979 & Si II       & 0.0018 0.0017 & 0.0018 0.0017 & & 0.0018 0.0017\\
6300 & [O~I]    & 0.0060 0.0055 & 0.0063 0.0057 & 0.0045 0.0042 & 0.0044 0.0040\\
6312 & [S~III]  & 0.0109 0.0100 & 0.0115 0.0105 & 0.0115 0.0108 & 0.0123 0.0113\\
6347 & Si II      & 0.0039 0.0036 & 0.0036 0.0033 & 0.0026 0.0025 & 0.0031 0.0028\\
6363 & [O~I]   & 0.0022 0.0020 & 0.0025 0.0023 & & 0.0014 0.0013 \\
6371 & Si II      & 0.0030 0.0028 & 0.0026 0.0024 & & 0.0036 0.0033\\
6548 & [N~II]  & 0.347 0.315 & 0.282 0.254 & 0.304 0.283 & 0.275 0.249\\
6563 & H I       & 3.188 2.895 & 3.210 2.886 & 3.097 2.885 & 3.198 2.890\\
6583 & [N~II]  & 0.913 0.828 & 0.913 0.820 & 0.833 0.775 & 0.805 0.727\\
6678 & He I     & 0.0265 0.0239 & 0.0264 0.0236 & 0.0305 0.0283 & 0.0300 0.0269\\
6716 & [S~II]  & 0.126 0.113 & 0.129 0.115 & 0.122 0.113 & 0.120 0.108\\
6731 & [S~II]  & 0.147 0.132 & 0.143 0.127 & 0.127 0.118 & 0.123 0.110\\
7065 & He I    & 0.0285 0.0252 &  & 0.0392 0.0358 & \\
7136 & [Ar~III] & 0.0838 0.0739& & 0.0924 0.0843 & \\
\enddata
\end{deluxetable}

\begin{deluxetable}{lcccc}
\tabletypesize{\scriptsize}
\tablecaption{Observed and Extinction Corrected Line Ratios-2
\label{ratios2}}
\tablewidth{0pt}
\tablenum{5}
\tablehead{
\colhead{Region} &
\colhead{} &
\colhead{I1I2-S180}  &
\colhead{I1I2-JW831} &
\colhead{M1M2M3-JW831}\\
\colhead{$\lambda$~(\AA)} &
\colhead{Ion} &
\colhead{F$_{\lambda}$~~~~~~ I$_{\lambda}$} &
\colhead{F$_{\lambda}$~~~~~~ I$_{\lambda}$} &
\colhead{F$_{\lambda}$~~~~~~ I$_{\lambda}$}}
\startdata
3869 & [Ne~III] & 0.0365 0.0377 & & \\
4070 & [S~II]    &  0.0233 0.0229 & & \\
4102 & H I         &  0.222 0.228 & &  \\
4340 & H I         &  0.434 0.433 & 0.420 0.455 & 0.433 0.466 \\
4363 & [O~III]   &  0.0068 0.0069 &  &  \\
4471 & He I       &  0.0336 0.0341 & 0.0310 0.0321 & 0.0278 0.0284\\
4658 & [Fe~III] &  0.0098 0.0099 & 0.0092 0.0093 & 0.0082 0.0083\\
4861 & H I         & 1.000 1.000 & 1.000  1.000 & 1.000 1.000\\
4922 & He I       &  0.0081 0.0081 & 0.0053 0.0053 & 0.0079 0.0079\\
4959 & [O~III]   &  0.318 0.318 & 0.287 0.285 & 0.283 0.282\\
5007 & [O~III]   &  0.867 0.863 & 0.877 0.867 & 0.848 0.842\\
5042 & Si II       &                & 0.0063 0.0063 & 0.0022 0.0022 \\
5056 & Si II       &                 & 0.0044 0.0043 & 0.0042 0.0041 \\
5199 & [N~I]    & 0.0068 0.0067 & 0.0050 0.0049 & 0.0069 0.0068 \\
5270 & [Fe~III] &                  & 0.0060 0.0058 & 0.0057 0.0056 \\
5518 & [Cl~III] &  0.0033 0.0032 & 0.0050 0.0048 & 0.0038 0.0037\\
5538 & [Cl~III] &  0.0033 0.0032 & 0.0042 0.0040 & 0.0038 0.0037 \\
5755 & [N~II]   &  0.0075 0.0073 & 0.0070 0.0066 & 0.0073 0.0070 \\
5876 & He I      &  0.103 0.0993 & 0.0995 0.0925 & 0.0883 0.0842 \\
5979 & Si II       &                        &              & 0.0019 0.0018
\\
6300 & [O~I]    &  0.0052 0.0050 & 0.0032 0.0029 & 0.0023 0.0022\\
6312 & [S~III]  &  0.0106 0.0101 & 0.0119 0.0108 & 0.0116 0.0109\\
6347 & Si II      &  0.0033 0.0032 & 0.0035 0.0032 & 0.0024 0.0022 \\
6363 & [O~I]   &  0.0014 0.0013 & 0.0012 0.0011 & 0.0010 0.0009\\
6371 & Si II      & 0.0031 0.0030 & 0.0029 0.0026 & 0.0020 0.0019 \\
6548 & [N~II]  &  0.352 0.335 & 0.325 0.290 & 0.330 0.306 \\
6563 & H I       &  3.052 2.901 & 3.256 0.2909 & 3.120 2.892 \\
6583 & [N~II]  &  0.849 0.807 & 0.895 0.855 & 0.952 0.882 \\
6678 & He I     &  0.0310 0.0294 & 0.0278 0.0246 & 0.0234 0.0216 \\
6716 & [S~II]  & 0.123 0.120  & 0.154 0.136 & 0.153 0.141 \\
6731 & [S~II]  & 0.117 0.111 & 0.154 0.136 & 0.138 0.127 \\
7065 & He I    & 0.0286 0.0269 & &   \\
7136 & [Ar~III] & 0.0893 0.0837 & & \\
\enddata
\end{deluxetable}

\begin{deluxetable}{lccccccc}
\tabletypesize{\scriptsize}
\tablecaption{Observed and Extinction Corrected Line Ratios-3 Near Position
V2*
\label{ratios3}}
\tablewidth{0pt}
\tablenum{6}
\tablehead{
\colhead{Region} &
\colhead{} &
\colhead{JW873-NE}  &
\colhead{JW887-E} &
\colhead{V2} &
\colhead{JW887-W} &
\colhead{JW887-WW} &
\colhead{JW873-SW}\\
\colhead{$\lambda$~(\AA)} &
\colhead{Ion} &
\colhead{F$_{\lambda}$~~~~~~ I$_{\lambda}$} &
\colhead{F$_{\lambda}$~~~~~~ I$_{\lambda}$} &
\colhead{F$_{\lambda}$} &
\colhead{F$_{\lambda}$} &
\colhead{F$_{\lambda}$} &
\colhead{F$_{\lambda}$}}
\startdata
4340 & H I         &  0.443 0.453 & 0.448 0.460 & 0.438 & 0.466 & 0.469 &
0.456 \\
4363 & [O~III]   &                             &                       &
0.0158  \\
4471 & He I       &  0.0282 0.0287 & 0.0270 0.0275 & 0.0200 & 0.0115 &
0.0130 & 0.0091 \\
4658 & [Fe~III] &  0.0082 0.0083 & 0.0072 0.0073 & 0.0052  & 0.0088 & 0.0097
& 0.0073 \\
4861 & H I & 1.000 1.000           & 1.000  1.000   & 1.000         & 1.000
& 1.000 & 1.000 \\
4922 & He I       &  0.0055 0.0055 & 0.0048 0.0048 & 0.0057 & 0.0019 &
0.0027 & 0.0035 \\
4959 & [O~III]   &  0.490 0.488 & 0.505 0.503 & 0.410 & 0.197 & 0.282 &
0.175 \\
4986 & [Fe~III] &  &  &  & 0.0034 & 0.0041 & 0.0049 \\
5007 & [O~III]   &  1.622 1.613 & 1.640 1.629 & 1.223 & 0.579 & 0.819 &
0.518 \\
5042 & Si II       &             &      & 0.0079 & 0.026 & & 0.0035 \\
5056 & Si II       &             &      & 0.0046 &0.0030 & & 0.0018 \\
5199 & [N~I]    &   0.0305 0.0301 & 0.0286 0.0281 & 0.0598 & 0.0270 & 0.0150
& 0.0278 \\
5262 & [Fe~II] & 0.0025 0.0025 & 0.0021 0.0021 & 0.0045 & 0.0023 & & 0.0023
\\
5270 & [Fe~III] & 0.0063 0.0062 & 0.0066 0.0065 & 0.0079 & 0.055 & 0.0047 &
0.0054  \\
5518 & [Cl~III] & 0.0040 0.0039 & 0.0057 0.0044 & 0.0041 &0.0039 & 0.0047 &
0.0040 \\
5538 & [Cl~III] & 0.0030 0.0029 & 0.0057 0.0054 & 0.0039 & 0.0031 & 0.0038 &
0.0025  \\
5755 & [N~II]   & 0.0103 0.0100 & 0.0109 0.0105 & 0.0103 &0.0153 & 0.0152 &
0.0142 \\
5876 & He I      &  0.0925 0.0871 & 0.0977 0.0933 & 0.0675 &0.0314 & 0.0443
& 0.0300 \\
5979 & Si II       &  0.0029 0.0028 & 0.0022 0.0021 & 0.0048 & 0.0024 &
0.0020 & 0.0027 \\
6046 & O I         & 0.0019 0.0018 & 0.0014 0.0013 & 0.0038 & & & 0.0014 \\
6300 & [O~I]    & 0.0112 0.0107 & 0.0130 0.0123 & 0.0246 &0.0433 & &
0.0293\\
6312 & [S~III]  & 0.0144 0.0137 & 0.0152 0.0144 & 0.0135  &0.0136 & &
0.0125\\
6347 & Si II      &  0.0038 0.0036 & 0.0043 0.0041 & 0.0061 &0.0042 & 0.0026
& 0.0031 \\
6363 & [O~I]   & 0.0036 0.0034 & 0.0043 0.0041 & 0.0077&0.0136 & 0.006 &
0.0088  \\
6371 & Si II      & 0.0034 0.0032 & 0.0033 0.0031 & 0.0055 & 0.0036 & 0.0020
& 0.0033 \\
6548 & [N~II]  &  0.366 0.346 & 0.356 0.344 & 0.375 &0.488 & 0.413 & 0.477\\
6563 & H I       &   3.106 2.938 & 3.065 2.870 & 2.821& 2.816 & 2.734 &
2.780  \\
6583 & [N~II]  &  1.065 0.994 & 0.972 0.910 & 1.083 &1.468 & 1.150 & 1.457\\
6678 & He I     &  0.0255 0.0241 & 0.0267 0.0249 & 0.0204 &0.0075 & 0.0098 &
0.0071\\
6716 & [S~II]  &  0.212 0.200 & 0.211 0.196 & 0.282 & 0.510 & 0.221 & 0.459
\\
6731 & [S~II]  &  0.176 0.169 & 0.190 0.177 & 0.227 & 0.422 & 0.175 & 0.362
\\
\enddata
\tablecomments{*Where only the observed flux ratios (F$_{\lambda}$) are
shown, no extinction could be determined since the \Ha /\Hb\  ratio was less
than the theoretically expected value.}
\end{deluxetable}

\begin{deluxetable}{lcccc}
\tabletypesize{\scriptsize}
\tablecaption{Observed and Extinction Corrected Line Ratios-4 2009 Observations
\label{ratios4}}
\tablewidth{0pt}
\tablenum{7}
\tablehead{
\colhead{Region} &
\colhead{} &
\colhead{M4}  &
\colhead{V1} &
\colhead{V3}\\
\colhead{$\lambda$~(\AA)} &
\colhead{Ion} &
\colhead{F$_{\lambda}$~~~~~~ I$_{\lambda}$} &
\colhead{F$_{\lambda}$~~~~~~ I$_{\lambda}$} &
\colhead{F$_{\lambda}$~~~~~~ I$_{\lambda}$}}
\startdata
3869 & [Ne~III] & 0.0405 0.0412 & 0.0439 0.0443 & 0.1050 0.1091 \\
4070 & [S~II]    &  0.0157 0.0159 & 0.0251 0.0253 & 0.0304 0.0314 \\
4102 & H~I         &  0.2307 0.2340 & 0.2418 0.2436 & 0.2204 0.2273  \\
4340 & H~I         &  0.4470 0.4516 & 0.4550 0.4574 & 0.4213 0.4309 \\
4363 & [O~III]   &  0.0020 0.0020 & 0.0048 0.0048 & 0.0116 0.0119 \\
4471 & He~I       &  0.0256 0.0258 & 0.0120 0.0120 & 0.0271 0.0276 \\
4658 & [Fe~III] &   0.0084 0.0084 & 0.0089 0.0089 & 0.0076 0.0077 \\
4861 & H~I &      1.000 1.000           & 1.000 1.000   & 1.000 1.000     \\
4922 & He~I       &   0.0057 0.0057 & 0.0033 0.0033 & 0.0046 0.0046 \\
4959 & [O~III]   &   0.2029 0.2025 & 0.1926 0.1924 & 0.6034 0.6010 \\
5007 & [O~III]   &   0.6269 0.6253 & 0.5917 0.5925 & 1.8118 1.8010 \\
5199 & [N~I]    &    0.0084 0.0083 & 0.0198 0.0198 & 0.0269 0.0265 \\
5270 & [Fe~III] &   0.0052 0.0051 & 0.0053 0.0053 & 0.0032 0.0031 \\
5518 & [Cl~III] &    0.0045 0.0045 & 0.0030 0.0030 & 0.0052 0.0051 \\
5538 & [Cl~III] &    0.0029 0.0029 & 0.0024 0.0024 & 0.0036 0.0035 \\
5755 & [N~II]   &    0.0085 0.0084 & 0.0141 0.0140 & 0.0102 0.0097 \\
5876 & He~I      &    0.0791 0.0778 & 0.0377 0.0374 & 0.1080 0.1042  \\
5979 & Si~II       &    0.0018 0.0018 & 0.0025 0.0025 & 0.0039 0.0037 \\
6300 & [O~I]    &    0.0085 0.0083 & 0.0108 0.0107 & 0.0251 0.0239 \\
6312 & [S~III]  &     0.0111 0.0109 & 0.0121 0.0120 & 0.0138 0.0131 \\
6347 & Si~II      &     0.0018 0.0018 & 0.0024 0.0024 & 0.0032 0.0030 \\
6363 & [O~I]   &     0.0024 0.0024 & 0.0028 0.0028 & 0.0084 0.0080 \\
6371 & Si~II      &     0.0014 0.0014 & 0.0019 0.0019 & 0.0017 0.0016 \\
6548 & [N~II]  &     0.3425 0.3337 & 0.4480 0.4420 & 0.2829 0.2672 \\
6563 & H~I       &      2.9673 2.89     & 2.9292 2.89 & 3.0625 2.89  \\
6583 & [N~II]  &      1.1022 1.0730 & 1.3934 1.3890 & 0.8438 0.7960\\
6678 & He~I     &       0.0205 0.0199 & 0.0108 0.0106 & 0.0320 0.0301\\
6716 & [S~II]  &       0.1669 0.1622 & 0.3105 0.3059 & 0.1759 0.1652 \\
6731 & [S~II]  &       0.1394 0.1355 & 0.2387 0.2352 & 0.1509 0.1417\\
7065 & He~I    &        0.0188 0.0182 & 0.0129 0.0127 & 0.0401 0.0373 \\
7136 & [Ar~III] &      0.0617 0.0596 & 0.0346 0.0340 & 0.1145 0.1062 \\
\enddata
\end{deluxetable}

\begin{deluxetable}{lcccccc}
\tabletypesize{\scriptsize}
\tablecaption{Electron Densities and Temperatures Derived from Optical Lines}
\tablewidth{0pt}
\tablenum{8}
\tablehead{
\colhead{Sample Name} &
\colhead{Distance} &
\colhead{\Ne [S~II]} &
\colhead{\Te [N~II]} &
\colhead{\Te [O~III]} \\

& \colhead{(arcmin)}  & \colhead{(\cmq)} & \colhead{(K)} & \colhead{(K)}

}
\startdata
I4-S120        	 & 	2.91	 & 	714	 & 	8270	 & 	   8550\\
I3-JW831                      	 & 	3.51	 & 	586	 & 	8280	 & 	 10150\\
I2I3-S150                     	 & 	3.48	 & 	478	 & 	8070	 & 	  ---      \\
I2-JW831                  	 & 	3.95	 & 	434	 & 	8160	 & 	 ---         \\
I1I2-S180	 & 	4.5	 & 	290	 & 	8310	 & 	 10630 \\
I1I2-JW831	 & 	4.47	 & 	404	 & 	7950	 & 	 ---         \\
M1M2M3-JW831        	 & 	5.64	 & 	256	 & 	8090	 & 	 ---         \\
M4 	 & 	7.04	 & 	168	 & 	8230	 & 	 8280 \\
V1 	 & 	8.82	 & 	87	 & 	8930	 & 	 10780 \\
V2-JW873-NE     	 & 	10.59	 & 	181	 & 	8890	 & 	 ---         \\
V2-JW887-E         	 & 	10.94	 & 	259	 & 	9260	 & 	 ---     \\
V2-Combined        	 & 	10.44	 & 	131	 & 	8880	 & 	 13400 \\
V2-JW887-W        	 & 	9.86	 & 	158	 & 	9030	 & 	 ---    \\
V2-JW887-WW    	 & 	9.47	 & 	114	 & 	8970	 & 	 ---  \\
V2-JW873-SW     	 & 	10.34	 & 	111	 & 	8830	 & 	     ---   \\
V3                          	 & 	12.08	 & 	197	 & 	9580	 & 	 10100  \\
\enddata
\end{deluxetable}

\begin{deluxetable}{ccccccr}    
\tabletypesize{\scriptsize}
\tablecaption{Electron Densities from [S {\sc iii}] Infrared Lines}
\tablewidth{0pt}
\tablenum{9}
\tablehead{
\colhead {Chex}
& \colhead{$N_e$ [S \sc{iii}]} \\
& \colhead{(\cmq)}
}

\startdata
I4 & 1041                  \\                                               
I3 & 755\\
I2 & 698\\
I1 & 572\\
M1 & 487\\
M2 & 403\\
M3 & 378\\
M4 & 365\\
V1-1 & 276\\
V1-2 & 273\\
V1-3 & 249\\
V2-1 & 337\\
V2-2 & 335\\
V2-3 & 297\\
V3-1 & 143\\
V3-2 & 105\\
V3-3 & 74\\
\enddata

\end{deluxetable}

\begin{deluxetable}{lccccccccccc} 
\rotate
\setlength{\tabcolsep}{0.04in}
\tabletypesize{\scriptsize}
\tablecaption{Derived Parameters for M42}
\tablewidth{0pt}
\tablenum{10}
\tablehead{
 \colhead{Chex}
& \colhead{{\underbar{Ne$^+$}}}
& \colhead{\underbar{Ne$^{++}$}}
& \colhead{\underbar{S$^{++}$}}
& \colhead{\underbar{S$^{3+}$}}
& \colhead{\underbar{Ne$^{++}$}}
& \colhead{\underbar{S$^{3+}$}}
& \colhead{\underbar{Ne~$^a$}}
&\colhead{\underbar{Fe$^{++}$}}
&\colhead{\underbar{Fe$^{+}$}}
&\colhead{\underbar{Fe$^{++}$}}
& \colhead{\underbar{Si$^{+}$}}
\\

& \colhead{H$^+$}
& \colhead{H$^+$}
& \colhead{H$^+$}
& \colhead{H$^+$}
& \colhead{Ne$^+$}
& \colhead{S$^{++}$}
& \colhead{S}
& \colhead{Fe$^+$}
& \colhead{H}
& \colhead{H}
& \colhead{Ne}
\\
& \colhead{$(\times$10$^{-6}$$)$}
& \colhead{$(\times$10$^{-6}$$)$}
& \colhead{$(\times$10$^{-6}$$)$}
& \colhead{$(\times$10$^{-8}$$)$}
& \colhead{$(\times$10$^{-3}$$)$}
& \colhead{$(\times$10$^{-3}$$)$}
& & 
&\colhead{$(\times$10$^{-6}$$)$}
& \colhead{$(\times$10$^{-6}$$)$}
& \colhead{$(\times$10$^{-2}$$)$}
}

\startdata

I4        &        101        $\pm$        9        &        8.59        $\pm$        0.76        &        6.66        $\pm$        0.59        &        13.1        $\pm$        1.2        &        84.8        $\pm$        0.7        &        19.7        $\pm$        0.4        &        16.2        $\pm$        0.1        &        1.68        $\pm$        0.26        &        0.584        $\pm$        0.049        &        0.980        $\pm$        0.131        &        5.10        $\pm$        0.40        \\
I3        &        99        $\pm$        9        &        6.70        $\pm$        0.60        &        6.58        $\pm$        0.59        &        8.27        $\pm$        0.75        &        67.4        $\pm$        0.5        &        12.6        $\pm$        0.3        &        15.9        $\pm$        0.1        &        1.38        $\pm$        0.14        &        0.573        $\pm$        0.032        &        0.791        $\pm$        0.067        &        4.44        $\pm$        0.27        \\
I2        &        100        $\pm$        9        &        4.05        $\pm$        0.35        &        6.59        $\pm$        0.57        &        4.74        $\pm$        0.43        &        40.7        $\pm$        0.4        &        7.18        $\pm$        0.19        &        15.6        $\pm$        0.1        &        1.80        $\pm$        0.11        &        0.523        $\pm$        0.024        &        0.941        $\pm$        0.037        &        3.14        $\pm$        0.18        \\
I1        &        95        $\pm$        9        &        3.19        $\pm$        0.29        &        6.50        $\pm$        0.59        &        3.40        $\pm$        0.34        &        33.7        $\pm$        0.7        &        5.23        $\pm$        0.23        &        15.0        $\pm$        0.3        &        3.33        $\pm$        0.17        &        0.311        $\pm$        0.014        &        1.037        $\pm$        0.027        &        2.73        $\pm$        0.14        \\

M1        &        96        $\pm$        8.3        &        4.05        $\pm$        0.35        &        6.60        $\pm$        0.57        &        3.66        $\pm$        0.34        &        42.2        $\pm$        0.4        &        5.54        $\pm$        0.19        &        15.1        $\pm$        0.1        &        3.43        $\pm$        0.16        &        0.286        $\pm$        0.012        &        0.982        $\pm$        0.016        &        2.26        $\pm$        0.12        \\
M2        &        89.1        $\pm$        8.6        &        2.96        $\pm$        0.28        &        6.21        $\pm$        0.59        &        2.51        $\pm$        0.26        &        33.2        $\pm$        0.7        &        4.04        $\pm$        0.18        &        14.8        $\pm$        0.3        &        3.14        $\pm$        0.16        &        0.306        $\pm$        0.014        &        0.962        $\pm$        0.023        &        2.74        $\pm$        0.12        \\
M3        &        85.5        $\pm$        9        &        2.07        $\pm$        0.22        &        6.23        $\pm$        0.65        &        1.62        $\pm$        0.20        &        24.3        $\pm$        0.4        &        2.60        $\pm$        0.18        &        14.0        $\pm$        0.2        &        3.92        $\pm$        0.17        &        0.289        $\pm$        0.012        &        1.133        $\pm$        0.019        &        2.99        $\pm$        0.10        \\
M4        &        100        $\pm$        9        &        1.52        $\pm$        0.14        &        6.84        $\pm$        0.62        &        1.64        $\pm$        0.27        &        15.2        $\pm$        0.3        &        2.40        $\pm$        0.33        &        14.8        $\pm$        0.1        &        4.00        $\pm$        0.21        &        0.303        $\pm$        0.015        &        1.214        $\pm$        0.023        &        2.98        $\pm$        0.07        \\

V1-1        &        101        $\pm$        9        &        1.34        $\pm$        0.13        &        5.89        $\pm$        0.54        &        0.940        $\pm$        0.100        &        13.2        $\pm$        0.4        &        1.60        $\pm$        0.09        &        17.4        $\pm$        0.1        &        2.04        $\pm$        0.14        &        0.423        $\pm$        0.024        &        0.864        $\pm$        0.031        &        3.94        $\pm$        0.11        \\
V1-2        &        105        $\pm$        9        &        1.44        $\pm$        0.13        &        6.04        $\pm$        0.54        &        1.48        $\pm$        0.22        &        13.7        $\pm$        0.2        &        2.45        $\pm$        0.29        &        17.6        $\pm$        0.1        &        2.08        $\pm$        0.09        &        0.396        $\pm$        0.012        &        0.822        $\pm$        0.028        &        3.93        $\pm$        0.10        \\
V1-3        &        109        $\pm$        11        &        1.40        $\pm$        0.14        &        6.40        $\pm$        0.63        &        1.84        $\pm$        0.48        &        12.8        $\pm$        0.3        &        2.88        $\pm$        0.69        &        17.3        $\pm$        0.2        &        1.95        $\pm$        0.16        &        0.475        $\pm$        0.028        &        0.926        $\pm$        0.053        &        4.04        $\pm$        0.12        \\
V2-1        &        103        $\pm$        10        &        0.602        $\pm$        0.061        &        3.86        $\pm$        0.37        &        1.15        $\pm$        0.13        &        5.86        $\pm$        0.22        &        2.98        $\pm$        0.17        &        26.7        $\pm$        0.2        &        0.656        $\pm$        0.027        &        0.773        $\pm$        0.015        &        0.507        $\pm$        0.018        &        6.69        $\pm$        0.11        \\
V2-2        &        110        $\pm$        10        &        0.588        $\pm$        0.053        &        4.15        $\pm$        0.37        &        1.19        $\pm$        0.14        &        5.35        $\pm$        0.09        &        2.87        $\pm$        0.23        &        26.5        $\pm$        0.3        &        0.799        $\pm$        0.031        &        0.729        $\pm$        0.008        &        0.583        $\pm$        0.022        &        6.51        $\pm$        0.09        \\
V2-3        &        98        $\pm$        9        &        0.517        $\pm$        0.049        &        3.73        $\pm$        0.33        &        1.00        $\pm$        0.17        &        5.26        $\pm$        0.18        &        2.68        $\pm$        0.38        &        26.5        $\pm$        0.2        &        0.755        $\pm$        0.107        &        0.677        $\pm$        0.013        &        0.512        $\pm$        0.072        &        6.41        $\pm$        0.10        \\
V3-1        &        $\ge$        40.0                &        $\ge$        1.11                &        $\ge$        1.33                &                ---                &        27.9        $\pm$        1.8        &        $\le$        6.9                &        $\ge$        30.8                &        $\le$        0.0851                &        $\le$        2.05                &                ---                &        30.5        $\pm$        1.3        \\
V3-2        &        $\ge$        47.4                &        $\ge$        1.27                &        $\ge$        1.60                &        $\ge$        2.67                &        26.8        $\pm$        1.7        &        16.7        $\pm$        2.9        &        29.9        $\pm$        1.0        &        $\le$        0.149                &        $\le$        2.38                &                ---                &        30.3        $\pm$        1.2        \\
V3-3        &        $\ge$        55.6                &        $\ge$        1.39                &        $\ge$        1.91                &                ---                &        25.1        $\pm$        1.7        &        $\le$        22.5                &        $\ge$        29.2                &        $\le$        0.207                &        $\le$        2.64                &                ---                &        27.5        $\pm$        1.1        \\
\enddata

\raggedright{~~~~$^a$~~(Ne$^{+}$+Ne$^{++}$)/(S$^{++}$+S$^{3+}$). See text for correction for S$^+$.}

\end{deluxetable}

\end{document}